\input harvmac

\input amssym
\input epsf

\def\unit{\relax{\rm 1\kern-.26em I}}
\def\nada{\relax{\rm 0\kern-.30em l}}
\def\tilde{\widetilde}

\def\alphadot{{\dot \alpha}}


\def \la {\langle}
\def \ra {\rangle}
\def \pa {\partial}

\def \eps {\epsilon}


\noblackbox
\def\IL{\relax{\rm I\kern-.18em L}}
\def\IH{\relax{\rm I\kern-.18em H}}
\def\IR{\relax{\rm I\kern-.18em R}}
\def\IC{\relax\hbox{$\inbar\kern-.3em{\rm C}$}}
\def\IZ{\relax\ifmmode\mathchoice
{\hbox{\cmss Z\kern-.4em Z}}{\hbox{\cmss Z\kern-.4em Z}} {\lower.9pt\hbox{\cmsss Z\kern-.4em Z}}
{\lower1.2pt\hbox{\cmsss Z\kern-.4em Z}}\else{\cmss Z\kern-.4em Z}\fi}

\def\CN {{\cal N}}

\def\CF {{\cal F}}

\def\CO {{\cal O}}

\def\CA{{\cal A}}


\def\CN {{\cal N}}

\def\CO {{\cal O}}

\font\manual=manfnt \def\dbend{\lower3.5pt\hbox{\manual\char127}}

\def\IZ{\relax\ifmmode\mathchoice
{\hbox{\cmss Z\kern-.4em Z}}{\hbox{\cmss Z\kern-.4em Z}} {\lower.9pt\hbox{\cmsss Z\kern-.4em Z}}
{\lower1.2pt\hbox{\cmsss Z\kern-.4em Z}}\else{\cmss Z\kern-.4em Z}\fi}
\def\half {{1\over 2}}

\def\bar{\overline}

\def\pa{\partial}

\def\rt2{\sqrt{2}}
\def\irt2{{1\over\sqrt{2}}}

\def\hat{\widehat}
\def\slashchar#1{\setbox0=\hbox{$#1$}           
   \dimen0=\wd0                                 
   \setbox1=\hbox{/} \dimen1=\wd1               
   \ifdim\dimen0>\dimen1                        
      \rlap{\hbox to \dimen0{\hfil/\hfil}}      
      #1                                        
   \else                                        
      \rlap{\hbox to \dimen1{\hfil$#1$\hfil}}   
      /                                         
   \fi}

\def\foursqr#1#2{{\vcenter{\vbox{
    \hrule height.#2pt
    \hbox{\vrule width.#2pt height#1pt \kern#1pt
    \vrule width.#2pt}
    \hrule height.#2pt
    \hrule height.#2pt
    \hbox{\vrule width.#2pt height#1pt \kern#1pt
    \vrule width.#2pt}
    \hrule height.#2pt
        \hrule height.#2pt
    \hbox{\vrule width.#2pt height#1pt \kern#1pt
    \vrule width.#2pt}
    \hrule height.#2pt
        \hrule height.#2pt
    \hbox{\vrule width.#2pt height#1pt \kern#1pt
    \vrule width.#2pt}
    \hrule height.#2pt}}}}
\def\psqr#1#2{{\vcenter{\vbox{\hrule height.#2pt
    \hbox{\vrule width.#2pt height#1pt \kern#1pt
    \vrule width.#2pt}
    \hrule height.#2pt \hrule height.#2pt
    \hbox{\vrule width.#2pt height#1pt \kern#1pt
    \vrule width.#2pt}
    \hrule height.#2pt}}}}
\def\sqr#1#2{{\vcenter{\vbox{\hrule height.#2pt
    \hbox{\vrule width.#2pt height#1pt \kern#1pt
    \vrule width.#2pt}
    \hrule height.#2pt}}}}
\def\square{\mathchoice\sqr65\sqr65\sqr{2.1}3\sqr{1.5}3}

\def\figin{\epsfcheck\figin}\def\figins{\epsfcheck\figins}
\def\epsfcheck{\ifx\epsfbox\UnDeFiNeD
\message{(NO epsf.tex, FIGURES WILL BE IGNORED)}
\gdef\figin##1{\vskip2in}\gdef\figins##1{\hskip.5in}
\else\message{(FIGURES WILL BE INCLUDED)}%
\gdef\figin##1{##1}\gdef\figins##1{##1}\fi}
\def\DefWarn#1{}
\def\figinsert{\goodbreak\midinsert}
\def\ifig#1#2#3{\DefWarn#1\xdef#1{fig.~\the\figno}
\writedef{#1\leftbracket fig.\noexpand~\the\figno}%
\figinsert\figin{\centerline{#3}}\medskip\centerline{\vbox{\baselineskip12pt \advance\hsize by
-1truein\noindent\footnotefont{\bf Fig.~\the\figno:\ } \it#2}}
\bigskip\endinsert\global\advance\figno by1}

\lref\GiddingsGJ{
  S.~B.~Giddings and R.~A.~Porto,
  ``The Gravitational S-matrix,''
Phys.\ Rev.\ D {\bf 81}, 025002 (2010).
[arXiv:0908.0004 [hep-th]].
}

\lref\NachtmannMR{
  O.~Nachtmann,
  ``Positivity constraints for anomalous dimensions,''
Nucl.\ Phys.\ B {\bf 63}, 237 (1973)..
}

\lref\HeemskerkPN{
  I.~Heemskerk, J.~Penedones, J.~Polchinski and J.~Sully,
  ``Holography from Conformal Field Theory,''
JHEP {\bf 0910}, 079 (2009).
[arXiv:0907.0151 [hep-th]].
}

\lref\FitzpatrickZM{
  A.~L.~Fitzpatrick, E.~Katz, D.~Poland and D.~Simmons-Duffin,
  ``Effective Conformal Theory and the Flat-Space Limit of AdS,''
JHEP {\bf 1107}, 023 (2011).
[arXiv:1007.2412 [hep-th]].
}

\lref\NachtmannMR{
  O.~Nachtmann,
  ``Positivity constraints for anomalous dimensions,''
Nucl.\ Phys.\ B {\bf 63}, 237 (1973).
}

\lref\BanksBJ{
  T.~Banks and G.~Festuccia,
  ``The Regge Limit for Green Functions in Conformal Field Theory,''
JHEP {\bf 1006}, 105 (2010).
[arXiv:0910.2746 [hep-th]].
}

\lref\Eden{
  R.~J.~Eden, P.~V.~Landshoff, D.~I.~Olive and J.~C.~Polkinghorne,
  ``The analytic S matrix''
}

\lref\PolyakovGS{
  A.~M.~Polyakov,
  ``Nonhamiltonian approach to conformal quantum field theory,''
Zh.\ Eksp.\ Teor.\ Fiz.\  {\bf 66}, 23 (1974).
}

\lref\HofmanAR{
  D.~M.~Hofman and J.~Maldacena,
  ``Conformal collider physics: Energy and charge correlations,''
JHEP {\bf 0805}, 012 (2008).
[arXiv:0803.1467 [hep-th]].
}

\lref\AldayMF{
  L.~F.~Alday and J.~M.~Maldacena,
  ``Comments on operators with large spin,''
JHEP {\bf 0711}, 019 (2007).
[arXiv:0708.0672 [hep-th]].
}

\lref\AldayZY{
  L.~F.~Alday, B.~Eden, G.~P.~Korchemsky, J.~Maldacena and E.~Sokatchev,
  ``From correlation functions to Wilson loops,''
JHEP {\bf 1109}, 123 (2011).
[arXiv:1007.3243 [hep-th]].
}

\lref\DolanDV{
  F.~A.~Dolan and H.~Osborn,
  ``Conformal Partial Waves: Further Mathematical Results,''
[arXiv:1108.6194 [hep-th]].
}

\lref\ElShowkHT{
  S.~El-Showk, M.~F.~Paulos, D.~Poland, S.~Rychkov, D.~Simmons-Duffin and A.~Vichi,
  ``Solving the 3D Ising Model with the Conformal Bootstrap,''
[arXiv:1203.6064 [hep-th]].
}

\lref\WilsonJJ{
  K.~G.~Wilson and J.~B.~Kogut,
  ``The Renormalization group and the epsilon expansion,''
Phys.\ Rept.\  {\bf 12}, 75 (1974)..
}

\lref\LangZW{
  K.~Lang and W.~Ruhl,
  ``The Critical O(N) sigma model at dimensions 2 < d < 4: Fusion coefficients and anomalous dimensions,''
Nucl.\ Phys.\ B {\bf 400}, 597 (1993)..
}

\lref\HoffmannDX{
  L.~Hoffmann, L.~Mesref and W.~Ruhl,
  ``Conformal partial wave analysis of AdS amplitudes for dilaton axion four point functions,''
Nucl.\ Phys.\ B {\bf 608}, 177 (2001).
[hep-th/0012153].
}

\lref\ElShowkHT{
  S.~El-Showk, M.~F.~Paulos, D.~Poland, S.~Rychkov, D.~Simmons-Duffin and A.~Vichi,
  ``Solving the 3D Ising Model with the Conformal Bootstrap,''
[arXiv:1203.6064 [hep-th]].
}

\lref\GrossUN{
 D. Gross, {\it unpublished} 
}

\lref\CallanPU{
  C.~G.~Callan, Jr. and D.~J.~Gross,
  ``Bjorken scaling in quantum field theory,''
Phys.\ Rev.\ D {\bf 8}, 4383 (1973)..
}

\lref\MackJE{
  G.~Mack,
 ``All Unitary Ray Representations of the Conformal Group SU(2,2) with Positive Energy,''
Commun.\ Math.\ Phys.\  {\bf 55}, 1 (1977)..
}

\lref\GrinsteinQK{
  B.~Grinstein, K.~A.~Intriligator and I.~Z.~Rothstein,
  ``Comments on Unparticles,''
Phys.\ Lett.\ B {\bf 662}, 367 (2008).
[arXiv:0801.1140 [hep-ph]].
}

\lref\RattazziPE{
  R.~Rattazzi, V.~S.~Rychkov, E.~Tonni and A.~Vichi,
  ``Bounding scalar operator dimensions in 4D CFT,''
JHEP {\bf 0812}, 031 (2008).
[arXiv:0807.0004 [hep-th]].
}

\lref\CostaCB{
  M.~S.~Costa, V.~Goncalves and J.~Penedones,
  ``Conformal Regge theory,''
[arXiv:1209.4355 [hep-th]].
}

\lref\PappadopuloJK{
  D.~Pappadopulo, S.~Rychkov, J.~Espin and R.~Rattazzi,
  ``OPE Convergence in Conformal Field Theory,''
[arXiv:1208.6449 [hep-th]].
}

\lref\AdamsSV{
  A.~Adams, N.~Arkani-Hamed, S.~Dubovsky, A.~Nicolis and R.~Rattazzi,
  ``Causality, analyticity and an IR obstruction to UV completion,''
JHEP {\bf 0610}, 014 (2006).
[hep-th/0602178].
}

\lref\FreyhultMY{
  L.~Freyhult and S.~Zieme,
  ``The virtual scaling function of AdS/CFT,''
Phys.\ Rev.\ D {\bf 79}, 105009 (2009).
[arXiv:0901.2749 [hep-th]].
}

\lref\PelissettoEK{
  A.~Pelissetto and E.~Vicari,
  ``Critical phenomena and renormalization group theory,''
Phys.\ Rept.\  {\bf 368}, 549 (2002).
[cond-mat/0012164].
}

\lref\VasilievBA{
  M.~A.~Vasiliev,
  ``Higher spin gauge theories: Star product and AdS space,''
In *Shifman, M.A. (ed.): The many faces of the superworld* 533-610.
[hep-th/9910096].
}

\lref\MaldacenaRE{
  J.~M.~Maldacena,
  ``The Large N limit of superconformal field theories and supergravity,''
Adv.\ Theor.\ Math.\ Phys.\  {\bf 2}, 231 (1998).
[hep-th/9711200].
}

\lref\GubserBC{
  S.~S.~Gubser, I.~R.~Klebanov and A.~M.~Polyakov,
  ``Gauge theory correlators from noncritical string theory,''
Phys.\ Lett.\ B {\bf 428}, 105 (1998).
[hep-th/9802109].
}

\lref\WittenQJ{
  E.~Witten,
  ``Anti-de Sitter space and holography,''
Adv.\ Theor.\ Math.\ Phys.\  {\bf 2}, 253 (1998).
[hep-th/9802150].
}

\lref\SundborgWP{
  B.~Sundborg,
  ``Stringy gravity, interacting tensionless strings and massless higher spins,''
Nucl.\ Phys.\ Proc.\ Suppl.\  {\bf 102}, 113 (2001).
[hep-th/0103247].
}

\lref\DolanUT{
  F.~A.~Dolan and H.~Osborn,
  ``Conformal four point functions and the operator product expansion,''
Nucl.\ Phys.\ B {\bf 599}, 459 (2001).
[hep-th/0011040].
}

\lref\CornalbaXK{
  L.~Cornalba, M.~S.~Costa, J.~Penedones and R.~Schiappa,
  ``Eikonal Approximation in AdS/CFT: From Shock Waves to Four-Point Functions,''
JHEP {\bf 0708}, 019 (2007).
[hep-th/0611122].
}

\lref\CornalbaXM{
  L.~Cornalba, M.~S.~Costa, J.~Penedones and R.~Schiappa,
  ``Eikonal Approximation in AdS/CFT: Conformal Partial Waves and Finite N Four-Point Functions,''
Nucl.\ Phys.\ B {\bf 767}, 327 (2007).
[hep-th/0611123].
}

\lref\CornalbaZB{
  L.~Cornalba, M.~S.~Costa and J.~Penedones,
  ``Eikonal approximation in AdS/CFT: Resumming the gravitational loop expansion,''
JHEP {\bf 0709}, 037 (2007).
[arXiv:0707.0120 [hep-th]].
}

\lref\DerkachovPH{
  S.~E.~Derkachov and A.~N.~Manashov,
  ``Generic scaling relation in the scalar phi**4 model,''
J.\ Phys.\ A {\bf 29}, 8011 (1996).
[hep-th/9604173].
}

\lref\KlebanovJA{
  I.~R.~Klebanov and A.~M.~Polyakov,
  ``AdS dual of the critical O(N) vector model,''
Phys.\ Lett.\ B {\bf 550}, 213 (2002).
[hep-th/0210114].
}

\lref\SezginRT{
  E.~Sezgin and P.~Sundell,
  ``Massless higher spins and holography,''
Nucl.\ Phys.\ B {\bf 644}, 303 (2002), [Erratum-ibid.\ B {\bf 660}, 403 (2003)].
[hep-th/0205131].
}
\lref\EpsteinBG{
  H.~Epstein, V.~Glaser and A.~Martin,
  ``Polynomial behaviour of scattering amplitudes at fixed momentum transfer in theories with local observables,''
Commun.\ Math.\ Phys.\  {\bf 13}, 257 (1969)..
}

\lref\FortinCQ{
  J.~-F.~Fortin, B.~Grinstein and A.~Stergiou,
  ``Scale without Conformal Invariance in Four Dimensions,''
[arXiv:1206.2921 [hep-th]].
}

\lref\FortinHN{
  J.~-F.~Fortin, B.~Grinstein and A.~Stergiou,
  ``A generalized c-theorem and the consistency of scale without conformal invariance,''
[arXiv:1208.3674 [hep-th]].
}

\lref\FortinHC{
  J.~-F.~Fortin, B.~Grinstein, C.~W.~Murphy and A.~Stergiou,
  ``On Limit Cycles in Supersymmetric Theories,''
[arXiv:1210.2718 [hep-th]].
}

\lref\NakayamaED{
  Y.~Nakayama,
  ``Is boundary conformal in CFT?,''
[arXiv:1210.6439 [hep-th]].
}

\lref\NakayamaND{
  Y.~Nakayama,
  ``Supercurrent, Supervirial and Superimprovement,''
[arXiv:1208.4726 [hep-th]].
}

\lref\AntoniadisGN{
  I.~Antoniadis and M.~Buican,
  ``On R-symmetric Fixed Points and Superconformality,''
Phys.\ Rev.\ D {\bf 83}, 105011 (2011).
[arXiv:1102.2294 [hep-th]].
}

\lref\LutyWW{
  M.~A.~Luty, J.~Polchinski and R.~Rattazzi,
  ``The $a$-theorem and the Asymptotics of 4D Quantum Field Theory,''
[arXiv:1204.5221 [hep-th]].
}

\lref\GiombiWH{
  S.~Giombi and X.~Yin,
  ``Higher Spin Gauge Theory and Holography: The Three-Point Functions,''
JHEP {\bf 1009}, 115 (2010).
[arXiv:0912.3462 [hep-th]].
}

\lref\KomargodskiVJ{
  Z.~Komargodski and A.~Schwimmer,
  ``On Renormalization Group Flows in Four Dimensions,''
JHEP {\bf 1112}, 099 (2011).
[arXiv:1107.3987 [hep-th]].
}

\lref\KomargodskiXV{
  Z.~Komargodski,
  ``The Constraints of Conformal Symmetry on RG Flows,''
JHEP {\bf 1207}, 069 (2012).
[arXiv:1112.4538 [hep-th]].
}

\lref\KehreinIA{
  S.~K.~Kehrein,
  ``The Spectrum of critical exponents in phi**2 in two-dimensions theory in D = (4-epsilon)-dimensions: Resolution of degeneracies and hierarchical structures,''
Nucl.\ Phys.\ B {\bf 453}, 777 (1995).
[hep-th/9507044].
}

\lref\BraunRP{
  V.~M.~Braun, G.~P.~Korchemsky and D.~Mueller,
  ``The Uses of conformal symmetry in QCD,''
Prog.\ Part.\ Nucl.\ Phys.\  {\bf 51}, 311 (2003).
[hep-ph/0306057].
}

\lref\HeemskerkPN{
  I.~Heemskerk, J.~Penedones, J.~Polchinski and J.~Sully,
  ``Holography from Conformal Field Theory,''
JHEP {\bf 0910}, 079 (2009).
[arXiv:0907.0151 [hep-th]].
}

\lref\RuhlBW{
  W.~Ruhl,
  ``The Goldstone fields of interacting higher spin field theory on AdS(4),''
[hep-th/0607197].
}

\lref\AdamsSV{
  A.~Adams, N.~Arkani-Hamed, S.~Dubovsky, A.~Nicolis and R.~Rattazzi,
  ``Causality, analyticity and an IR obstruction to UV completion,''
JHEP {\bf 0610}, 014 (2006).
[hep-th/0602178].
}

\lref\PhamCR{
  T.~N.~Pham and T.~N.~Truong,
  ``Evaluation Of The Derivative Quartic Terms Of The Meson Chiral Lagrangian From Forward Dispersion Relation,''
Phys.\ Rev.\ D {\bf 31}, 3027 (1985)..
}

\lref\FGG{
  S.~Ferrara, A.~F.~Grillo and R.~Gatto,
  ``Tensor representations of conformal algebra and conformally covariant operator product expansion,''
Annals Phys.\  {\bf 76}, 161 (1973).
}

\lref\BPZ{
  A.~A.~Belavin, A.~M.~Polyakov and A.~B.~Zamolodchikov,
  ``Infinite Conformal Symmetry in Two-Dimensional Quantum Field Theory,''
Nucl.\ Phys.\ B {\bf 241}, 333 (1984).
}

\lref\HeslopDU{
  P.~J.~Heslop,
  ``Aspects of superconformal field theories in six dimensions,''
JHEP {\bf 0407}, 056 (2004).
[hep-th/0405245].
}

\lref\DolanTT{
  F.~A.~Dolan and H.~Osborn,
  ``Superconformal symmetry, correlation functions and the operator product expansion,''
Nucl.\ Phys.\ B {\bf 629}, 3 (2002).
[hep-th/0112251].
}

\lref\PappadopuloJK{
  D.~Pappadopulo, S.~Rychkov, J.~Espin and R.~Rattazzi,
  ``OPE Convergence in Conformal Field Theory,''
[arXiv:1208.6449 [hep-th]].
}

\lref\FerraraEJ{
  S.~Ferrara, C.~Fronsdal and A.~Zaffaroni,
  ``On N=8 supergravity on AdS(5) and N=4 superconformal Yang-Mills theory,''
Nucl.\ Phys.\ B {\bf 532}, 153 (1998).
[hep-th/9802203].
}

\lref\DidenkoTV{
  V.~E.~Didenko and E.~D.~Skvortsov,
  ``Exact higher-spin symmetry in CFT: all correlators in unbroken Vasiliev theory,''
[arXiv:1210.7963 [hep-th]].
}

\lref\MaldacenaSF{
  J.~Maldacena and A.~Zhiboedov,
  ``Constraining conformal field theories with a slightly broken higher spin symmetry,''
[arXiv:1204.3882 [hep-th]].
}

\lref\MaldacenaJN{
  J.~Maldacena and A.~Zhiboedov,
  ``Constraining Conformal Field Theories with A Higher Spin Symmetry,''
[arXiv:1112.1016 [hep-th]].
}

\lref\BriganteGZ{
  M.~Brigante, H.~Liu, R.~C.~Myers, S.~Shenker and S.~Yaida,
  ``The Viscosity Bound and Causality Violation,''
Phys.\ Rev.\ Lett.\  {\bf 100}, 191601 (2008).
[arXiv:0802.3318 [hep-th]].
}

\lref\GirardelloPP{
  L.~Girardello, M.~Porrati and A.~Zaffaroni,
  ``3-D interacting CFTs and generalized Higgs phenomenon in higher spin theories on AdS,''
Phys.\ Lett.\ B {\bf 561}, 289 (2003).
[hep-th/0212181].
}

\lref\HeemskerkPN{
  I.~Heemskerk, J.~Penedones, J.~Polchinski and J.~Sully,
  ``Holography from Conformal Field Theory,''
JHEP {\bf 0910}, 079 (2009).
[arXiv:0907.0151 [hep-th]].
}

\lref\CornalbaZB{
  L.~Cornalba, M.~S.~Costa and J.~Penedones,
  ``Eikonal approximation in AdS/CFT: Resumming the gravitational loop expansion,''
JHEP {\bf 0709}, 037 (2007).
[arXiv:0707.0120 [hep-th]].
}

\lref\DolanDV{
  F.~A.~Dolan and H.~Osborn,
  ``Conformal Partial Waves: Further Mathematical Results,''
[arXiv:1108.6194 [hep-th]].
}

\lref\DolanUT{
  F.~A.~Dolan and H.~Osborn,
  ``Conformal four point functions and the operator product expansion,''
Nucl.\ Phys.\ B {\bf 599}, 459 (2001).
[hep-th/0011040].
}

\lref\GiombiKC{
  S.~Giombi, S.~Minwalla, S.~Prakash, S.~P.~Trivedi, S.~R.~Wadia and X.~Yin,
  ``Chern-Simons Theory with Vector Fermion Matter,''
Eur.\ Phys.\ J.\ C {\bf 72}, 2112 (2012).
[arXiv:1110.4386 [hep-th]].
}

\lref\AharonyJZ{
  O.~Aharony, G.~Gur-Ari and R.~Yacoby,
  ``d=3 Bosonic Vector Models Coupled to Chern-Simons Gauge Theories,''
JHEP {\bf 1203}, 037 (2012).
[arXiv:1110.4382 [hep-th]].
}

\lref\KovtunKW{
  P.~Kovtun and A.~Ritz,
  ``Black holes and universality classes of critical points,''
Phys.\ Rev.\ Lett.\  {\bf 100}, 171606 (2008).
[arXiv:0801.2785 [hep-th]].
}

\lref\FroissartUX{
  M.~Froissart,
  ``Asymptotic behavior and subtractions in the Mandelstam representation,''
Phys.\ Rev.\  {\bf 123}, 1053 (1961)..
}

\lref\MartinRT{
  A.~Martin,
  ``Unitarity and high-energy behavior of scattering amplitudes,''
Phys.\ Rev.\  {\bf 129}, 1432 (1963)..
}

\lref\FrankfurtMC{
  L.~Frankfurt, M.~Strikman and C.~Weiss,
 ``Small-x physics: From HERA to LHC and beyond,''
Ann.\ Rev.\ Nucl.\ Part.\ Sci.\  {\bf 55}, 403 (2005).
[hep-ph/0507286].
}

\lref\PolchinskiJW{
  J.~Polchinski and M.~J.~Strassler,
  ``Deep inelastic scattering and gauge / string duality,''
JHEP {\bf 0305}, 012 (2003).
[hep-th/0209211].
}

\lref\BarnesBM{
  E.~Barnes, E.~Gorbatov, K.~A.~Intriligator, M.~Sudano and J.~Wright,
  ``The Exact superconformal R-symmetry minimizes tau(RR),''
Nucl.\ Phys.\ B {\bf 730}, 210 (2005).
[hep-th/0507137].
}

\lref\PolandWG{
  D.~Poland and D.~Simmons-Duffin,
  ``Bounds on 4D Conformal and Superconformal Field Theories,''
JHEP {\bf 1105}, 017 (2011).
[arXiv:1009.2087 [hep-th]].
}

\lref\LiendoHY{
  P.~Liendo, L.~Rastelli and B.~C.~van Rees,
  ``The Bootstrap Program for Boundary $CFT_d$,''
[arXiv:1210.4258 [hep-th]].
}

\lref\PolyakovXD{
 A.~M.~Polyakov,
 ``Conformal symmetry of critical fluctuations,''
JETP Lett.\  {\bf 12}, 381 (1970), [Pisma Zh.\ Eksp.\ Teor.\ Fiz.\  {\bf 12}, 538 (1970)].
}

\lref\FitzpatrickDM{
  A.~L.~Fitzpatrick and J.~Kaplan,
  ``Unitarity and the Holographic S-Matrix,''
JHEP {\bf 1210}, 032 (2012).
[arXiv:1112.4845 [hep-th]].
}

\lref\CostaDW{
  M.~S.~Costa, J.~Penedones, D.~Poland and S.~Rychkov,
  ``Spinning Conformal Blocks,''
JHEP {\bf 1111}, 154 (2011).
[arXiv:1109.6321 [hep-th]].
}

\lref\OsbornVT{
  H.~Osborn,
  ``Conformal Blocks for Arbitrary Spins in Two Dimensions,''
Phys.\ Lett.\ B {\bf 718}, 169 (2012).
[arXiv:1205.1941 [hep-th]].
}

\lref\LangGE{
  K.~Lang and W.~Ruhl,
  ``Critical O(N) vector nonlinear sigma models: A Resume of their field structure,''
[hep-th/9311046].
}

\lref\KazakovKM{
  D.~I.~Kazakov,
  ``The Method Of Uniqueness, A New Powerful Technique For Multiloop Calculations,''
Phys.\ Lett.\ B {\bf 133}, 406 (1983)..
}

\lref\FitzpatrickYX{
  A.~L.~Fitzpatrick, J.~Kaplan, D.~Poland and D.~Simmons-Duffin,
  ``The Analytic Bootstrap and AdS Superhorizon Locality,''
[arXiv:1212.3616 [hep-th]].
}


\rightline{PUPT-2433}
\rightline{WIS/21/12-DEC-DPPA}
\vskip-10pt

\Title{
} {\vbox{\centerline{Convexity and Liberation at Large Spin}
}}

\centerline{Zohar Komargodski$^{1,2}$ and Alexander Zhiboedov$^{3}$}
\bigskip
\centerline{ $^{1}$ {\it Weizmann Institute of Science, Rehovot
76100, Israel}}
  \centerline{$^{2}${\it
Institute for Advanced Study, Princeton, NJ 08540, USA}}
 \centerline{$^{3}$ {\it Department of Physics, Princeton University, Princeton, NJ 08544, USA}}

\vskip12pt
     
\noindent 

 We consider several aspects of unitary higher-dimensional conformal field theories~(CFTs). We first study massive deformations that trigger a flow to a gapped phase. 
Deep inelastic scattering in the gapped phase leads to a convexity property of dimensions of spinning operators of the original CFT. 
We further investigate the dimensions of spinning operators via the crossing equations in the light-cone limit. We find that, in a sense, CFTs become free at large spin and $1/s$ is a weak coupling parameter. The spectrum of CFTs enjoys additivity: if two twists $\tau_1$, $\tau_2$ appear in the spectrum, there  are operators whose twists are arbitrarily close to $\tau_1+\tau_2$. 
We characterize how $\tau_1+\tau_2$ is approached at large spin by solving the crossing equations analytically. We find the precise form of the leading correction, including the prefactor. We compare with examples where these observables were computed in perturbation theory, or via gauge-gravity duality, and find  complete agreement. The  crossing equations show that certain operators have a convex spectrum in twist space. We also observe a connection between convexity and the ratio of dimension to charge. Applications include the 3d Ising model, theories with a gravity dual, SCFTs,  and patterns of higher spin symmetry breaking.

\vskip 7pt

\Date{December 2012}

\listtoc \writetoc
\vskip .5in \noindent

\newsec{Introduction}

Even though conformal field theories (CFTs) have been studied for many years, very little is known about the general structure of their operator spectrum and three-point functions.
The interest in conformal field theories stems from their role in the description of phase transitions~\PolyakovXD\ and from their relation to renormalization group flows.\foot{A priori there may also be renormalization group flows where one finds  scale invariant theories at the end points without the special conformal generators. The feasibility of this scenario has been discussed recently, for example, in~\refs{\AntoniadisGN\LutyWW\FortinCQ
\FortinHN\NakayamaND\FortinHC-\NakayamaED}. See also references therein. To date, scale invariant but non-conformal theories have not been constructed.} In addition, significant motivation to study conformal field theories is derived from the connection between quantum gravity in AdS and conformal field theories~\refs{\MaldacenaRE, \GubserBC,\WittenQJ}.  

A classic result is that, in unitarity CFTs, operators obey the so-called unitarity bounds~\MackJE\ (see also~\GrinsteinQK). For example, the dimension of primary symmetric traceless tensors of spin $s\geq 1$ satisfies the following inequality
\eqn\unitbound{
\Delta \geq d-2 + s~.
}
(Here and below $d$ is the dimension of space-time.) 

Primary operators that saturate the bound are known to be conserved, namely, they satisfy the equation $\pa_{\mu} J^{\mu}_{\ ...} = 0$ inside any correlation function. Similarly, scalar operators ($s=0$) are known to satisfy $\Delta\geq {d-2\over 2}$. If this inequality is saturated, a second order differential equation holds true $\square \CO=0$.

The unitarity bounds above are traditionally derived from reflection positivity of Euclidean correlation functions. It has been known for a while that considering processes in Lorentz signature can lead to further constraints on quantum field theory, and often such constraints do not seem to admit a straightforward derivation in Euclidean space. Simple examples based on an application of the optical theorem are discussed in~\refs{\PhamCR,\AdamsSV,\KomargodskiVJ, \KomargodskiXV}. 

In this note we will use some properties of quantum field theory that are particularly easy to understand in Minkowski space in order to derive various constraints.  In addition, we will find it useful to embed our conformal field theories in renormalization group flows that lead to a gapped phase. 

Before we plunge into a detailed summary of our results, we would like to review quickly some basic notions in conformal field theory, mostly in order to set the terminology we will be using throughout this note. Given scalar operators $\CO_1(x)$, $\CO_2(x)$ we can consider their OPE 
\eqn\ope{\CO_1(x)\CO_2(0)\sim \sum_s\sum_{k=1}^{\alpha(s)} C_s^kx^{-\Delta_1-\Delta_2+\Delta_s^k-s} x^{\mu_1}x^{\mu_2}\cdots x^{\mu_s}\CO^{k}_{\mu_1,\mu_2,...,\mu_s}(0) ~. }
Clearly, only symmetric traceless representations of the Lorentz group can appear.  The sum over $s$ is the sum over spin and the sum over $k$ is the sum over all the operators that have spin $s$. The sum over $k$ is a finite sum.  One can further re-package the sum above in terms of representations of the conformal group by combining the contributions of all the operators that are  derivatives of some primary operator. (The coefficients $C_s^k$ are fixed by conformal symmetry in terms of the coefficients of primary operators.)

In Euclidean space it is natural to consider $x^2\rightarrow 0$ which means that all the components of $x^\mu$ go to zero. In this case the terms that dominate the sum~\ope\ are the operators $\CO^k_{\mu_1,...,\mu_s}$ with the smallest dimensions $\Delta_s^k$.

In Minkowski signature we can send $x^2\rightarrow 0$ while being on the light cone. For convenience, we can take light-cone coordinates on a two-dimensional plane  $x^-,x^+$, and set $x^i=0$.  Then we consider the light-cone limit $x^+\rightarrow 0$ while $x^-$ is finite. The operators that dominate then are $\CO_{-,-,...,-}$ and their contribution to the OPE goes like $x^{-\Delta_1-\Delta_2+\Delta_s^k-s}$, hence the strength with which they contribute is dictated by the twist  
$\tau_s^k\equiv \Delta_s^k-s$. We denote $\tau_s^*\equiv\min_k(\tau_s^k)$, i.e. $\tau_s^*$ is the minimal twist that appears in the OPE for a given spin. Another useful notation is $\tau_{\min}\equiv \min_s{\tau_s^*}$, that is, the overall minimal twist that appears in the OPE, {\it excluding the unit operator}, which has twist zero. 

The twists of conformal primaries are constrained by~\unitbound. Traceless symmetric conformal primaries of spin $s\geq 1$ must have twist at least $d-2$ and a scalar conformal primary must have twist at least ${d\over 2}-1$. The unit operator has twist zero. The energy momentum tensor and conserved currents have twist $d-2$.

When two points are light-like separated, the group $SL(2,R)$ preserves the corresponding light-ray. It is therefore natural to consider representations of it. The representations  are classified by collinear primaries of dimension $\Delta$ and spin $s_{+-}$ in the two dimensional plane. We can act on this collinear primary any number of times with $\del_-$ in order to generate the complete collinear representation. The twist of a collinear primary is defined as $\tau^{{\rm coll}}\equiv \Delta-s_{+-}$.

A conformal primary $\CO_{\mu_1,...,\mu_n}$ can be decomposed into (generally infinitely many) collinear primaries. In this decomposition, the collinear primary with the smallest collinear twist is clearly $\CO_{-,...,-}$, and this minimal collinear twist coincides with the conformal twist. 

It would be useful for our purposes to rewrite the OPE~\ope\ in terms of collinear representations. From the discussion above we see that it takes the form as $x^+\rightarrow 0$
\eqn\ope{\eqalign{&\CO_1(x^+,x^-)\CO_2(0,0)\cr&\sim (x^+)^{-\half(\Delta_1+\Delta_2)}\sum_{\tau^{{\rm coll}},s_{+-},...} \tilde C_s^k \   (x^+)^{\half\tau^{{\rm coll}}}\CF(x^-,\del_-)\CO^{k}_{\underbrace{-,-,...,-}_{\half(s+s_{+-})},\underbrace{+,+...,+}_{\half(s-s_{+-})}}(0) ~.}}
The functions $\CF$ are determined from representation theory, see~\BraunRP\ for a review.  
The important message in~\ope\ is that the approach to the light cone is controlled by the collinear twists.  

Having introduced the basic terminology that we will employ, let us now summarize the constraints we discuss in this note. 

First, we merely extend the observations of Nachtmann~\NachtmannMR\ about the QCD sum rules to an argument pertaining to arbitrary CFTs. In~\NachtmannMR\ it was pointed out that the Deep Inelastic Scattering (DIS) QCD sum rules imply certain convexity properties of the high energy limit of QCD. We discuss such constraints on the operators in the OPE in the context of a generic CFTs. We emphasize the assumptions involved in this argument.

To this end, we consider an RG flow where the CFT of interest is the UV theory and in the IR we have a gapped phase. We consider an experiment with the operator of interest, ${\cal O}(x) $, playing the role of the usual electromagnetic current in DIS. Assuming Regge asymptotics of the amplitude in the gapped phase (namely, the amplitude is bounded by some power of $s$)
\eqn\Reggeassum{
\lim_{s \to \infty} A (s,t =0) \leq s^{N-1}
}
one is led to a convexity property of the minimal twists $\tau_s^*$ of operators appearing in the ${\cal O}(x) {\cal O}^\dagger(0)$ OPE 
\eqn\convexiintro{{\tau^*_{s_3}-\tau^*_{s_1}\over s_3-s_1}\leq {\tau^*_{s_2}-\tau^*_{s_1}\over s_2-s_1}~,\qquad s_3 > s_2 > s_1 \geq s_c~.}
Here $s_c$ is the minimal spin above which the convexity property starts. From~\Reggeassum\ we know that $s_c$ is finite and certainly does not exceed $N$. 
The argument leading to the convexity property~\convexiintro\ depends on the following three assumptions:
\item{a)} unitarity;
\item{b)} the CFT can flow to a gapped phase;
\item{c)} polynomial boundedness of the DIS cross section in the gapped phase~\Reggeassum.

\medskip

Assumption c) may seem as the least innocuous. However, there are formal arguments connecting it to causality and other basic properties of QFT. See for example~\EpsteinBG, and a more recent discussion in~\GiddingsGJ. In addition, we do not know of field theories violating assumption b) (such theories are known to exist in the realm of critical phenomena, but, to our knowledge, there are no concrete examples of unitary Lorentz invariant theories that cannot be deformed to a gapped phase). Hence, convexity of the function $\tau_s^*$ is established under very general assumptions.
Moreover, in all the examples we checked, the convexity above holds true starting from  $s_c=2$. We do not  prove here that this is the case in most generality.

A second topic we study here is the crossing symmetry (i.e. the bootstrap equations) in Lorentzian signature.
The bootstrap equations were introduced long ago by~\refs{\FGG, \PolyakovGS, \BPZ}. It has been demonstrated recently, starting from~\RattazziPE, that one can go about and systematically bound numerically solutions to these formidable equations. Sometimes these numerical bounds seem to be saturated by actual CFTs, as has been demonstrated, for example, in~\ElShowkHT. These methods of bounding solutions to the bootstrap equations are easily described in Euclidean space, and the consistency of the procedure follows from the fact that the contribution of the operators with high dimension to the OPE in Euclidean space is exponentially suppressed~\PappadopuloJK.

By contrast, as we recalled above, the structure of the operator product expansion in Minkowski space is such that we can probe directly operators with very high spin and small twist. We therefore use the bootstrap methods in Lorentzian signature to constrain operators with high dimension and high spin, but with low twist.\foot{Different applications of the bootstrap equations in Lorentzian signature were discussed in~\refs{\CornalbaXK,\CornalbaXM,\CornalbaZB }. In these papers the properties of operators with both high spin and high twist were analyzed.} 

Studying the light-cone OPE and bootstrap equations in the appropriate limit, we conclude that starting from any two primaries with twists $\tau_1$ and $\tau_2$, operators with twists arbitrarily close to $ \tau_1 + \tau_2$ are necessarily present. This can be regarded as an additivity property of the twist spectrum of general CFT. In the derivation of this result, it is important to assume that $d>2$. Indeed, this additivity property is definitely violated in two-dimensional models, such as the minimal models. The condition $d>2$ comes from the fact that we need to separate the contribution of the unit operator from everything else in the light-cone OPE.

In particular, this means that the minimal twists appearing in the OPE of an operator $\CO_1$ with any other operator $\CO_2$  are bounded from above as follows (in $d>2$):
\eqn\upperbound{
\lim_{s \to \infty} \tau^*_{s} \leq  \tau_{\CO_1}+\tau_{ \CO_2}~.
}

We conclude that there should not be unitary solutions of the crossing equations with $\tau^*_{s} > \tau_{\CO_1}+\tau_{ \CO_2}$. This could be useful to bear in mind when trying to bound solutions to the crossing equations along the lines of~\RattazziPE.

Additivity implies that for large enough spin there are operators whose twists are arbitrary close to $\tau_{\CO_1}+\tau_{ \CO_2}$. We refer to these operators as double-twist operators. We also denote these operators symbolically as $\CO_1\del^s\CO_2$.  Similar argument shows that operators with twists asymptoting to $\tau_{\CO_1}+\tau_{\CO_2}+2n$ for any integer $n$ are present as well. Those can be denoted symbolically as $\CO_1\del^s\square^n\CO_2$. We do not discuss the case $n\neq  0$ in as much detail as $n=0$.
In addition, it also makes sense to talk about multi-twist operators.

Given that their twists approach $\tau_{\CO_1}+\tau_{ \CO_2}$ for large enough spin, it is interesting to ask how precisely this limit of $\tau_{\CO_1}+\tau_{ \CO_2}$ is reached.  
It has been argued in~\AldayMF\ that for sufficiently large spin we can parameterize the twists of these operators as follows
\eqn\approach{\tau(s)=\tau_{\CO_1}+\tau_{\CO_2}-{c_{\tau_{\min}}\over s^{\tau_{\min}}}+\cdots~.} 
  Here $\tau_{\min}$ is the twist of $\CO_{\min}$ which is the operator of smallest twist exchanged by $\CO_1$ and $\CO_2$.\foot{By this we mean it appears both in the OPE of $\CO_1\CO_1^\dagger$ and $\CO_2 \CO_2^\dagger$. } (As in the definition before, we exclude from this the unit operator.) In many cases we would expect this to be the energy momentum tensor, thus $\tau_{\min}=d-2$. 

The essence of the argument of~\AldayMF\ for~\approach\ is as follows. One maps conformally the CFT from flat space to an $AdS_3 \times S^{d-3}$ background. By choosing special coordinates in $AdS_3$ one can think about the four-point functions above as being described by a two-dimensional gapped theory embedded in $AdS_3$. The energy in this theory is the twist, and~\approach\ corresponds to the leading interaction between separated localized excitations where the separation is governed by the spin.   

Here we arrive at the same conclusion~\approach\ by studying the crossing equation in flat Minkowski space. Furthermore, by using the bootstrap equations, we compute the coefficient $c_{\tau_{\min}}$ in~\approach\ analytically in every $d>2$ CFT. The simplest case is when we consider the OPE of an operator with its Hermitian conjugate $\CO(x)\CO^\dagger(0)$. Then the coefficient governing the asymptotic twists of double-twist operators is 
\eqn\correctionform{\eqalign{
c_{\tau_{\min}} &=  {\Gamma (\tau_{\min} +  2 s_{\min}) \over 2^{s_{\min}-1} \Gamma \left({\tau_{\min} + 2 s_{\min} \over 2}\right)^2} {\Gamma (\Delta_\CO)^2 \over \Gamma (\Delta_\CO  - {\tau_{\min} \over 2} )^2} f^2~, \cr
f^2 &=  { C_{{\cal O} {\cal O}^{\dagger} {\cal O}_{\tau_{\min}}}^2 \over  C_{{\cal O} {\cal O} } C_{{\cal O}^{\dagger} {\cal O}^{\dagger} }  C_{{\cal O}_{\tau_{\min}}{\cal O}_{\tau_{\min}}}}~,
}}
where $s_{\min}$ and $\tau_{\min}$ are the spin and the twist of the minimal twist operator (other than the unit operator) appearing in the OPE of $\CO$ with its conjugate. Hence, $c_{\tau_{\min}}$ is fixed by some universal factors and the two- and three-point functions of the operators $\CO$ and $\CO_{\tau_{\min}}$. As we remarked, in many models one would expect $\CO_{\tau_{\min}}$ to be the stress tenors, in which case the three-point function above is fixed in terms of  $\Delta_\CO$ and the two-point function of the stress tensor (which is some measure of the number of degrees of freedom of the theory).

Note that in unitary CFTs we have  $\tau_{\min} \leq 2 \Delta_\CO$, and an equality can only be reached for free fields.
In general, there could be several operators of the same minimal twist $\tau_{\min}$. In this case $c$ is the sum over all of these operators. It is worth mentioning that~\correctionform\ provides another point of view on convexity. Since all the leading contributions to $c$ are manifestly positive, the spectrum of double-twist operators at large enough spins is convex. This statement is slightly different from the one outlined in~\convexiintro, which is about the minimal twists in reflection positive OPEs.\foot{In this note when we say ``reflection positive OPE'' we mean an OPE of the type $\CO(x)\CO(0)^\dagger$.}
 Here the statement is about double-twist operators in reflection positive OPEs.

We also consider non-reflection positive OPEs, such as those of a charged operator (with charge $q$ under some $U(1)$ global symmetry, with the generalization to non-Abelian symmetries being straightforward) $\CO$ with itself, $\CO(x)\CO(0)$. The double-twist operators ${\cal O} \pa^s {\cal O}$ with twists that approach $2\Delta_\CO$ are necessarily present.  Here we can compute the corrections to the twists at large spin $\tau_s-2\Delta_\CO$ and we find a formula very similar to~\correctionform\ (we will describe it in detail in the text) with one small but important difference, a factor of $(-1)^s$. Hence, the conserved current and the energy momentum tensor, both of which have twist $d-2$ contribute with opposite signs. We assume those are the minimal twist operators in the OPE $\CO\CO^\dagger$.  The spectrum is non-concave at large enough spin if and only if 
\eqn\convexnonre{\Delta\geq {d-1\over 2}|q|~,}
in some normalization  for the two-point function of the $U(1)$ current that we define carefully in the main text. Hence convexity at large spin of such operators is related to whether or not the operator $\CO$ satisfies a BPS-like bound.  

In SUSY theories, for chiral primaries,~\convexnonre\ is saturated (the charge $q$ is that of the superconformal $U(1)_R$ symmetry) and the leading corrections to the anomalous twists vanish. For chiral operators that are not superconformal primaries,~\convexnonre\ is satisfied in unitary theories. Hence, the spectrum of twists of the double-twist operators ${\cal O} \pa^s {\cal O}$ is convex.  It would be interesting to understand more generally whether convexity 
at large spin is a general phenomenon or not. The arguments in this note show that a large class of OPEs is consistent with being convex.\foot{Throughout the paper, convexity is to be understood as non-concavity.}

If we combine~\convexiintro\ and~\upperbound\ (i.e. convexity and additivity) we are led to rather stringent predictions concerning interesting models. In particular, we predict that in the 3d critical $O(N)$ models (which are highly relevant for 3d phase transitions) there is a set of almost conserved currents of spin $s \geq 4$ with the dimension $\Delta_s = s + \tau_s^*$ with 
\eqn\prediction{
1 \leq \tau^*_s \leq 2 \Delta_{\sigma}~,
}
where $\Delta_{\sigma}$ is the dimension of the spin field $\sigma_i$. Furthermore, we can argue using additivity that 
\eqn\predictioni{\lim_{s \to \infty} \tau_s^* = 2 \Delta_{\sigma}~,}
and we can characterize in detail how this limit is approached via~\approach\ and \correctionform.
One can use the numerical results from the literature \PelissettoEK\ to arrive at concrete predictions. Furthermore, the function $\tau^*_s$ is monotonically rising and convex as a function of the spin. 

In the case of the 3d Ising model, it is known that there is a spin $4$ operator with the twist $\tau_4^* \sim 1.02$, which agrees with our expectations. Using convexity, this allows us to further narrow down the possible window for higher spin currents and predict that in the 3d Ising model (which describes, among others, boiling water)
\eqn\predising{\eqalign{
&1.02 \leq \tau_s^* \leq 1.037, ~~~ s \geq 6~, \cr
&\lim_{s \to \infty} \tau_s^* = 2 \Delta_{\sigma} \approx 1.037.
}}
Recall that $\tau_s^*$ is again monotonic and convex as a function of the spin.  In addition, the upper bound above is in fact saturated for $s\rightarrow \infty$ (this follows from additivity and the fact that the set of operators of the Ising model is continuously connected to free field theory). 
We will present several other examples in the text.

We also discuss the relation of some of our results to small breaking of higher spin symmetry and, through AdS/CFT, to quantum gravity in AdS. 

For the parity preserving Vasiliev theory in $AdS_4$~\VasilievBA\  with higher spin breaking boundary conditions,~\predictioni\ implies for the masses of the higher spin bosons in the bulk
\eqn\masses{
m^2_s \approx 4 \gamma_{\sigma} s+ O(1),  ~ ~ ~ s \gg 1
}
here $\gamma_{\sigma}$ is the anomalous dimension of the spin field in a non-singlet $O(N)$ model and the result is exact in $N$.

The outline of the paper is as follows. In section 2 we discuss DIS and establish~\convexiintro. In section 3 we consider the bootstrap equations in Lorentz signature and derive~\upperbound,\approach,\correctionform. We also review the argument of~\AldayMF\ in this section. In sections 4,5 we consider examples and applications. In section 6 we describe some open problems and conclude. Five appendices complement the main text with technical details.

{\it Note: After this work had been completed, we became aware of a related work by A.~Liam~Fitzpatrick, Jared~Kaplan, David~Poland, and David~Simmons-Duffin~\FitzpatrickYX. }

\newsec{Deep Inelastic Scattering}

{In this section we explain how one can derive inequalities about operator dimensions in CFT by considering RG flows that start in the ultraviolet from the CFT and end up in the gapped phase. } The basic ingredients are as follows. 

\ifig\figone{A relevant perturbation of a CFT that leads to a gapped phase. } {\epsfxsize1.6in\epsfbox{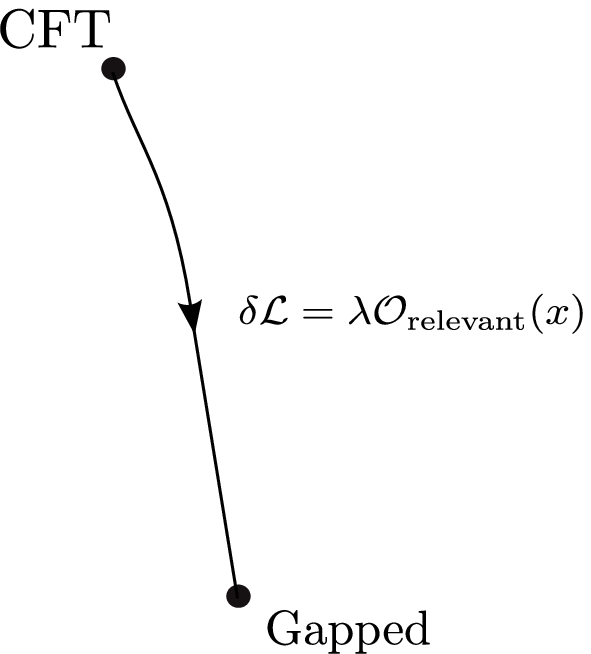}}

Take any CFT (in any number of space-time dimensions, $d$) and assume it can be perturbed by a relevant deformation to a gapped phase (see \figone). Then we take the lightest particle in the Hilbert space, and denote this particle by $|P\rangle$. Its mass is denoted by $M$. 

There may exist CFTs which have no relevant operators whatsoever (``self-organizing'' CFTs), or CFTs which have relevant operators but can never flow to a gapped phase. We are not aware of such examples, and we will henceforth assume that flowing to a gapped phase is possible. 
In the critical $O(N)$ models one can make the following heuristic argument to support this scenario. 

The critical $O(N)$ models can be reached if one tunes appropriately the relevant perturbations of free N bosons. On the other hand, if the mass parameter in the UV is large enough, the model clearly flows to a gapped phase. Now we can decrease the mass gradually until the RG flow hovers very close to the critical $O(N)$ model. In this case we can describe the late part of the flow as a perturbation of the $O(N)$ model by the energy operator, $\int d^3 x\epsilon(x)$.  Indeed, it is a primary relevant operator in the critical $O(N)$ models.\foot{This is known to be certainly true at large $N$ and at some small values of $N$. It seems natural that this would be indeed the case for any $N$.} If there are no phase transitions as a function of this mass, then it means that the critical $O(N)$ models, perturbed by the energy operator, flow to a gapped phase. This is also consistent with the fact that the energy operator corresponds to dialing the temperature away from the critical temperature. 

\ifig\figtwo{When we add a large mass $m$ in the ultraviolet, the model clearly flows to a gapped phase. We can gradually reduce the mass and tune the flow such that it passes close to the critical model. The flow after this stage can be interpreted as a perturbation of the critical $O(N)$ model by the energy operator, $\epsilon(x)$.  } {\epsfxsize2.7in\epsfbox{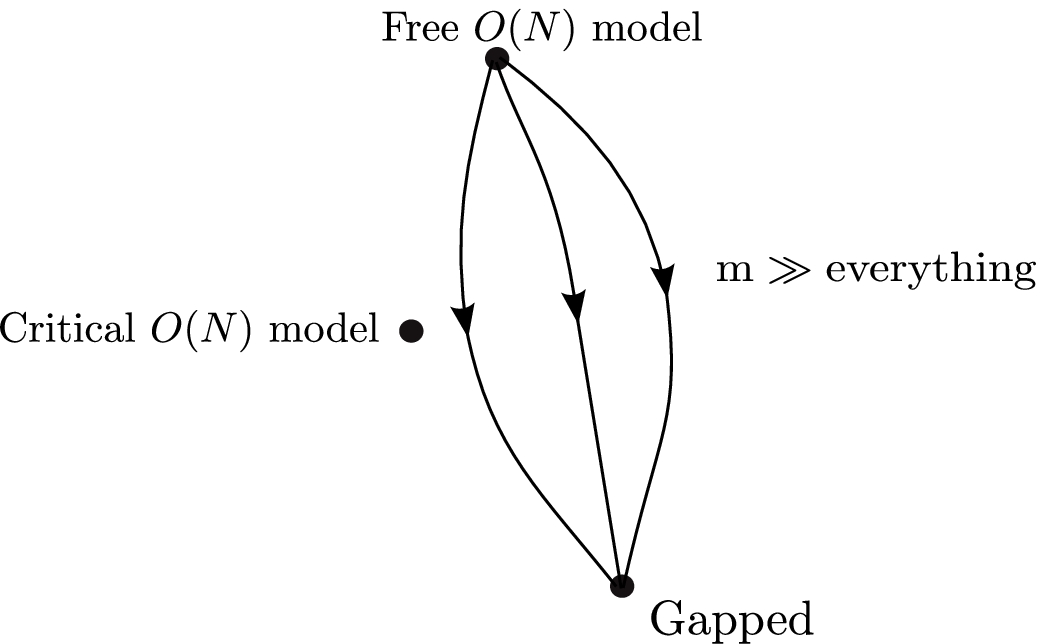}}

This heuristic argument is summarized in \figtwo. In the following we discuss general CFTs, assuming they can be deformed to a gapped phase by adding relevant operators (and maintaining unitarity).

\subsec{General Theory}

Deep Inelastic Scattering (DIS) allows to probe the internal structure of matter. One bombards some state with very energetic (virtual) particles and examines the debris.  Of particular interest is the total cross section for this process, as a function of various kinematical variables.

\ifig\figthree{A lepton emits a virtual, space-like, photon that strikes a hadron. As a result, the hadron generally breaks up into a complicated final state. } {\epsfxsize1.8in\epsfbox{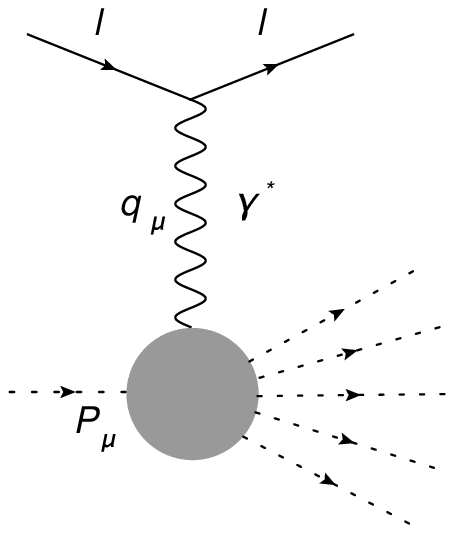}}

 Traditionally, the most common setup to consider is that of \figthree. 
A lepton emits a virtual, space-like, photon that strikes a hadron. 

To study the inner structure of $|P\rangle$ we can try to shoot various particles at it. If the theory (i.e. the CFT and the relevant deformations) preserves a global symmetry, we can couple the conserved currents to external gauge fields and repeat the experiment of \figthree. (We can introduce arbitrarily weakly interacting charged matter particles and make the gauge field dynamical in order to have a perfect  analog of \figthree.) A more universal probe that exists in any local QFT is the EM tensor. We can couple it to a background graviton and consider the same experiment. We can also perform DIS with any other operator in the theory. 
Let us start by reviewing the kinematics of DIS in the case of  a scalar background field $J(x)$. In other words, a background field that is coupled to some scalar operator in the theory $\CO(x)$ via $\int d^dx J(x)\CO(x)$. ($J(x)$ and $\CO(x)$ are taken to be real.)

\ifig\figfour{The total inclusive cross section can be extracted from the imaginary part of a Compton-like scattering where the momenta of the outgoing states are identical to the momenta of the incoming states.} {\epsfxsize4.5in\epsfbox{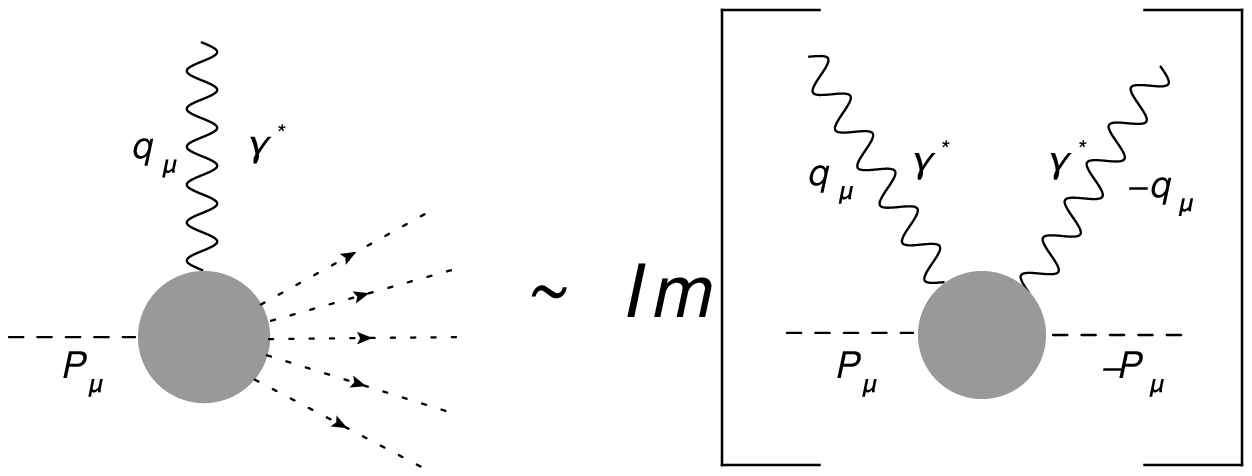}}

The optical theorem allows to extract the inclusive amplitude for DIS via the analog of Compton scattering (see  \figfour). The amplitude in our ``scalar DIS'' setup is 
\eqn\scalarDIS{A(q_\mu,P_\mu)\equiv \int d^dy e^{iqy} \langle P| T\left( \CO(y)\CO(0)  \right) |P\rangle ~, }
where we have denoted the momentum of the target $|P\rangle$ by $P_\mu$. We will only discuss the case of $q^2<0$, i.e. space-like momentum for the virtual particle. 
The above amplitude obviously depends on the mass scales of the theory (we set $P^2=1$ for convenience), and the two invariants $q^2, \nu \equiv 2q\cdot P$.

\ifig\figfive{ The analytic structure in the $\nu$ plane.} {\epsfxsize2.5in\epsfbox{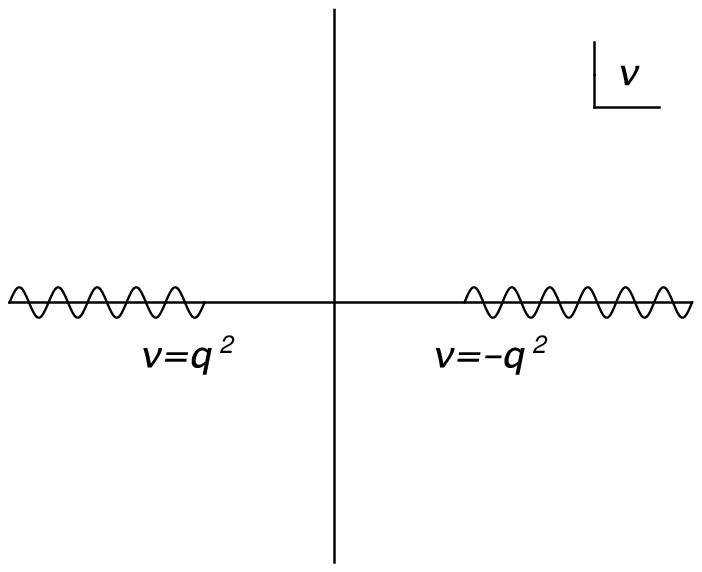}}

We can promote $\nu$ to a complex variable. Since we have assumed the particle $|P\rangle$ is the lightest in the theory, the above amplitude has a branch cut for $\nu\geq -q^2$ and $\nu\leq q^2$. The $\nu$ plane is depicted in \figfive. The optical theorem connects the discontinuity along the branch cut of~\scalarDIS\ to the square of the DIS amplitude. (There are no other cuts since $q$ is space-like.)

To get a handle on~\scalarDIS, we can invoke the OPE expansion. In the case that $\CO(x)$ is real, only even spins contribute. (Odd spins would be allowed if  ${\cal O}^{+} \neq {\cal O}$.) 
\eqn\OPEexpansion{ \CO(y)\CO(0)=\sum_{s=0,2,4,...}\ \sum_{\alpha\in \CI_s  }f_{s}^{(\alpha)} (y)y^{\mu_1}...y^{\mu_s} \CO^{(\alpha)}_{\mu_1...\mu_s}(0)~.}
 Here $s$ labels the spins, and we only write primary fields and ignore descendants since they do not contribute to the correlation function under consideration. The set $\CI_s$ corresponds to the collection of operators of spin $s$. 
 
Conformal symmetry  at small $y$ fixes  the OPE coefficients in~\OPEexpansion\  to be
 \eqn\smallx{ f_{s}^{(\alpha)}(y)=C_{s}^{(\alpha)}y^{\tau^{(\alpha)}_{s}-2\Delta_\CO} \left(1+\cdots\right)~,}
where $\tau^{(\alpha)}_{s}=\Delta_{\CO_{\mu_1...\mu_s}^{(\alpha)}}-s$ is the twist of the operator. The~$\cdots$ denote corrections suppressed by positive powers of $y^2$. These corrections depend on the relevant operators.  Parameterizing these corrections would be unnecessary for our purposes. 

The expectation values of $\CO^{\alpha}_{\mu_1...\mu_s}$ in the state $P$ are parameterized as follows 
\eqn\VEVs{\langle P|\CO^{\alpha}_{\mu_1...\mu_s}(0)|P\rangle=\CA_{s}^{(\alpha)}\left(P_{\mu_1}P_{\mu_2}\cdots P_{\mu_s}-{\rm traces}\right)~,} with the $\CA^{(\alpha)}_s$ some dimensionful coefficients. 

We can now insert the expansion~\OPEexpansion\  into~\scalarDIS\ and find 
\eqn\DISamp{A(q_\mu,P_\mu)\equiv \sum_{s=0,2,4,...}\ \sum_{\alpha\in \CI_s  }\CA_{s}^{(\alpha)}\left(\left(P\cdot {\del\over \del q}\right)^s-{\rm traces} \right)\tilde f_{s}^{(\alpha)} (q) ~.}
Here ``traces'' stands for terms of the form $P^2 \left(P\cdot {\del\over \del q}\right)^{s-2}\left({\del\over \del q}\cdot {\del\over \del q}\right)$ etc.

At this point it is natural to switch to the variable $x=-q^2/\nu$, and write the answer in terms of $x, q^2$. The OPE allows to control easily the leading terms as $-q^2\rightarrow\infty$. Hence, for any given power of $x$ we keep only those terms that are leading in the limit of large~$-q^2$. 

The leading terms take the form 
\eqn\DISampi{A(x,q^2)\equiv \sum_{s=0,2,4,...}\CA_{s}^*C_{n}^*x^{-s}(q^2)^{-\half\tau^*_{s}+\Delta_\CO-d/2} ~.}
Among the set of operators of given spin $s$, $\CI_s$, we select the one that has the smallest twist and denote the corresponding coefficients and twists with an asterisk. In the limit of large $-q^2$ the ``traces'' appearing in~\DISamp\  can be dropped.\foot{The reason is as follows. The nature of the expansion is that we keep all powers of $x$ and for each power we only retain the most dominant power of $q^2$. The trace terms are negligible, because, for instance, the first ``trace term'' scales as $x^{-s+2}(q^2)^{-\half\tau^*_{s}+\Delta_\CO-d/2-2}$. This has to be compared to the contribution from the spin $s-2$ operator,  $x^{-s+2}(q^2)^{-\half\tau^*_{s-2}+\Delta_\CO-d/2}$. We see that the ``trace terms'' can be indeed consistently neglected, for example, if the smallest twist of a spin $n$ operator, $\tau_s^*$, is monotonically increasing as a function of the spin. We will justify this assumption self-consistently later. This subtlety is apparently overlooked in various places. Another way to avoid this issue is to use the spin projection trick \NachtmannMR.} 

We see that the OPE expansion is useful for finite $\nu$ and large $q^2$ (i.e. large $x$ and large $q^2$). The physically relevant configuration is, however, large $q^2$ and $x\in [0,1]$. These two regimes can be related by the usual contour deformations in the complex plane.  

To this end we promote $x$ to a complex variable and recall that the branch cut extends over $x\in [-1,1]$. Being careless for a moment about the behavior near $x\sim 0$, we write sum rules by following the usual trick of pulling the contour from infinite $x$ to the branch cut. We define~$\mu_s(q^2)$ as the s-th moment 
\eqn\sumrules{ \mu_s(q^2)=\int_0^1 dx x^{s-1}{\rm Im} A(x,q^2)~,}
and from~\DISamp\ we infer the sum rule for large $-q^2$ 
\eqn\asympmu{\mu_s(q^2)\rightarrow (q^2)^{-\half\tau^*_{s}+\Delta_\CO-d/2}\CA_{s}^*C_{s}^*~.}

The finiteness of these moments depends on the small $x$ behavior. 
Equivalently, it depends on the behavior of the amplitude for fixed $q^2$ and large $\nu$. This is the Regge limit. We would like to take a conservative approach to the problem of determining the asymptotics in the Regge limit. We simply assume the amplitude is bounded polynomially\foot{For on-shell scattering processes, we have the bound of Froissart-Martin~\refs{\FroissartUX,\MartinRT} $\sigma\leq C(\log(s))^2$, where $C$ is some dimensionful coefficient. In deep inelastic scattering we are dealing with an amplitude involving an off-shell particle (for example, a virtual photon), so the argument of Froissart-Martin does not carry over. In the context of QCD, one can use various phenomenological approaches to model the small $x$ behavior. For example, see the review~\FrankfurtMC. The result of these phenomenological models is that the amplitude grows only as a power of a logarithm with $1/x$. This may be more general than QCD, for example, it would be interesting to understand the small $x$ asymptotics in $\CN=4$~\PolchinskiJW. We thank M.~Lublinsky and A.~Schwimmer for discussions about the Regge limit. }
\eqn\polbound{\lim_{x\rightarrow 0 }A(x,q^2) \leq x^{-N +1 }~,}
for some integer $N$. Polynomial boundedness is discussed in~\EpsteinBG. A recent discussion and more references to the original literature where polynomial boundedness is discussed can be found in~\GiddingsGJ.   In this case the contour manipulation leading to~\sumrules\ can be justified only for $s \geq N$. 
This has to be borne in mind in the following, where we derive some consequences of~\sumrules\ using convexity properties. The first place we are aware of where convexity properties appear in this context is~\NachtmannMR.

\subsec{Convexity}

We will discuss various simple inequalities that the moments $\mu_s(q^2)$ have to satisfy. These properties follow simply from unitarity 
\eqn\unitarity{{\rm Im} A(x,q^2) \geq 0~,}
which is nothing but saying that the cross section for the process of DIS is nonnegative. In fact, ${\rm Im} A(x,q^2)$ must be nonzero at least around some points, otherwise, the scattering is trivial. 

The general inequalities stated below can be proven with the method we are using only for $s \geq N$ with some finite $N$. However, they may or may not be true for smaller spins as well. We will denote the spin from which these inequalities become true by $s_c$. We see that $s_c$ is finite and it is at most $N$. 

From~\sumrules\ it is clear (because of~\unitarity) that $\mu_{s}>\mu_{s+1}$, which together with the sum rule~\asympmu\ leads to \eqn\nondec{\tau^*_s\leq \tau^*_{s+1}~.} Hence, the smallest twist as a function of the spin is a nondecreasing function. There is a stronger conclusion that follows from~\sumrules,\asympmu,\unitarity. That is, take three ordered even integers $s_1<s_2<s_3$, then 
\eqn\convex{\left({\mu_{s_3}\over \mu_{s_1}}\right)^{s_2-s_1}\geq \left({\mu_{s_2}\over \mu_{s_1}}\right)^{s_3-s_1}~.}
This implies for the twists 
\eqn\convexi{{\tau^*_{s_3}-\tau^*_{s_1}\over s_3-s_1}\leq {\tau^*_{s_2}-\tau^*_{s_1}\over s_2-s_1}~.}
Hence, the twist is a monotonic, convex (more precisely, non-concave) function of the spin.\foot{Let us first prove this in the case $s_3=s_2+2=s_1+4$, and we denote $s\equiv s_1$. In this case,~\convex\ is true due to the simple identity $$0<\half\int_{(x,y)\in [0,1]^2} dx dy x^{s-1}y^{s-1}(x^2-y^2)^2 {\rm Im} A(x,q^2) {\rm Im} A(y,q^2)=\mu_{s}\mu_{s+4}-\mu_{s+2}^2~.$$
This establishes local convexity, namely, $$\tau^*_{s+4}-\tau^*_{s}\leq 2\left(\tau^*_{s+2}-\tau^*_{s}\right)~.$$ From this, global convexity~\convexi\ follows trivially. One can also show that without further assumptions on~${\rm Im} A(x,q^2)$ (or some extra input) there is no stronger inequality that follows from~\sumrules,\asympmu,\unitarity.}
We remind again that these inequalities start from some finite spin, $s_c$. In all the examples we checked they are in fact true for all spins above two (at spin two we have the energy momentum tensor, which has the minimal possible twist allowed by unitarity, $d-2$).

\ifig\figsix{ The generic structure of the spectrum of minimal twists in a CFT. } {\epsfxsize2.5in\epsfbox{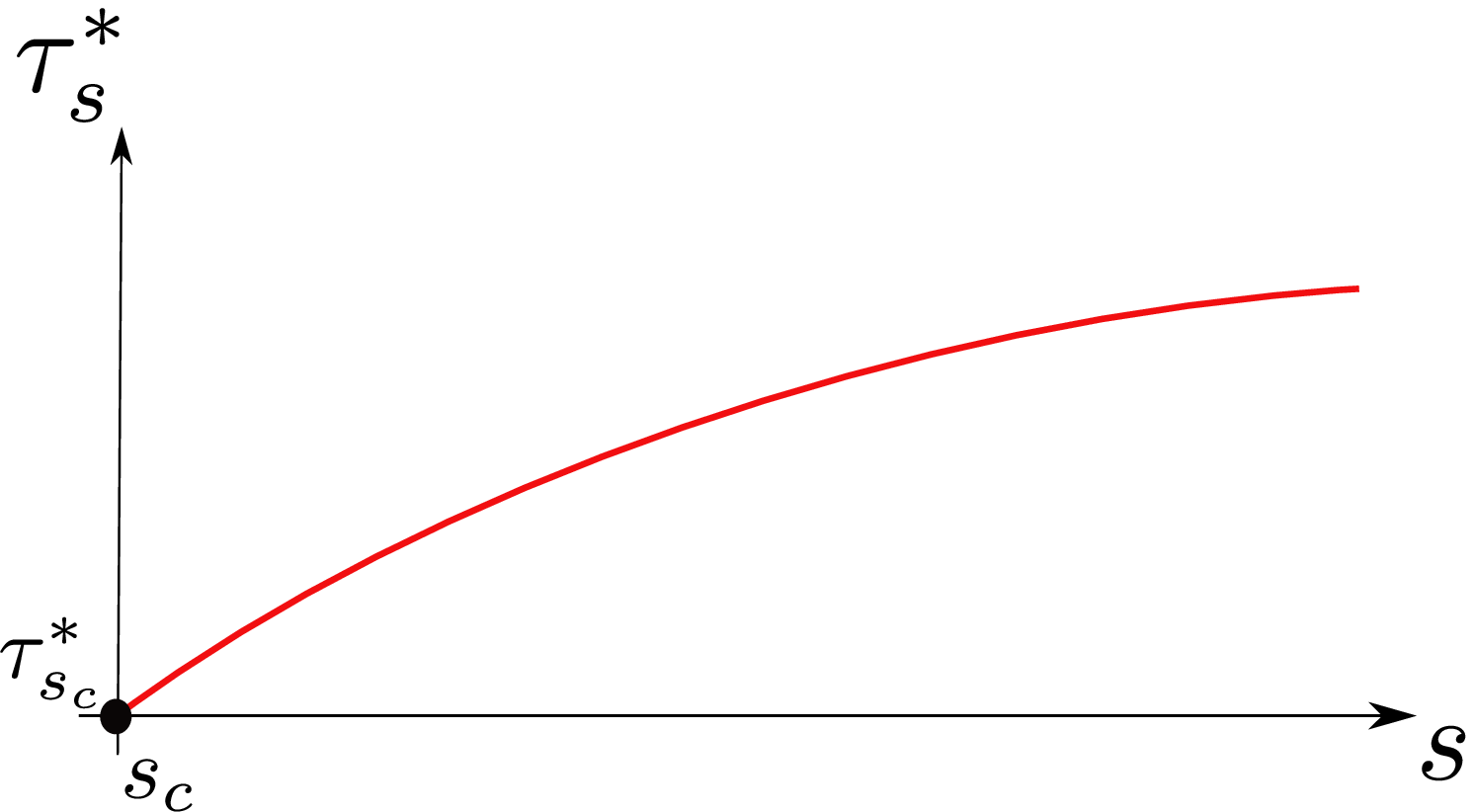}}

The typical resulting spectrum of minimal twists is depicted in \figsix. This is consistent with the known examples of CFTs in $d>2$. The situation is slightly subtler in $d=2$. In two-dimensional CFTs there are infinitely many operators with vanishing twist (these are found in the Verma module of the unit operator). Hence, the minimal twist is zero for all spins. This is however consistent with our analysis here, as we have only shown that $\tau_s^*$ is 
non-concave.  (When all the minimal twists are identical, unitarily~\unitarity\ simply translates to inequalities between the coefficients appearing in~\asympmu.)

So far we have established convexity and monotonicity (starting from some finite spin). It is also natural to ask what is the asymptotic limit of the minimal twist as the spin goes to infinity. We will see that $\tau^*_s$ has a finite and determined limit as $s \rightarrow\infty$. This is the topic of the next section.

\newsec{Crossing Symmetry and the Structure of the OPE}

In this section we analyze the crossing equations in Minkowski space. Using a conformal transformation we can always put any four points on a plane. On this plane it is natural to introduce light-like coordinates, such that the metric on $R^{d-1,1}$ takes the form $d s^2 = d x^+ d x^-+(dx^i)^2$.
We study the crossing equations in the limit when one of the points approaches the light cone of another point. We have explained in the introduction that one should distinguish between the collinear twist $\Delta - s_{+-}$ and the  conformal twist $\Delta - s$. 

Our analysis consists of two parts. First, we analyze how the unit operator constrains general CFTs via the crossing equations. 
We conclude that  in any unitary CFT in $d>2$, given any two operators with collinear twists $\tau^{{\rm coll}}_1$ and $\tau^{{\rm coll}}_2$, we either have in the theory operators with collinear twist precisely $\tau^{{\rm coll}}_1 + \tau^{{\rm coll}}_2$, or there exist operators whose twists are arbitrarily close to $\tau^{{\rm coll}}_1 + \tau^{{\rm coll}}_2$. 

Combined with convexity (which holds in reflection positive OPEs), this implies that in any unitary CFT the minimal twist obeys the following inequality\foot{In some sense, a predecessor of this statement is the so-called Callan-Gross theorem \CallanPU . The discussion of~\CallanPU\ is in the context of weakly coupled Lagrangian theories.}
\eqn\ineqbott{
\lim_{s \to \infty} \tau_{s}^{*} \leq 2 \tau_{{\cal O}}~.
}

So far we have discussed additivity in the space of collinear twists. We then give an argument of why in fact additivity holds in the space of the ordinary conformal twists. Namely, 
{\it given any two operators of twists $\tau_1$ and $\tau_2$, there will be an operator with twist arbitrary close to $\tau_1 + \tau_2$.} This sharper statement motivates the symbolic notation ${\cal O}_{1} \pa^s {\cal O}_{2} $ introduced above.

We present several arguments for this stronger statement, and we also emphasize that it can be understood in the analysis of Alday and Maldacena~\AldayMF. The arguments we give and the discussion in~\AldayMF\ provide very compelling reasons for making this stronger claim. We leave the task of constructing a detailed proof of this statement to the future.\foot{In the setup of~\AldayMF\ this would amount to understanding better some analytic continuations. In our setup this presumably amounts to understanding in more detail the structure of the decomposition of a conformal primary into collinear primaries. }

Thus, the spectrum of any CFT has an additivity property in twist space. It makes sense to talk about double- and multi-twist operators.
We denote the double-twist operators appearing in the OPE of $\CO_1$ and $\CO_2$  symbolically as ${\cal O}_{1} \pa^s {\cal O}_{2} $. The three-point functions $\la \CO_1 \CO_2 \left( {\cal O}_{1} \pa^s {\cal O}_{2} \right) \ra$ are given by the generalized free fields values to leading order in ${1 \over s}$.
By analyzing terms that are sub-leading in the small $z$ limit we conclude that operators which can be symbolically denoted as ${\cal O}_{1} \pa^s \square^n {\cal O}_{2} $ appear as well, and for large spin their twist asymptotes to $\tau_1+\tau_2+2n$. Again the three-point functions  $\la \CO_1 \CO_2  \left( {\cal O}_{1} \pa^s \square^n {\cal O}_{2} \right) \ra$ approach the generalized free fields values to leading order in ${1 \over s}$. In this paper we do not discuss the case $n\neq 0$ in detail.
 
We then include in the crossing equations the operator with the  minimal twist after the unit operator. Using the crossing equations we characterize how the limiting twist $\tau_1+\tau_2$ is approached as the spin is taken to infinity. The corrections to the twist, $\tau_s-\tau_1-\tau_2$, go to zero at large $s$  as some power of $s$ that we  characterize in great detail below.  For simplicity, we mostly concentrate on the case where  ${\cal O}_1$ and ${\cal O}_2$ are scalar operators.

\subsec{The Duals of the Unit Operator}

As a warm-up, consider any large $N$ vector model, and a four-point function of some flavor-neutral operators. (With minor differences the same holds  for adjoint large $N$ theories.) We decompose the four-point function into its disconnected and connected contributions 
\eqn\fourlargeN{\eqalign{
\la {\cal O}(x_1)  {\cal O}(x_2)  {\cal O}(x_3)  {\cal O}(x_4) \ra &= \la {\cal O}(x_1)  {\cal O}(x_2) \ra \la  {\cal O}(x_3)  {\cal O}(x_4) \ra + \la {\cal O}(x_1)  {\cal O}(x_3) \ra \la  {\cal O}(x_2)  {\cal O}(x_4) \ra  \cr
&+ \la {\cal O}(x_1)  {\cal O}(x_4) \ra \la  {\cal O}(x_2)  {\cal O}(x_3)\rangle +\la {\cal O}(x_1)  {\cal O}(x_2)  {\cal O}(x_3)  {\cal O}(x_4) \ra_{conn}.
}}
For any configuration of points and large enough $N$, this correlator is dominated by the disconnected pieces. The connected contribution is suppressed by $1\over N$ compared to the disconnected pieces.  This fact by itself guarantees the existence of the double trace operators, which, as a matter of definition, we can denote ${\cal O} \pa^{s} {\cal O}$, with the scaling dimensions $\Delta_s = 2 \Delta_{{\cal O}} + s + O({1 \over N})$. 

To see that these operators are present and that their dimensions are as above, we simply expand the disconnected pieces in a specific channel. This exercise is reviewed in appendix B.
Of course, it is well known that such operators indeed exist in all the large $N$ theories, and the notation ${\cal O} \pa^{s} {\cal O}$ makes sense because the dimensions of these operators are very close to the sum of dimensions of the ``constituents.''

We would like to prove that a similar structure exists in arbitrary $d>2$ CFT, where for asymptotically large spins we find operators whose twists approach $\tau_1+\tau_2$. 

This universal property of higher-dimensional CFTs can be seen from crossing symmetry and the presence of a twist gap. 
By the twist gap we mean the non-zero twist difference between the unit operator that is always present in reflection positive OPEs and any other operator in the theory. This fact follows from unitarity.

To establish this we consider a four-point correlation function in the Lorentzian domain.\foot{We thank Juan Maldacena for many useful discussions on this topic.} The main idea is to consider the consequences in the s-channel of the existence of the unit operator in the t-channel. In Euclidean space, the consequences of the unit operator in the dual channel were recently discussed in~ \PappadopuloJK. It turns out to constrain the high energy asymptotics of the integrated spectral density. Here we would like to exploit the presence of the unit operator in the Lorentzian domain. 

For simplicity we consider the case of four identical real scalar operators ${\cal O}$ of dimension $\Delta$ and then generalize the discussion. The argument in the most general case goes through essentially verbatim. 

\ifig\figseven{ We consider all four points to lie on a plane, with the following light cone $(z, \bar{z})$ coordinates: $x_1= (0,0)$, $x_2  = (z,\bar{z})$, $x_3  = (1,1)$,  $x_4  = (\infty,\infty)$. We consider the light-cone OPE in the small $z$ fixed $\bar{z}$ channel and then explore its asymptotic as $\bar{z} \to 1$. } {\epsfxsize1.4in\epsfbox{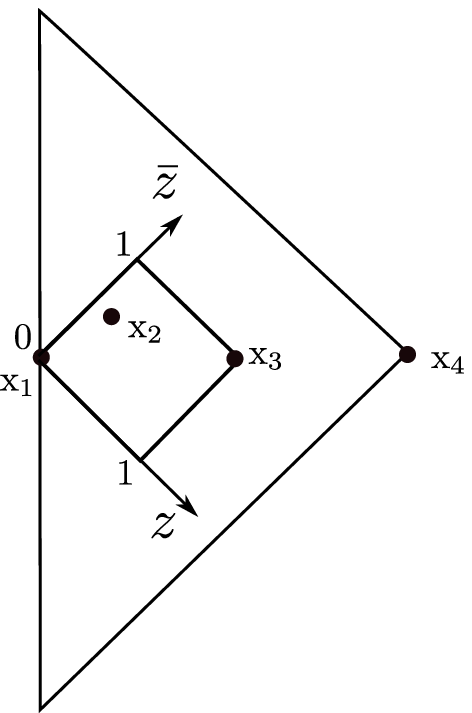}}

Consider the four-point function
\eqn\leadcorrPre{\eqalign{
\la {\cal O}(x_{1})  {\cal O}(x_{2})  {\cal O}(x_{3})  {\cal O}(x_{4}) \ra &= { {\cal F} (z, \bar{z}) \over ( x_{12}^2 x_{34}^2)^{\Delta} }~, \cr
u ={x_{12}^2 x_{34}^2 \over x_{13}^2 x_{24}^2} = z \bar{z}~,~~~ v &={x_{14}^2 x_{23}^2 \over x_{13}^2 x_{24}^2} = (1-z) (1-\bar{z})~.
}} 
Let us introduce two variables which will parameterize the approach to different light cones  $\sigma, \beta \in (0, \infty)$. In this domain all points are space-like separated. The cross ratios are related to $\sigma$ and $\beta$ as follows 
\eqn\crossr{\eqalign{
z &= e^{-2 \beta}, \cr
\bar{z} &= 1 - e^{-2\sigma}.
}}
 Note that $z$ and $\bar z$ are two independent real numbers. The reason for the notation we are using is that in Euclidean signature $z$ is a complex number and $\bar z$ is its conjugate. A convenient choice of a four-point function realizing the notation above is depicted in \figseven.

Below we present two versions of the argument. In the first version we  exploit crossing symmetry in Minkowski space. We use some basic properties of the light-cone OPE and show how crossing symmetry can lead to certain constraints on the operators spectrum of the theory.  We find that crossing symmetry leads to a certain ``duality'' between fast spinning operators with very large dimensions and the low-lying operators of the theory.  Here we compute the first two orders in the large spin expansion that we introduce. The second version of the argument relies on the picture of~\AldayMF\ for fast spinning operators. In this nice, intuitive, picture many facts that we derive from the bootstrap equations have a straightforward interpretation in terms of particles that interact weakly with each other in some gapped theory. We begin by presenting the approach based on the bootstrap equations, and then re-interpret some of the facts we find using the ideas of~\AldayMF.

\subsec{An Argument Using the Light-cone OPE}

 We would like to consider the s-channel light-cone OPE expansion $z \to 0^+$ and $\bar{z}$ fixed. As we have reviewed in the introduction,  this expansion is governed by the twists of the operators. More precisely, it is governed by the collinear twist $\Delta - s_{-+}$. Any primary operator can be decomposed into irreducible representations of the collinear conformal group and one can easily see that, for any operator, the minimal collinear twist coincides with the usual conformal twist $\Delta - s$. An operator with collinear twist $\tau^{{\rm coll}}$ contributes ${\cal F}(z , \bar z) \sim z^{{\tau^{{\rm coll}} \over 2 }}\CF(\bar z)$. The function $\CF(\bar z)$ is a partial wave of the collinear conformal group. We quote some properties of it when needed.
As usual, by re-summing the light-cone OPE expansion we can recover the correlation function at any fixed $z$ and $\bar z$. 

In the notation we defined after~\leadcorrPre, the light-cone OPE expansion in the s-channel corresponds to a large $\beta$, fixed $\sigma$, expansion
\eqn\lightcone{
{\cal F} (\beta, \sigma) = \sum_{i} e^{- \tau_{i}^{coll} \beta} f_{i} (\sigma) = \int_0^{\infty} d \tau e^{- \tau^{{\rm coll}} \beta} f(\sigma, \tau).
}
We assume this OPE to be convergent in the region where all points are spacelike separated, namely $\sigma, \beta \in (0, \infty)$. 

The s-channel unit operator gives the most important contribution for finite $\sigma$ and large $\beta$. 
Let us now consider a different regime. We keep $\beta$ fixed and start increasing $\sigma$. In this way our coordinate $\bar z$ approaches 1 and hence gradually becomes light-like separated from the insertion $x_3$. If we take $\sigma\gg \beta$ the four-point function is dominated by the t-channel OPE, and the most dominant term comes from the unit operator in the t-channel. 
We can estimate the behavior of~\lightcone\ in this limit:  
\eqn\limitanot{
{\cal F} (\beta, \sigma) \sim e^{- 2 \tau_{{\cal O}} \beta} e^{2 \tau_{{\cal O}} \sigma}\left( 1 + O(e^{-\sigma} ) \right)~.
}
Operators other than the unit operator in the t-channel contribute to the exponentially suppressed terms. 
Here we are using the fact that there is a twist gap in $d>2$ CFTs. 

One can naively conclude that \limitanot\ implies that the spectral density is non-zero at $\tau^{{\rm coll}} = 2 \tau_{{\cal O}}$. This is, however, not necessarily true. Imagine we have a series of operators with twists $2 \tau_{{\cal O}} + a_i$ such that $a_{i} \to 0$ when $i \to \infty$. Then for arbitrary large $\sigma$ there will be $a_i < e^{- \tau_{\min} \sigma}$ such that their twist will be effectively $2 \tau_{{\cal O}}$ to the precision we are probing the correlation function. Thus, equation~\limitanot\  only implies that the spectral density $ f(\sigma, \tau)$ is non-zero in any neighborhood of $\tau^{{\rm coll}} = 2 \tau_{{\cal O}}$. 
This shows that there are operators with collinear twist in any neighborhood of $\tau^{{\rm coll}} = 2 \tau_{{\cal O}}$. 

The discussion above concerned with the collinear twists. As mentioned in the introduction to this section, we claim a slightly stronger result -- that there are operators whose conformal twists are arbitrarily close to  $\tau = 2 \tau_{{\cal O}}$. 

Let us explain the motivations for this stronger claim. First of all, as we will review later, this follows from the construction of~\AldayMF. Second, if $\tau = 2 \tau_{{\cal O}}$ had not been the minimal collinear twist in the decomposition of a conformal primary into collinear primaries, then the conformal twist of the corresponding conformal primary would have been $2\tau_\CO-n$ for some nonzero integer $n$. That  conflicts with unitarity for some choices of $\CO$.   Finally, a heuristic argument:  different collinear primaries that follow from a single conformal primary are obtained by applying $\del_z$ to the collinear primary with the lowest collinear twist. However, since we are interested in the behavior for small $1 - \bar z$ (this is the $e^{2\tau_\CO \sigma}$ in~\limitanot), higher collinear twists are expected to produce the same (or weaker)  singularity as they roughly transport ${\cal O} $ in the $z$ direction. 

We conclude that there must be operators with twists arbitrarily close to $2\tau_\CO$. This follows from the presence of the unit operator in the t-channel. We can say more about these operators. An easy lemma to prove is that we must have infinitely many operators with twists $2\tau_\CO$ (or arbitrarily close to it). The argument is simply that conformal blocks behave logarithmically as $\bar z\rightarrow 1$ and hence any finite number of them cannot reproduce the exponential growth of~\limitanot.

Let us also comment about the three-point functions of these operators. To fix them we expand the leading term ${ z^{\Delta} \over (1 - \bar z)^{\Delta}}$ in terms of s-channel collinear conformal blocks. This can be done using the generalized free fields solution of the crossing equations. The $\bar z \to 1$ limit is dominated by the large spin operators. To reproduce the leading $\bar z \to 1$ asymptotics correctly, the three-point functions $\la \CO_1 \CO_2 \left( {\cal O}_{1} \pa^s {\cal O}_{2} \right) \ra$ should coincide with the ones of generalized free fields to the leading order in ${1 \over s}$. 

So far we analyzed the consequences of matching the ${ z^{\Delta} \over (1 - \bar z)^{\Delta}}$ piece from the unit operator contribution to the s-channel OPE. However, we can consider the complete dependence on $z$ coming from the unit operator in the t-channel, ${ z^{\Delta} \over (1-z)^{\Delta} (1 - \bar z)^{\Delta}}$. Clearly, the primary operators discussed so far are not enough. Again the theory of generalized free fields reproduces the necessary function by  means of additional primary operators of the form $\CO \pa^s \square^n \CO$. By taking the large spin limit of this solution (equivalently, $\bar z\rightarrow 1$) we conclude that operators with twists $2 \tau_{\CO} + 2 n$ and 
three-point functions approaching those of generalized free fields are always present in the spectrum, of any CFT.

All these facts naturally combine into the following picture (that we will henceforth use): CFTs are free at large spin -- they are given by generalized free fields. More precisely, in the OPE of two arbitrary operators $\CO_1(x)$ and $\CO_2(x)$, at large enough spin, there are operators whose conformal twists are arbitrarily close to $\tau_{1}+\tau_2$. We  denote these operators symbolically by  ${\cal O}_1 \pa^s {\cal O}_2$. Their twists behave as  $ \tau_{1} + \tau_2+O({1 \over s^{\alpha} } )$ where the power correction will be discussed in the next subsection  ($\alpha$ is some positive number). These operators, ${\cal O}_1 \pa^s {\cal O}_2$, are referred to as {\it double-twist operators}. Their three-point functions asymptote the ones of the theory of generalized free fields. Similar remarks apply to $\CO \pa^s \square^n \CO$.

In principle, one can have either of the following two scenarios:

\item{a)} Operators with the twist $\tau =  \tau_{1}+\tau_2$ are present in the spectrum.

\item{b)} The point $\tau =  \tau_{1}+\tau_2$ is a limiting point of the spectrum at infinite spin.

By option b) we mean that in a general CFT there is a set of operators with large enough spin such that their twist is $ \tau_{1}+\tau_2 - O({1 \over s^{\alpha} })$ with $\alpha>0$. 
In this way, for arbitrary $\eps>0$, there will be operators $X$ with twist $ \tau_{1}+\tau_2 - \tau_{X} < \eps$. They will be responsible for the behavior~\limitanot.\foot{At one loop this was observed, for example, in \DerkachovPH,\KehreinIA.  Here we see this is a general property of any CFT above two dimensions.}

Strictly speaking, so far we only discussed scalar external operators. However, the generalization to operators with spin is trivial in this case. Indeed, the usual complication of having many different structures in three point functions for operators with spin is irrelevant for us. The reason is that our problem is effectively two dimensional, and by considering, for example, operators with all the indices along the $z$ direction $\la {\cal O}_{z...z} \tilde{{\cal O}}_{z...z} \tilde{{\cal O}}_{z...z} {\cal O}_{z...z}\ra$ the argument above goes through. 

Indeed, in the case of four identical real operators, by keeping only unit operators in the s- and t-channel we get
\eqn\genfree{\eqalign{
\la {\cal O}_{z...z} {\cal O}_{z...z} {\cal O}_{z...z} {\cal O}_{z...z}\ra &= {z^{2 s} \over (z \bar z)^{\Delta + s}} + {(1-z)^{2 s} \over  \left[ (1 - z) (1 - \bar z) \right]^{\Delta + s}} \cr
&=  {z^{2 s} \over (z \bar z)^{\Delta + s}}  \left[1 + \left({z \over 1 - z}\right)^{\Delta - s} \left({\bar z \over 1- \bar z}\right)^{\Delta + s} \right].
}}
Again we see that $z^{\Delta - s}$ corresponds to the $\tau^{{\rm coll}} = 2 \tau_{\CO} = 2 (\Delta - s)$ operators in the s-channel. The crucial simplification is that our perturbation theory is effectively 2d and, thus, all the usual complications of three-point functions of operators with spin are absent.

Notice that our conclusion are obviously incorrect in $d=2$. There are many $2d$ theories where the double-twist operators are absent. As we have mentioned above, this is due to the fact that in two dimensions there is no twist gap above the unit operator.  

This is also related to why some theories in $2d$ can be very simple: the absence of the twist gap provides a way out from  this additivity property. Then we can have a very simple twist spectrum. For example, in minimal models the twist spectrum takes the form $\tau_i + n$, where $n$ is an integer that corresponds to the presence of the Virasoro descendants, and $\tau_i$ is some finite collection of real numbers. Such a spectrum is generally inconsistent in higher dimensions.

As a simple illustration of how this result is evaded in two dimensions, we can consider the four-point function of spin fields in the 2d Ising model~\BPZ . In this case the only Virasoro primaries have
quantum numbers $(0,0)$, $({1 \over 2}, {1 \over 2})$, and $({1 \over 16}, {1 \over 16})$ for the spin field. One can check explicitly that the leading piece in the $\bar z \to 1$ limit has the the leading twist contribution $z^{{\tau \over 2} }$ with $\tau = 0$,  consistently with the known spectrum. The spectral density around $2\tau_\sigma$ vanishes.

\subsec{Relation to Convexity}

\ifig\figeight{The general form of the minimal twist spectrum in an arbitrary CFT in $d>2$. } {\epsfxsize2.0in\epsfbox{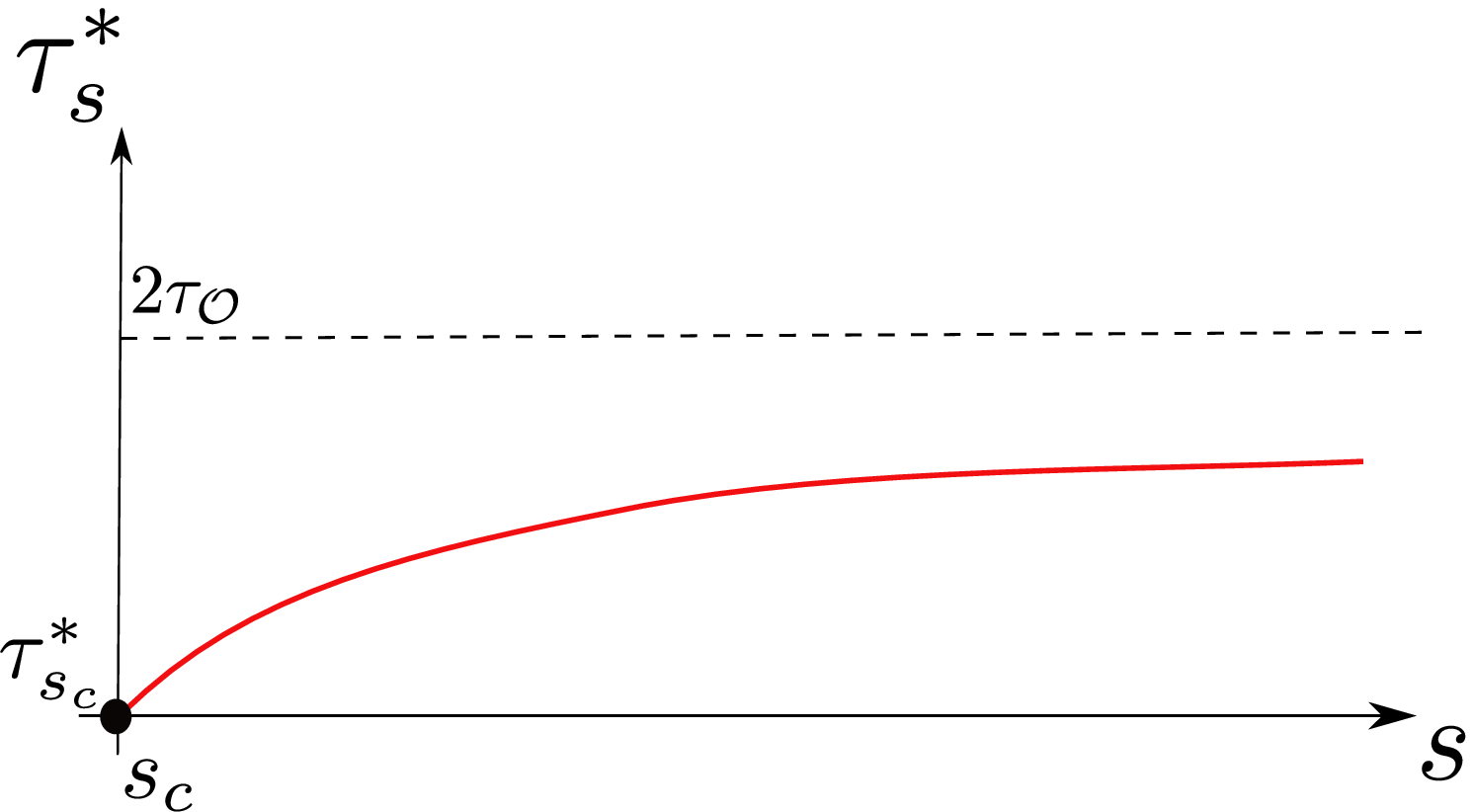}}

The discussion above allows us to put an upper bound on the minimal twist of the large spin operators appearing in the OPE of any operator with its Hermitian conjugate in $d>2$ CFTs. The upper bound is, thus, $2 \tau_\CO$. This leads to a compelling picture of the spectrum of general CFTs if we combine this observation with the main claims of section~2,~\nondec,\convexi. $\tau^*_s$ is thus a nondecreasing convex function that asymptotes to a finite number, not exceeding $2\tau_\CO$. See~\figeight. In addition, as we emphasized above, there must be some operators populating the region of twists around $2\tau_\CO$. These may be operators of non-minimal twist in the OPE.

An interesting direct application of these results is for the Euclidean bootstrap algorithm of~\RattazziPE. If we assume that $s_c = 2$ (which is consistent with all the examples we know), then we see that {\it imposing the gap condition for operators with $s>2$,  $\tau_{s}^{*} >  \tau_1+\tau_2$ should not have any solutions (in unitary conformal field theories) }.  

\subsec{The Leading Finite Spin Correction}

Let us, for simplicity, focus on the case of four scalar operators of the type $\langle \CO\CO^\dagger\CO\CO^\dagger\rangle$ where the dimension of $\CO$ is $\Delta$. (The case of non-identical scalar operators is considered in appendix B.) The crossing equation in this case takes the form
\eqn\crossingequationAAA{
{\cal F} (z, \bar z) = \left( {z \bar z \over (1 - z) (1 - \bar z)} \right)^{\Delta} {\cal F} (1-z, 1- \bar z)~.
}

So far we have focused on the limit where we probe the t-channel OPE, $\bar z\rightarrow 1$ with $z$ finite. We have taken into account the contribution of the unit operator in this limit. Let us now consider the next operator that contributes in this limit. This operator has the smallest twist among the operators appearing in the OPE (excluding the unit operator). We denote this smallest twist by  $\tau_{\min}$.  We denote the dimension and spin of this operator by $\Delta_{min}$, $s_{\min}$ such that $\Delta_{\min}-s_{\min}=\tau_{\min}$. The twist gap $\tau_{\min}>0$ in theories that live in more than two space-time dimensions guarantees that the contribution of such operators in~\limitanot\ is exponentially suppressed by $e^{-\sigma \tau_{\min}}$ compared to the unit operator contribution. 

It would be very important for our purpose to determine the contribution of this operator in more detail. For this we need the conformal block corresponding to this operator. The general conformal block entering the s-channel is denoted by $g_{\Delta, s} (z , \bar z)$. Since in this case we are doing it for the t-channel we need to evaluate it with the arguments $g_{\Delta, s} (1-z , 1-\bar z)$. 

One useful property of conformal blocks that we need is their behavior near the light-cone. In the limit $\bar z\rightarrow 0$ with fixed $z$ the conformal block is given by $\bar z^{\tau\over 2} \CF(z)$, where $\CF$ is simply related to the hypergeometric function $_2F_1$. This is true in any number of dimensions. We can consider $\CF(z)$ as $z$ approaches 1. We find the following leading asymptotic form~\DolanDV
\eqn\contr{
g_{\Delta, s} (z , \bar z) \rightarrow - {\Gamma (\tau +  2s) \over (-2)^{s} \Gamma \left({\tau + 2 s \over 2}\right)^2} \bar z^{ { \tau \over 2 } } \left(\log \left(1-z\right)+\CO(1)\right) .
}
In~\contr\ we have taken the limit $\bar z\rightarrow 0$ first and subsequently we let $z$ approach $1$.

Combining all the factors we arrive at the following result for~\lightcone\ in the limit  $\sigma\gg\beta$
\eqn\finalanswerforF{ 
\CF(z,\bar z)=\left({z\bar z\over (1-z)(1-\bar z)}\right)^{\Delta_{{\cal O}}} \left(1- \tilde f^2{\Gamma (\tau_{\min} +  2 s_{\min}) \over (-2)^{s_{\min}} \Gamma \left({\tau_{\min} + 2 s_{\min} \over 2}\right)^2} \left(1-\bar z\right)^{ { \tau_{\min} \over 2 } } \log \left(z\right) \right) +\cdots~.
}
Here $\cdots$ stand for operators with higher twist than $\tau_{\min}$ and for various subleading contributions from the $\CO_{min}$ conformal block. The contributions from operators with higher twist than $\tau_{\min}$ are further suppressed by powers of $1-\bar z$ (this translates to exponential suppression in $\sigma$ in the limit $\sigma \gg 1$ and $\beta$ fixed).  $f^2$ is the usual coefficient appearing in the conformal block decomposition. It is fixed by the three point function $\langle\CO_{min}\CO\CO^\dagger\rangle$ and the two-point functions of these operators.

We rewrite~\finalanswerforF\ in terms of $\beta$ and $\sigma$. We get
\eqn\finalanswerforFi{ \CF(\beta,\sigma)=\left({e^{-2\beta}(1-e^{-2\sigma})\over (1-e^{-2\beta})e^{-2\sigma}}\right)^{\Delta_{{\cal O}}} \left(1+2\beta \tilde f^2{\Gamma (\tau_{\min} +  2 s_{\min}) \over (-2)^{s_{\min}} \Gamma \left({\tau_{\min} + 2 s_{\min} \over 2}\right)^2} e^{-\tau_{\min} \sigma}\right)+\cdots~.}
Now we need to interpret this result in the s-channel. Consider first the case of unit operator. Its contribution can be accounted for in terms of the s-channel conformal blocks of operators  that have twist $2 \Delta$ and large spin (as in the theory of generalized free fields). In fact, as we show in appendix B, by doing a saddle point analysis one can see that the leading contribution comes from operators with spin 
\eqn\dominantspin{
\log s_{{\rm dom}} = \sigma + \log {(2 \Delta - 1)^{{ 3 \over 2 }} (2 \Delta + 1)^{ {1 \over 2} } \over 4 \Delta} + O(e^{- \sigma})~.
}
The second derivative at the saddle point is parametrically large only when $\Delta\rightarrow\infty$. (In this case the formula simplifies to $\log s_{{\rm dom}} = \sigma + \log \Delta + O(e^{- \sigma})$.) Otherwise, the saddle point is not localized and its precise location is not very meaningful. It is however meaningful that when $\sigma$ becomes large the dominant spins are large approximately as~\dominantspin\ dictates, $s_{{\rm dom}}\sim e^{\sigma}$. Below we will ``re-sum'' all the contributions around the saddle point in order to get various detailed predictions. 

Thus, we are building a perturbation theory around the point $s = \infty$, which is dominated by the unit operator in the t-channel and double-twist operators in the s-channel. Note that this perturbation theory is very different from the one in~\HeemskerkPN, where the expansion is  for large $N$ theories. The expansion we are considering is valid in any CFT in $d>2$. 

In~\finalanswerforFi\ we have included the contribution of the first leading operator in the t-channel, after the unit operator. Unitarity guarantees that  $\tau_{\min} < 2 \Delta$. Notice that $\tau_{\min} \leq d-2$ (the upper bound comes from the energy momentum tensor, which is always present), while $\Delta \geq {d -2 \over 2}$. Thus, equality is only possible for free fields. This is important because it means that this contribution to $\CF(\beta,\sigma)$ from $\tau_{\min}$ is rising exponentially. Hence, it cannot be accounted for by adding finitely many operators in the s-channel, or changing finitely many three-point functions. The saddle point~\dominantspin\ suggests that this contribution from $\tau_{\min}$ should be accounted for by introducing small corrections to operators with very large spins.  

Taking into account the exponentially suppressed correction proportional to $e^{-\tau_{\min}\sigma}$ in~\finalanswerforFi, the twists of the operators propagating in the s-channel are now modified.  The modification is very small for large enough spin because of the relation~\dominantspin\ between the spin and $\sigma$.
Treating this term proportional to  $e^{-\tau_{\min}\sigma}$  as a perturbation, to leading order we can thus replace $e^{-\tau_{\min}\sigma}$ by $1/s^{\tau_{\min}}$, where $s$ is the spin. Then we can interpret this perturbation as a correction to the twist, $\delta\tau_s$, of such high spin operators by matching the dependence on $\beta$.\foot{In other words, we use $e^{-\beta\tau-\beta\delta\tau}=e^{-\beta\tau}\left(1-\beta\delta \tau+\cdots\right)$. } We find that for large $s$ 
\eqn\approachtolimit{\delta \tau_s=-{c_{\tau_{\min}}\over s^{\tau_{\min}}}+\CO\left({1\over s^{\tau_{\min}+\epsilon}} \right)~,}
where $c$ is some constant independent of the spin and $\epsilon>0$. If these double-twist operators are also the minimal twist operators in the OPE (as we will discuss in some examples), then we know from monotonicity of the twist that $c_{\tau_{\min}}>0$. We see that the limiting twist $2\tau_\CO$ is approached in a manner that is completely fixed by the operator with the lowest lying nontrivial twist in the problem. For example, it could be some low dimension scalar operator, or the energy momentum tensor (for which $\tau_{\min}=d-2$).

It is natural to ask what can be said about $c_{\tau_{\min}}$. Here we compute it for the case of external scalar operators of dimension $\Delta$. As explained in appendix B, the relevant contribution from the s-channel takes the form\foot{For the small spin cut-off we chose $\Lambda $. Since the sum is dominated by large spin operators in the limit we are considering, nothing will depend on $\Lambda$ in an important way.}
\eqn\lhs{
-{1 \over 2} \log z \ z^{\Delta} \times \lim_{\bar z \to 1} \sum_{s=\Lambda}^{\infty} {c_{\tau_{\min}} \over s^{\tau_{\min}} } c_s\bar z ^{\Delta + s} \ _{2} F_{1} (\Delta + s, \Delta + s, 2 s + 2 \Delta, \bar z)~.
}
First we switch from the sum to the integral $\sum_{s=0}^{\infty} \to \int d s$ and also make a change of variables motivated by the saddle point analysis of appendix B, $s \to {s \over \sqrt{1 - \bar z}}$. Denoting $\eps = 1 - \bar z$ and using for the $\eps \to 0$ limit
\eqn\result{
_{2} F_{1} ({h \over \sqrt{\eps}}, {h \over \sqrt{\eps}}, 2 {h \over \sqrt{\eps}}, 1 - \eps) \sim 4^{{h \over \sqrt{\eps}}} {\sqrt{h} K_{0} (2 h)  \over \sqrt{\pi} \eps^{1 \over 4}}~,
}
(which can be derived using the integral representation for the hypergeometric function) we get (in the small $\eps$ limit) that~\lhs\ is equal to
\eqn\smalleps{\eqalign{
&{c_{\tau_{\min}} \over (1-\bar z)^{\Delta - {\tau_{\min} \over 2}} } {4 \over \Gamma(\Delta)^2} \int_{0}^{\infty} d s \ s^{2 \Delta - \tau_{\min} - 1}  K_{0} (2 s) \cr
&= {c_{\tau_{\min}} \over (1-\bar z)^{\Delta - {\tau_{\min} \over 2}} } { \Gamma(\Delta - {\tau_{\min} \over 2})^2 \over \Gamma(\Delta)^2}~. 
}}

Thus, the crossing equation, i.e. the requirement that~\lhs\ is equal to~\finalanswerforF,  takes the form 
\eqn\crossingeq{
c_{\tau_{\min}}  { \Gamma(\Delta - {\tau_{\min} \over 2})^2 \over \Gamma(\Delta)^2}  = 2 \tilde f^2 {\Gamma (\tau_{\min} +  2 s_{\min}) \over (-2)^{s_{\min}} \Gamma \left({\tau_{\min} + 2 s_{\min} \over 2}\right)^2}~,
}
which can be solved for $c_{\tau_{\min}}$ to get 
\eqn\correctionformx{\eqalign{
c_{\tau_{\min}}  &=  {\Gamma (\tau_{\min} +  2 s_{\min}) \over 2^{s_{\min}-1} \Gamma \left({\tau_{\min} + 2 s_{\min} \over 2}\right)^2} {\Gamma (\Delta)^2 \over \Gamma (\Delta  - {\tau_{\min} \over 2} )^2} f^2, \cr
f^2 &= { C_{{\cal O} {\cal O}^{\dagger} {\cal O}_{\tau_{\min}}}^2 \over  C_{{\cal O} {\cal O} } C_{{\cal O}^{\dagger} {\cal O}^{\dagger} }  C_{{\cal O}_{\tau_{\min}}{\cal O}_{\tau_{\min}}}} = (-1)^{s_{\min}} \tilde f^2 ~.
}}
The additional factor $(-1)^{s_{\min}}$  in the second line of~\correctionformx\ comes about because $C_{{\cal O} {\cal O}^{\dagger} {\cal O}_{\tau_{\min}}}$ differs from $C_{{\cal O}^{\dagger} {\cal O} {\cal O}_{\tau_{\min}}}$ by a minus sign for odd spins.
This will be important below.

For convenience, above we have included the expression connecting $f^2$ with various two- and three-point functions (our conventions for two- and three-point functions are summarized in appendix A). Thus, we have computed the coefficient that controls the leading correction to the anomalous dimension of double-twist operators in any CFT $\tau_{{\cal O} \pa^s {\cal O}^\dagger } = 2 \tau_{{\cal O}} - {c_{\tau_{\min}} \over s^{\tau_{\min}}} +\cdots$. Let us emphasize several features of this formula: 

\item{-} It is manifestly invariant under redefinitions of the normalizations of operators;
\item{-} The formula is valid for any $d>2$ CFT and does not depend on the existence of any perturbative expansion parameters. Thus, it is an exact result. 
\item{-} The sign of all the corrections is positive, both for odd and even spins. In the next section we will consider more general situations where different contributions come with different signs. Here, since all the contributions are positive, the spectrum of twists of ${\cal O} \pa^s {\cal O}^\dagger$ is clearly convex for large enough spin, in any CFT.
\item{-} If the minimal twist operator $\CO_{min}$ is not unique, we just have to sum the different contributions $c = \sum c_{i}$;
\item{-} From unitarity we know that $\tau_{\min}<2\Delta$, and also $\tau_{\min}\leq d-2$. In section 5 we will see an example where all the contributions with $\tau_{\min}\leq 2\Delta$ cancel out. Then the leading contribution to the asymptotic anomalous twist will come from $\tau_{\min}>2\Delta$, and our formula reproduces it correctly.
\item{-} The formula should be modified to incorporate the case when the external operators are not scalars. We expect that this should be manageable because the problem is effectively two dimensional and all conformal blocks are known \OsbornVT. We do not pursue this here.

\medskip

\noindent This formula has several interesting consequences and applications that we study in the next section. The following subsection is dedicated to a brief review of the picture of~\AldayMF\  and its connection to the discussion above.

\subsec{An Argument in the Spirit of Alday-Maldacena}

The general results discussed in the previous subsections can be naturally understood at the qualitative level using the ideas  of~\AldayMF. Such ideas were also applied in~\AldayZY . 

One can think about the four-point correlation function in a $d$-dimensional CFT as a four-point function in a two-dimensional gapped theory. Denote the coordinates on this two-dimensional space by $(u, v)$.  The gap in the spectrum is the twist gap of the underlying $d$-dimensional CFT. More generally, the evolution in the $u$ direction is dictated by the twist of the underlying $d$-dimensional CFT. The operators could be thought as being inserted at $(\pm u_0, \pm v_0)$. 

When operators are largely separated in the $v$ direction the interaction is weak and corrections to the free propagation $e^{- 2 \tau_{{\cal O}} u_0 }$ are small. 
The corrections are governed by the separation $v_0$. 
Since the theory is gapped, the correction to the energy due to the exchange of a particle of ``mass'' $\tau_{{\rm exch}}$ takes the form $e^{- \tau_{{\rm exch}} v_0}$, which could be made arbitrarily small by taking large enough $v_0$. 

Since $\pa_{u}$ measures the twist we see that in this limit of large $v_0$ we have a propagating state with the twist arbitrarily close to $ 2 \tau_{{\cal O}}$. By the state-operator correspondence we have to identify it with some operators in the theory. This is the additivity property alluded to above. 

Moreover, \AldayMF\ identified how the $v_0$ separation is related to the spin of the operators. They suggested the relation $\log s_{{\rm dom}} \sim v_0 $. That allows to interpret the correction potential energy due to the leading interaction, $e^{- \tau_{{\rm exch}} v_0}$, in terms of a correction to the twist of high spin operators, $\delta \tau\sim {c \over s^{\tau_{exch}}}$. 
For large $v_0$ we the most important exchanged particle is the one with the smallest twist, hence $\tau_{exch}$ is identified with $\tau_{\min}$. This is the statement of~\approachtolimit.
The fact that at large spin we approach the generalized free field picture corresponds to the fact that at large separations excitations do not interact and propagate freely. (In other words, locality in the $v$ coordinate.)

We see that the free propagation in the Alday-Maldacena picture corresponds to the unit operator dominance in the OPE picture. Corrections due to the exchange of massive particles in the Alday-Maldacena picture correspond to the inclusion of the next terms in the light-cone OPE. Also in this picture the state-operator correspondence plays an important role.

The relation $\log s_{{\rm dom}} \sim v_0 $ of~\AldayMF\ is qualitatively correct, but there are very important corrections to it that we could determine precisely in our formalism~\dominantspin. (And as we explained, the saddle point itself is wide unless $\Delta\rightarrow\infty$, so one needs to re-sum the corrections around it. This is what we have essentially done in~\result,\smalleps.) Those corrections play a central role in the determination of the precise shift in the twists of fast spinning  operators. In the picture of~\AldayMF\ those corrections should result from taking into account the finite width of the wave function in the $v$ coordinate. This would be interesting to understand better, perhaps along the lines of~\AldayZY.

\newsec{Examples and Applications of (3.17)}

In this section we study the correction formula~\correctionformx\ in different regimes and circumstances, comparing it to known results when possible.

\subsec{Correction due to the Stress Tensor and Theories with Gravity Duals}

It is well known that in the case of the stress tensor the coupling of it to other operators is universal and depends only on the dimension of the operator 
and two-point function of the stress tensor. More precisely, in~\correctionformx\ $f^2 = {d^2 \Delta^2 \over (d-1)^2 c_{T}^2} $ so that we get
\eqn\stresstensorcorrection{
c_{stress} =  {d^2 \Gamma(d+2) \over 2 c_{T} (d-1)^2 \Gamma({d+2 \over 2})^2}  {\Delta^2 \Gamma (\Delta)^2 \over \Gamma (\Delta  - {d-2 \over 2} )^2}.
}

Let us consider a CFT where in the OPE of certain operators the minimal twist (after the unit operator) is realized by the stress tensor. Simple examples of such CFTs arise in the context of holography. 
In this case the two-point function is given by (see for example~\KovtunKW)
\eqn\twopgrav{
c_{T} = {d + 1 \over d - 1}  {  L^{d-1} \over 2 \pi G_{N}^{(d+1)} } {\Gamma (d+1) \pi^{{d \over 2}} \over \Gamma({d \over 2})^3 }~,
}
so that we get for the leading correction
\eqn\stresstensorcorrectionGR{
c_{stress} =  {4 \Gamma({d \over 2}) \over  \pi^{{d-2  \over 2}} (d-1)} { G_{N}^{(d+1)} \over L^{d-1}} { \Gamma (\Delta+1)^2 \over \Gamma (\Delta  - {d-2 \over 2} )^2}~.
}

We can apply this formula, for example,  in the case of ${\cal N} = 4$ SYM theory. We substitute $d = 4$ and ${ G_{N}^{(d+1)} \over L^{d-1}} = {\pi \over 2 N^2}$ so that $c_{T} = 40 N^2$ and we get that the correction due to the stress tensor exchange is given by
\eqn\stresstensorcorrectionGRNfour{
c_{stress}^{{\cal N} = 4} = {2 \Delta^2 (\Delta - 1)^2 \over 3 N^2}~.
}

One of the first computations of a four-point function at strong coupling was 2-2 dilaton scattering. in this case the anomalous dimensions of double trace operators are known~\HoffmannDX\ 
\eqn\strongcoupling{
\tau_{:{\cal O} \pa^s {\cal O}:} =8 - {96 \over (s+1)(s+6)} {1 \over N^2}~.
}
In the formula above $\CO$ stands for the operator dual to the dilaton. The only twist $2$ operator in the t-channel is the stress tensor. By plugging $\Delta = 4$ into our formula we correctly reproduce ${96 \over N^2}$!

For other computations of the anomalous dimensions of double-trace operators in  theories with gravity duals see \refs{\DolanTT,  \HeslopDU}.
In all those cases we found that the sign of the correction is consistent with convexity, and the leading power of $s$ is as predicted. 
However, often more than one operator of twist $2$ is exchanged in the t-channel and \stresstensorcorrectionGRNfour\ will not be the complete answer. To get the complete answer we just have to sum over finitely many contributions.

\subsec{The large $\Delta$ Limit}

It is curious to consider large $\Delta$ limit of \correctionformx . The factor ${ \Gamma(\Delta - {\tau_{\min} \over 2})^2 \over \Gamma(\Delta)^2}$ becomes $\Delta^{\tau_{\min}}$ and, thus, $c_{\tau_{\min}}$ has a simple dependence on $\Delta$,
$c_{\tau_{\min}} \sim \Delta^{\tau_{\min}} f^2$ where $f^2$ is the combination of three- and two-point functions written in~\correctionformx. The limit of large $\Delta$ is also nice in that the saddle point described in appendix~B becomes exact and simplifies as described after~\dominantspin. 

The point we would like to make in this subsection is that since for large $\Delta$ the saddle point~\dominantspin\ becomes exact, the idea of~\AldayMF, which we outlined in subsection 3.5, can be made precise.

According to the picture we reviewed in subsection 3.5, the scaling dimensions of the double-twist operators are related to the total energy of two static charges in some two-dimensional space interacting through a Yukawa-like potential
\eqn\correctionscaldim{
\Delta_1 + \Delta_2 + g q_1 q_2 e^{-m v_0}~,}
where $v_0$ is the separation between the particles. We assume for simplicity $\Delta_1=\Delta_2\equiv\Delta$. Now we use the fact that the saddle point is exact if $\Delta$ is large and the relation at the saddle point is $v_0 = \log {s \over \Delta}$. Additionally, from our formula for the leading correction to the dimensions of double-twist operators we read $g q_1 q_2 \sim { C_{{\cal O} {\cal O}^{\dagger} {\cal O}_{\tau_{\min}}}^2 \over  C_{{\cal O} {\cal O} } C_{{\cal O}^{\dagger} {\cal O}^{\dagger} }  C_{{\cal O}_{\tau_{\min}}{\cal O}_{\tau_{\min}}}}$.  The mass $m$ appearing in~\correctionscaldim\ corresponds to the minimal twist that we exchange (this is directly related to the fact that  the lightest particle that we exchange leads to the longest range interactions). 

These identifications are natural, for example,  the sign of the correction has a very simple interpretation. Gravity is always an attractive force, thus, the correction from the potential energy is always negative. This is reflected by the fact that for the energy momentum tensor we have $g \sim -{1 \over c_T }$ and $q \sim \Delta$.
However, if we have a $U(1)$ current the force could be either attractive or repulsive depending on the signs of the charges.

The identification $v_0 = \log {s \over \Delta}$ and the relation between $g q_1 q_2$ and various two- and three-point functions all receive nontrivial corrections in $\Delta^{-1}$.  We have accounted for all of them precisely in the previous section.

\subsec{The Case of Parametrically Small $\Delta - {\tau_{\min} \over 2}$}

The formula~\correctionformx\ goes to zero when  $\Delta = {\tau_{\min} \over 2}$. This is consistent because  such an equality could only be realized in free field theory (or, if one abandons unitarity it could be realized in models such as generalized free fields). And indeed, in free field theory (or generalized free fields) the answer is zero; there are no corrections to the twists of double-twist operators. 

However, the case of small $\Delta - {\tau_{\min} \over 2}$ is in fact more subtle.  For example, consider some weakly coupled CFT. Then if we choose the external state appropriately, the leading twist operators in the OPE would have $\tau_{\min}$ very close to $2\Delta$. But there would be infinitely many such operators. In order to get the right answer for the anomalous dimensions of fast spinning operators we would generally have to sum them all and be careful about performing the perturbation theory consistently. 

An example is the $\phi^4$ theory in $4-\epsilon$ dimensions. We consider the $\phi(x)\phi(0)$ OPE, which  includes many operators whose twists are very close to $2$. In order to obtain the right anomalous twists of fast spinning operators of the type $\phi \del^s\phi$ as discussed in section 3, we have to sum all of those operators up. In this case $\epsilon$ breaks a higher spin symmetry and so the infinitely many operators with twists close to $2$ can be regarded as due to a slightly broken higher spin symmetry. 

More generally, imagine we have a CFT which contains a small parameter $\eps$. Imagine we also know the twist  $\tau_{{\cal O} \pa^s {\cal O}}(s, \eps)$ of some double-twist operators exactly as a function of $\Delta$ and $s$. From the discussion above we know that at {\it sufficiently large spin} we expect it to have the form
\eqn\formofthetwist{
\tau_{{\cal O} \pa^s {\cal O}}(s, \eps)=2 \Delta - {c_{2}(\eps) \over s^{d-2}}  - {c_{4}(\eps) \over s^{d-2+\gamma_4(\eps)}} - ...
}
where $\gamma_4 (\eps)$ is the anomalous dimension of the minimal twist spin-four operator. In the above we have assumed that the minimal twists are realized by the energy-momentum tensor,  spin-four operators etc. Of course it could also be that $\CO^2$ has a twist smaller than $d-2$ (and hence dominates over the energy momentum tensor) but that won't change the point we would like to make. 
Clearly, if $\epsilon$ is small, the expansion~\formofthetwist\ is only useful for spins much larger than any other parameter in the theory (including $e^{1\over \gamma_4(\eps)}$ that would naturally arise in a situation like~\formofthetwist). 

If one decides to fix the spin and take $\epsilon$ arbitrarily small, then all the operators in~\formofthetwist\ become important. In this regime, the problem we are discussing is essentially that of solving perturbative CFTs via the bootstrap program. This is not our goal here, and we will not comment on it further. The interested reader should consult~\refs{\HeemskerkPN,\LiendoHY} for interesting recent progress in this direction. 

In any given CFT, the results~\approachtolimit,\correctionformx\ always apply for sufficiently large spin. For example, in the Ising model 
\eqn\almostconservedising{
\tau_{\sigma \pa^s \sigma}^{3d \ Ising} \sim 1.037 - {0.0028 \over s} + ...
}
where we have used here the numerical value for the central charge found in~\ElShowkHT. This is an analytic result  about the Ising model, but in practice its power is limited because one has to go to quite large spin $s$ in order for the next terms to be suppressed (the next term is controlled by the anomalous twist of the spin 4 operator, which is 0.02, and thus the formula~\almostconservedising\ is useful only for exponentially large spins). In the next section we will discuss some results about the Ising model that are more powerful from the practical point of view.

\subsec{The Charge to Mass Ratio is Related to Convexity}

As we explained above, the correction $c_{\tau_{\min}}$ in~\approachtolimit\ could have either sign depending on which operators propagates in the t-channel and depending on the precise form of the four-point function we study. 
We demonstrate this point here by considering four point functions of the form  
$\la  {\cal O}_i (x_1) {\cal O}_{j} (x_2)   {\cal O}_{\bar i}^{\dagger} (x_3) {\cal O}^{\dagger}_{\bar j} (x_4)  \ra $ where $\CO_i$ and ${\cal O}_{\bar i}$ are operators charged under global symmetries. In the expansion in the s-channel we necessarily encounter double-twist operators of the type $\CO\del^s\CO$ which are charged under some global symmetries. We cannot directly apply the results of section 2 for such operators since we do not have positivity of the cross section argument. It is thus interesting to ask whether such operators approach the limiting twist $2\Delta$ in a convex, flat, or concave manner. Here we will explain that under some very general circumstances this is, roughly speaking, fixed by the charge to dimension ratio of $\CO$. Let us set up the framework more precisely.

Imagine we have $N_f$ flavor currents $J_{K}^{\mu}$. And imagine also we have a set of scalar operators ${\cal O}_i$ charged under
this symmetry so that
\eqn\symmetryact{
[Q, {\cal O}_i] = - (T_{K})^j_{i} {\cal O}_j
}
This means that the OPE includes 
\eqn\OPEcoef{
J_{K}^{\mu} (x) {\cal O}_i (0) \sim {- i \over S_{d}} {x^{\mu} \over x^{d}} (T_{K})^j_{i} {\cal O}_j (0).
}
where $S_d$ is the surface area of a $d-1$ sphere.

Let us define the  two-point functions 
\eqn\twopoint{\eqalign{
\la {\cal O}_i (x) {\cal O}^{\dagger}_{\bar i} (0) \ra &= {g_{i \bar i} \over x^{2 \Delta}}~, \cr
\la J_{I}^{\mu} (x) J_{J}^{\mu} (0) \ra &= {\tau_{I J} \over S_{d}^2} {I^{\mu \nu} \over x^{2(d-1)}}.
}}
Using these and the OPE~\OPEcoef\ we can fix the three-point functions to be
\eqn\threep{
\la {\cal O}_i (x_1) {\cal O}^{\dagger}_{\bar j} (x_2) J_{I}^{\mu} (x_3) \ra =  {- i \over S_{d}} (T_{I})^j_{i} g_{j \bar j} {Z^{\mu} \over x_{12}^{2 \Delta - (d-2)} x_{13}^{d-2} x_{23}^{d-2}}
}
where $Z^{\mu} ={ x_{31}^{\mu}  \over x_{31}^2} - { x_{32}^{\mu}  \over x_{32}^2}$.

Now we consider the four-point function $\la  {\cal O}_i (x_1) {\cal O}_{j} (x_2)   {\cal O}_{\bar i}^{\dagger} (x_3) {\cal O}^{\dagger}_{\bar j} (x_4)  \ra $. Using ~\correctionformx\ we can account for the contributions of the conserved currents in the t-channel to $\delta\tau_s$. We find 
\eqn\anomaldimtr{
(\delta\tau_s)_{{\rm  currents}}\sim -  \sum_{I,J}{ C_{{\cal O}_{i} {\cal O}^{\dagger}_{\bar i} J^{I}}  C_{{\cal O}_{j} {\cal O}^{\dagger}_{\bar j} J^{J}} \over  C_{{\cal O}_{i} {\cal O}^{\dagger}_{\bar i} } C_{{\cal O}_{j} {\cal O}^{\dagger}_{\bar j} }  C_{J^{I} J^{J}}} = - \sum_{I,J} { (T_{I})^k_{i} g_{k \bar i}  (T_{J})^m_{j} g_{m \bar j} \over \tau_{I J} g_{i \bar i} g_{j \bar j}}.
}
Restoring the pre-factors from~\correctionformx\ (for a configuration of four insertions like we consider here there is an additional factor of $(-1)^s$) we get for the sum of the stress-tensor and repulsive flavor corrections (here we have assumed that the minimal twists that appear in the t-channel are the EM tensor and conserved currents)
\eqn\sumofcorrect{
\delta\tau_s \sim -{1\over s^{d-2}}\left[{d^2 \Delta^2 \over 4 (d-1)^2 c_{T}} {\Gamma (d+2) \over \Gamma ({d+2 \over 2})^2} - {\Gamma (d) \over 2 \Gamma ({d \over 2})^2}  \sum_{I,J} { (T_{I})^k_{i} g_{k \bar i}  (T_{J})^m_{j} g_{m \bar j} \over \tau_{I J} g_{i \bar i} g_{j \bar j}}\right] ~.
}

In general, the odd spin contributions are always repulsive in the context considered in this subsection and the even spin contributions (including scalar and stress tensor exchange) are attractive. Therefore, whether $\delta\tau_s$ is convex, flat, or concave for large spin is determined by the ratio of $\Delta/c_T$ compared to, roughly speaking, $q^2/\tau_{IJ}$ where $q$ is the charge of the operators (more generally, the representation). {\it Therefore, the question of convexity at large spin is determined by whether $\CO$ satisfies a BPS-like bound.}  The BPS-like bound that controls convexity for the simple case of stress tensor and a single $U(1)$ symmetry is
\eqn\convexnonresecf{{\Delta\over  \sqrt{c_T}}\geq {1\over \sqrt 2}{d-1\over \sqrt{d(d+1)}}{|q|\over \sqrt\tau}~.}

\ifig\fignine{ The correction due to the $U(1)$ current exchange has the opposite sign for the anomalous twists of ${\cal O} \pa^s {\cal O}$ and ${\cal O} \pa^s {\cal O}^{\dagger}$. At the level of the bootstrap equations, it corresponds to the fact that the three-point function $\la {\cal O} {\cal O}^{\dagger} J_s \ra$ is odd under a permutation of ${\cal O}$ and ${\cal O}^{\dagger}$ when $s$ is odd. (Here $J_s$ is an operator of spin $s$.)} {\epsfxsize3.5in\epsfbox{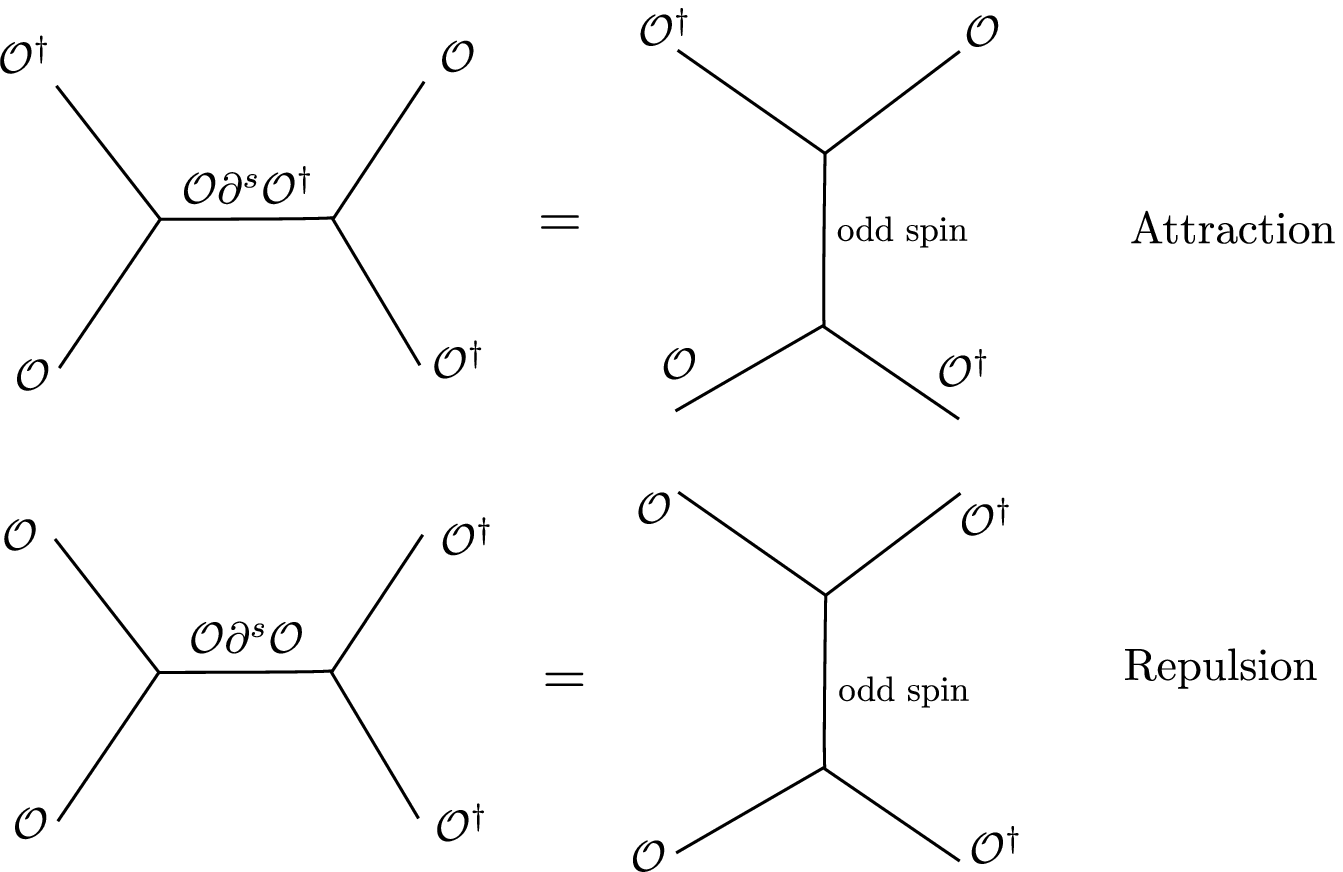}}

Let us now mention that if we consider instead the correlation function in the ordering $\la  {\cal O}_i (x_1)   {\cal O}_{\bar i}^{\dagger} (x_2) {\cal O}_{j} (x_3)  {\cal O}^{\dagger}_{\bar j} (x_4)  \ra $ and study the s-channel expansion, then we encounter double-twist operators of the type $\CO\del^s\CO^\dagger$. In this case, the t-channel expansions gives a different results from before. Because we flipped two operators (see~\fignine), all the even spins still contribute positively to $c$ but now the odd spins contribute positively as well. Hence, this essentially proves convexity in a new way (without alluding to an RG flow) for large enough spins! In fact, in some sense, it proves a slightly different result than the one in section 2, since here we do not need to assume that the double-twist operators are the minimal twist operators to get convexity -- the convexity of section 2 is always about the minimal twist operators.

As an example of the discussion above let us consider  an $\CN=1$ SCFT in four dimensions. 
In the case of SCFT we have a $U(1)_{R}$ global symmetry. The two-point function of the R-current is related via supersymmetry to the two-point function of the stress tensor (see, for example, formula (1.11) in~ \BarnesBM \foot{Our conventions are related to theirs by
${ \tau_{here} \over S_d^2} = {2 (d-1)(d-2)  \over (2 \pi)^d}\tau_{there}$ and ${c_{T; here} \over S_d^2} =  c_{T; there}$~.}). The relation in our conventions is
\eqn\ratio{
{c_{T} \over \tau} =  {d (d+1) \over 2}~.}

Now consider the correlation function $\la  {\cal O} (x_1) {\cal O} (x_2)   {\cal O}^{\dagger} (x_3) {\cal O}^{\dagger} (x_4)  \ra $ where $\CO$ is a chiral primary and $\CO^\dagger$ is an anti-chiral primary. We can use the formula~\sumofcorrect\ to evaluate the asymptotic corrections to the operators $\CO\del^s\CO$ appearing in the chiral$\times$chiral OPE. Using the relation~\ratio\ and the fact that for chiral primaries $\Delta={3r\over 2}$ where $r$ is the $R$-symmetry charge of $\CO$, we find that the correction~\sumofcorrect\ precisely vanishes. 
Furthermore, if $\CO$ carries some global symmetry quantum numbers other than the $R$-symmetry, we should include the contributions to $\delta\tau_s$ from the exchanges in the t-channel of the flavor supermultiplet (this is a linear multiplet of $\CN=1$). We find that this contribution again vanishes identically. 

One could ask why  these corrections to $\delta\tau_s$ vanish identically for chiral primaries.  In the OPE of two chiral primaries one finds short representations with twists precisely $2\Delta$ and long representations with twists larger than $2\Delta$. (There are no smaller twists.) The cancelation above may be related to the presence of infinitely many short representations in the OPE.
It would be nice to understand whether the OPE of two chiral primaries  contains infinitely many protected operators, as the computations above hint. For completeness, the structure of the OPE of two chiral primaries is reviewed in appendix~D.

Another curiosity that we would like to note is that  if $\CO$ is chiral but not a chiral primary, then $\Delta>{3r\over 2}$ and the exchange of the stress tensor dominates over the $R$-current. Thus, we have convexity again. We see that the general connection between the BPS-like bound~\convexnonresecf\ and convexity is realized very naturally in SUSY.

\newsec{Applications of Convexity}

In this section we consider different known examples and verify that the convexity and asymptotic form of the twists hold true.  
Moreover, we will see that in all known examples the convexity starts from spin 2 (thus $s_c =2$). 

\subsec{Free Theories and Weakly Coupled Theories}

It is easy to see that the propositions made in the previous sections are true in free theories. In this case we have
an infinite set of conserved currents $j_s$ with the minimal possible twist $\tau = d - 2 $. 

For a generic operator ${\cal O}$ we expect the three-point functions $\la {\cal O} {\cal O}^{\dagger} j_s \ra$ to be non-zero. This is suggested by the way higher spin symmetry is realized in the sector of conserved currents as well as from the way it acts on the fundamental field. 

\ifig\figten{The form of the minimal twist spectrum in free theories.} {\epsfxsize2.0in\epsfbox{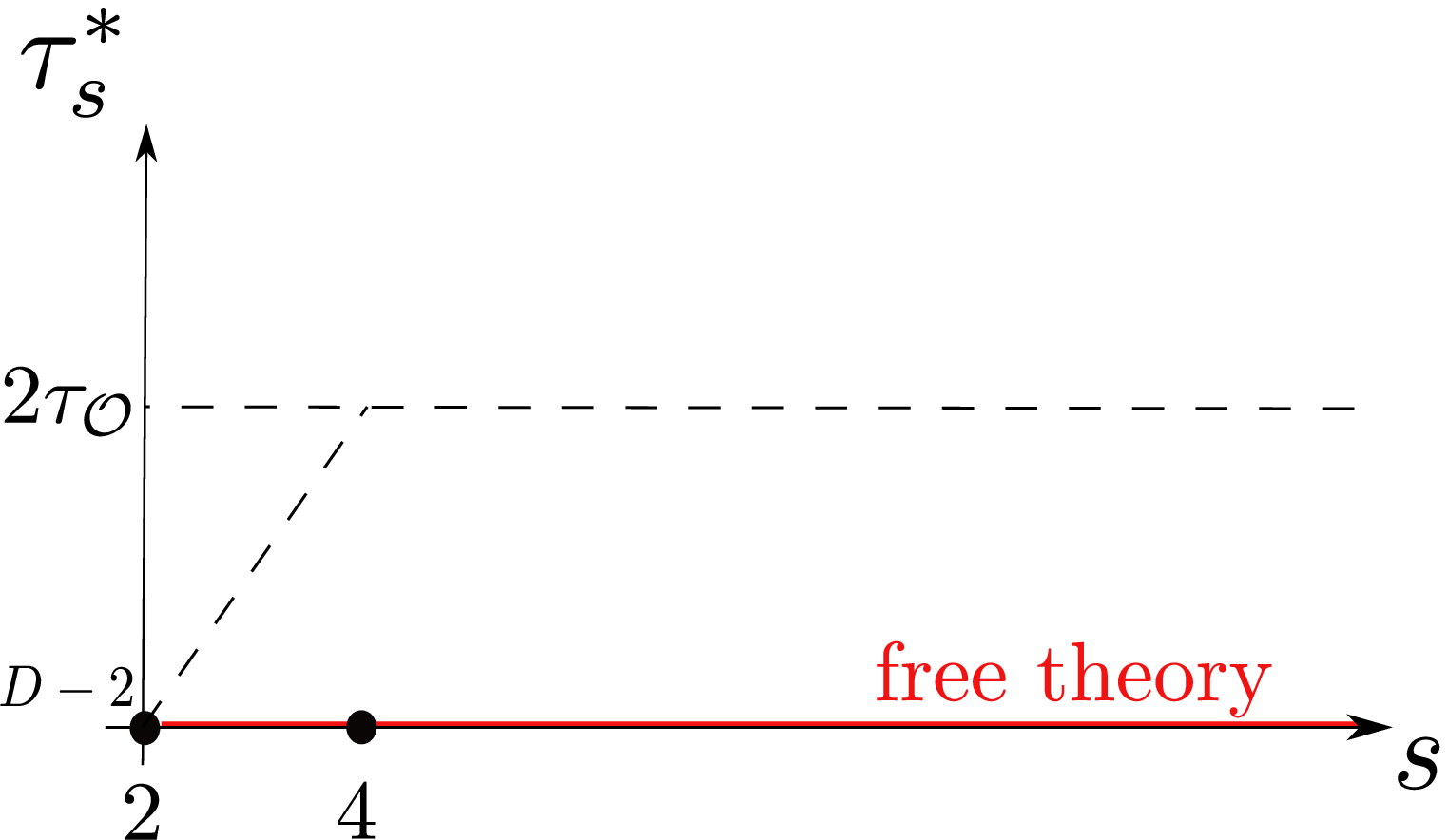}}

The conserved currents will be the leading twist operators. In this way convexity with $s_{c}=2$ is realized in the case of free theories. See \figten.

It is also trivial to see that there exist operators with twist arbitrarily close to $2\tau_\CO$. Indeed, for any composite operator $\CO$ in a free theory, in the OPE of this operator with itself there will appear the operators $\CO\del^s\CO^{\dagger}$ (written schematically). Those have twist precisely $2\tau_\CO$. If the operator $\CO$ is the elementary free field itself, then these operators are the higher spin currents. Their twist is precisely $d-2$, which coincides with twice the dimension (and hence twice the twist) of the elementary scalar field.

Weakly coupled CFTs inherit some of their  OPE structure from their higher spin symmetric parents. (The OPE coefficients and the dimensions change a little. Various short multiplets could combine, for example, the exactly conserved higher-spin currents undergo a ``Higgs mechanism'' and acquire a small anomalous twist. Also new operators that were not present in the free field theory could appear.) 
If the CFT is sufficiently weakly coupled, the  almost conserved currents will still be the minimal twist operators in the OPE of any operator with itself. This means that in this case our statement from section 2 boils down to the convexity of twists of almost conserved currents. The examples below are of this type.

Notice that in the case when we have a flux (e.g. in gauge theories) the situation is different. The almost conserved currents universally receive a correction to the twist of the form $\log s$. Thus, for large enough spin, their twist could be arbitrary large so that they cease to be the minimal twist operators. In this case some other operators (the double twist operators) ensure additivity and convexity. 

\subsec{The Critical $O(N)$ Models}

Let us consider the $O(N)$ critical point in $4 - \eps$ dimensions. 
We can use the results of~\WilsonJJ\ for the anomalous dimensions. We have
\eqn\currents{\eqalign{
\delta \tau_{\sigma_{i} \pa^s \sigma_{i}} - 2 \gamma_{\sigma} &= - \eps^2  {N+2 \over 2 (N+8)^2} {6 \over s (s+1)}~, \cr
\delta \tau_{\sigma_{(i} \pa^s \sigma_{j)}}- 2 \gamma_{\sigma} &=- \eps^2  {N+2 \over 2 (N+8)^2}  {6 (N+6)\over (N+2) s (s+1)}~,
}}
where $\gamma_{\sigma}$ is the anomalous dimension of the spin field. The operators $\sigma_{i} \pa^s \sigma_{i}$ are singlets under the global symmetry $O(N)$, and $\sigma_{(i} \pa^s \sigma_{j)}$ are symmetric traceless representations of $O(N)$. 

Let us make several comments about the formulae above. First of all, both formulae have a well-defined large $s \to \infty$ limit that implies $\delta\tau_s-2\gamma_\sigma\rightarrow 0$.\foot{To leading order $2 \gamma_{\sigma} = {N+2 \over 2 (N+8)^2} \eps^2$.} This means that the prediction~\ineqbott\ is saturated, as it {\it must} in weakly coupled CFTs where we consider operators that are bilinear in the fundamental field. 
Second, the leading correction around infinite spin in both cases is of the form ${1 \over s^2}$. This is what we expect due to the presence of the stress tensor and other currents which are conserved at this order. (To this order also $\sigma^2$ has twist $d-2$.) Third, the convexity starts from the energy momentum tensor, $s=2$. At higher orders in $\eps$, the higher spin-currents will lead to different typical exponents at large spin, 
${1 \over s^{\tau_{{\rm min}} (\eps)}}$. The dominant exponent at very large spin will still be due to the stress tensor, ${1 \over s^{2-\eps} }$. (The reason for that is that higher spin currents have a larger anomalous twist and the operator  $\sigma^2$ appears to have dimension, and thus twist, larger than $d-2$ as well. This information about $\sigma^2$ is suggested from the epsilon expansion, from the known results about low $N$ critical $O(N)$ models, and the results of large $N$.)

Alternatively, we can consider the case of large $N$ with the number of space-time dimensions being $d$. In this case we can use the results of the large $N$ expansion~\LangGE,\LangZW\
\eqn\currentslN{\eqalign{
\delta \tau_{\sigma_{i} \pa^s \sigma_{i}} &=2 \gamma_{\sigma}  {4 \over (d+2 s-4)(d+2 s -2)}\left[ (d+s-2)(s-1) - {1 \over 2} {\Gamma [d+1] \Gamma[s+1] \over 2 (d-1) \Gamma[d+s-3]} \right]~, \cr
\delta \tau_{\sigma_{(i} \pa^s \sigma_{j)}} &=2 \gamma_{\sigma} {4 (d+s-2) (s-1) \over (d+2 s-4) (d+2 s-2)}~, \cr
\gamma_{\sigma} &= {2 \over N} {\sin \pi {d \over 2} \over \pi} {\Gamma (d-2) \over \Gamma({d \over 2} - 2) \Gamma({d \over 2} + 1) }~. 
}}
In the large $s$ limit these formulae become
\eqn\currentslNb{\eqalign{
{\delta \tau_{\sigma_{i} \pa^s \sigma_{i}} - 2 \gamma_{\sigma} \over 2 \gamma_{\sigma}  }  &\sim - {\Gamma (d+1) \over 2 (d-1)} {1 \over s^{d-2}}- {d (d-2) \over 4} {1 \over s^{2}}~, \cr
{ \delta \tau_{\sigma_{(i} \pa^s \sigma_{j)}} - 2 \gamma_{\sigma} \over  2 \gamma_{\sigma}  } &\sim - {d (d-2) \over 4} {1 \over s^{2}}~.
}}
These results are consistent with the fact that at leading order in $N$ we have conserved currents operators with the twist $\tau = d - 2$, and we also see $1/s^2$ coming from the operator  $\sigma^2$. 

The difference in the coefficients of ${1 \over s^{d-2}}$ for the operators  $\sigma_{i} \pa^s \sigma_{i}$ and $\sigma_{(i} \pa^s \sigma_{j)}$ comes from the fact that different conserved higher spin currents contribute. We see that the sum over all the higher spin currents (each contributing to $1/s^{d-2}$) actually vanishes for the symmetric combination. 
However, the ${1 \over s^2}$ contribution is governed solely by the three point function $\la \sigma_i \sigma_{i} \sigma^2 \ra$ and, thus, the coefficient of $1/s^2$ is the same for both operators $\sigma_{i} \pa^s \sigma_{i}$ and $\sigma_{(i} \pa^s \sigma_{j)}$.\foot{The cancelation of $1/s^{d-2}$ in the second line of~\currentslNb\ is easily understood in our language. Consider the four-point function of spin fields $\langle \sigma_1\sigma_2\sigma_2\sigma_1 \rangle$. The only operator in the t-channel that can propagate is the interacting field $\sigma^2$. The higher spin currents cannot propagate because they behave essentially like free fields and so contribute zero to the t-channel diagrams. This shows that the operators $\sigma_1\del^s\sigma_2$ do not have a piece that goes like $1/s^{d-2}$ at large spin.} We can easily reproduce the coefficient of $1/s^2$ using our formula~\correctionformx\ and the known two- and three-point functions in this model (at large $N$). We find a precise match between the perturbative calculation~\currentslNb\ and our general methods using the bootstrap equations. The necessary details are collected in appendix E. 

Now we can turn to predictions using convexity. We consider arbitrary $N$ and $\eps$. Of particular interest are three-dimensional small $N$ models (for which there is no small parameter in the problem). They describe many second order phase transitions. For example, in the case of $N=1$ we have the 3d Ising model which describes the vapor-liquid critical point of water. 

All these theories are known to contain an almost free scalar operator (the spin field) $\Delta_{\sigma} \sim 0.5$, thus, the prediction is that 
all these theories contain operators of spin $4,6,...,\infty$ with twists
\eqn\twistpr{
1 < \tau_{s} < 2 \Delta_{\sigma}~.
}
This happens to be a rather stringent constraint on the spectrum of the leading twist higher spin operators. 
For example, in the case of the 3d Ising model, it is known that $\Delta_{\sigma} = 0.5182(3)$ (see e.g. \ElShowkHT). Thus, we would predict that there is an infinite set of operators with $1 < \tau_{s} < 1.037$. Moreover, based on other examples and on the known twist of the spin four operator that nicely satisfies~\twistpr\ ($\tau_4 \sim 1.02$), we expect that these operators are present starting from $s= 2$.

\ifig\figeleven{The form of the minimal twist spectrum implied by convexity in the 3D Ising model. The shaded region describes the allowed values for almost conserved currents of spin $s \geq 6$.} {\epsfxsize2.2in\epsfbox{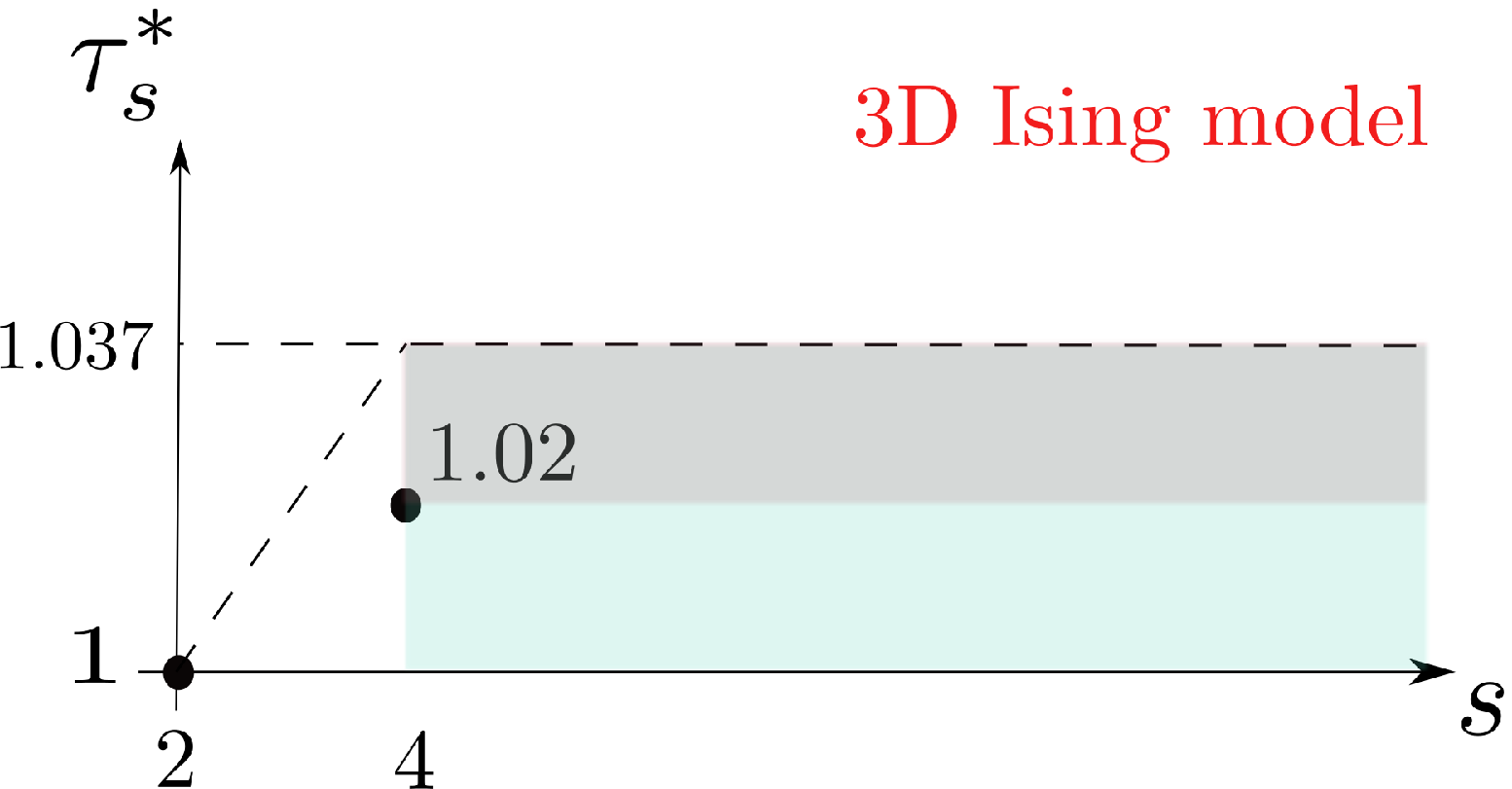}}

Using monotonicity and the known twist of the spin four operator, we can further make the prediction that there is an infinite set of almost conserved currents in the 3D Ising model starting from $s=6$ with the twists
\eqn\pred{\eqalign{
1.02 &< \tau_{s} < 1.037~,\qquad  s=6,8,...,\infty~.}}
This is illustrated in~\figeleven.

\subsec{Strongly Coupled Theories and AdS/CFT}

For theories with gravity duals, convexity could be tested via AdS/CFT.
One may be able to think about convexity as some hidden constraint on the low-energy effective actions in AdS in the spirit of \AdamsSV . We leave the exploration of this question for the future. A typical characteristic of theories with gravity duals is that these theories possess a large gap in the spectrum. See \HeemskerkPN\ for a systematic approach to analyzing such theories.

\ifig\figtwelve{The form of the minimal twist spectrum implied by convexity in a theory described by supergravity.} {\epsfxsize2.2in\epsfbox{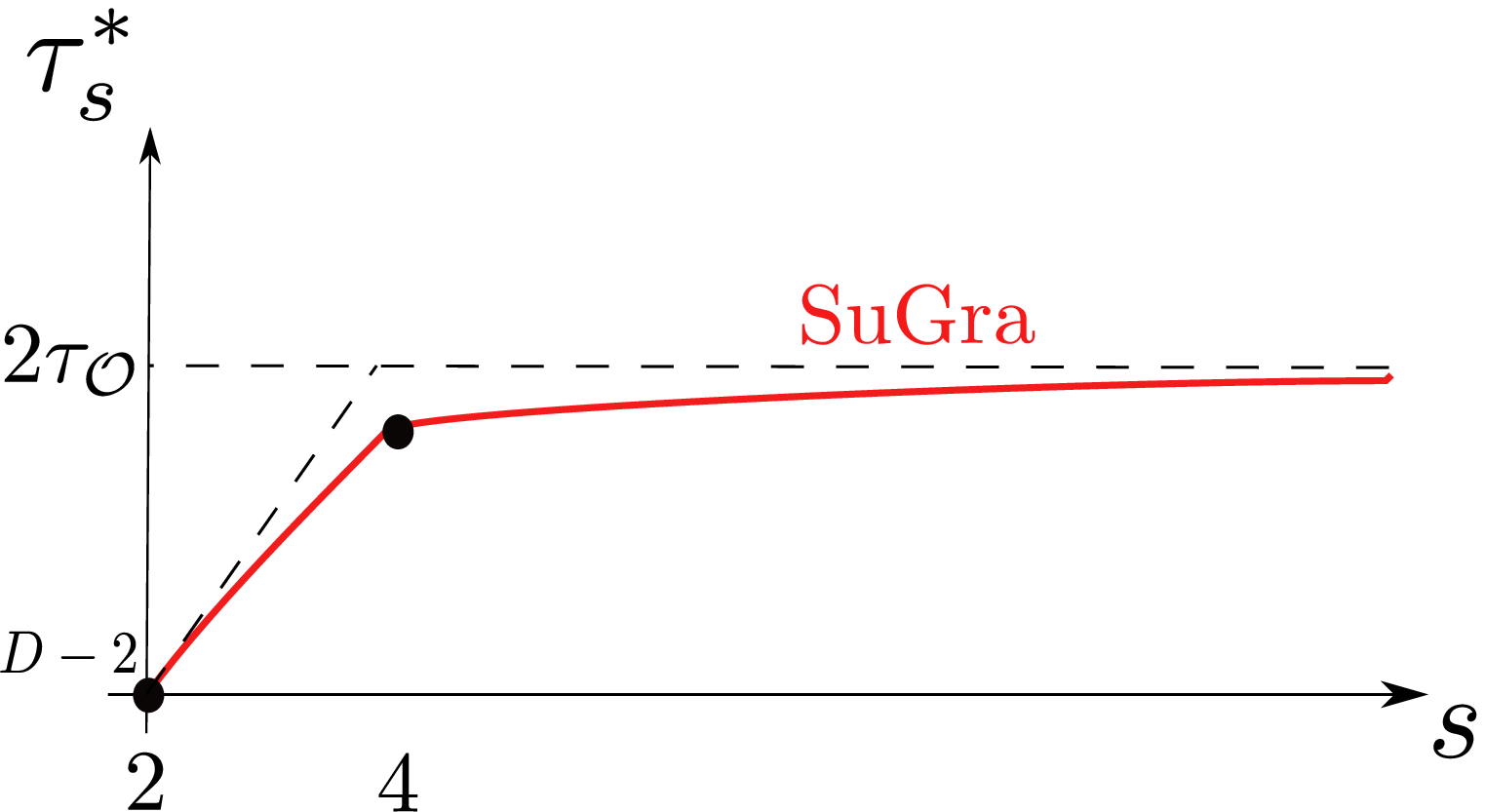}}

In this context, convexity is at  statement about the anomalous dimensions of various double-trace operators, which are the double-twist operators in this case. For very large spins convexity was proved using the crossing equations in section 3 while for spins of order~$\CO(1)$ we can use the argument of section 2. In the bulk gravity duals, this corresponds to the energy of various spinning two-particle states. Hence, convexity implies that the binding energy of fast spinning operators in the bulk is always negative (i.e. the particles have net attraction) and the binding energy approaches zero in a power law fashion. This is depicted schematically in \figtwelve, where $2\tau_\CO$ is the twist of the two-particle state at very large spin (hence, the binding energy goes to zero).
We emphasize that all these binding energies are of order $1/N$ or less. Hence, the properties delineated above are about various $1/N$ corrections that come from tree-level diagrams in supergravity.

\subsec{Higher Spin Symmetry Breaking}

Recently there was some progress in understanding the higher spin symmetric phase of the AdS/CFT correspondence~\refs{\KlebanovJA,\SezginRT,\GiombiWH}.
On the CFT side one can notice that such symmetries fix correlations functions up to a number~\MaldacenaJN.\foot{The expression for $n$-point correlation functions of currents was recently found in~\DidenkoTV.} 

However, in the case when higher spin symmetry is broken the situation is much less clear. On the boundary sometimes it is possible to use the constraints to get some general predictions \refs{\GiombiKC,\AharonyJZ, \MaldacenaSF}. But they are very limited and are not applicable in the general situation. In the bulk the situation is even less clear. It would be very interesting to understand higher spin symmetry breaking in the bulk better.

Convexity is a generic constraint on any possible higher spin symmetry breaking. On the boundary, slightly broken higher spin symmetry currents are the minimal twist operators discussed before. In the bulk, this maps to a statement about the masses of Higgsed gauge bosons.

The scaling dimensions of operators with spin $s$ are mapped to the masses of the dual states in the bulk \FerraraEJ
\eqn\massesbulk{
m^2 = C_2 (\Delta,s) - C_2 (d-2+s, s) = \delta \tau_s (\delta \tau_s + 2 s + d - 4) .
}
here $C_2$ is the quadratic Casimir of $SO(d,2)$ and $\delta \tau_s$ is the anomalous twist (when it is zero the particle is massless). 

Thus, convexity is some general constraint on the higher spin symmetry breaking in quantum gravities in AdS. It could be applied both to Vasiliev theories  \VasilievBA\ and tensionless limits of string theories  \SundborgWP . 

Consider as an example case of parity preserving Vasiliev theory in $AdS_4$ with higher spin breaking boundary conditions. We know from additivity that for large spins $\tau_s = d-2 + 2 \gamma_{\sigma} + O(s^{2-d})$ so we can approximate the masses of the higher spin bosons in the bulk as
\eqn\masses{
m^2_s \approx 4 \gamma_{\sigma} s+ O(1)~.
}
Here $\gamma_{\sigma}$ is the anomalous dimension of the spin field in the $O(N)$ model (the singlet $O(N)$ model was first considered in~\KlebanovJA). 

In the singlet $O(N)$ model, $\sigma$ is not part of the spectrum, but we can nevertheless define its anomalous dimension:
\eqn\slope{
\gamma_{\sigma} = \lim_{s \to \infty} {1 \over 4} {\pa m^2_s \over \pa s} = \lim_{s \to \infty} {1 \over 2} ( \tau_s - 1 )~.
}
Hence, it is defined from the limit of anomalous dimensions of almost conserved currents (alternatively, from the masses of bulk gauge bosons).

\newsec{Conclusions and Open Problems}


In this paper we investigated unitary CFTs in the Lorentzian domain. By flowing to the gapped phase and using positivity of a certain cross section,
we established a convexity property of the minimal twists in the ${\cal O}(x) {\cal O}^{\dagger}(0)$ OPE~\convexi.
Studying the crossing equations in the light-cone limit we found that there is an additivity property in the twist space and that the inequality~\ineqbott\  has to hold. Additivity means that in $d>2$ it is meaningful to talk about double- and multi-twist operators, in any CFT.
We also computed explicitly the leading large spin corrections to the twist of double-twist operators in the case of scalar operators and explained how to compute it for arbitrary operators. We discussed various double-twist operators and the leading correction to their twists at large spin. This led to a purely CFT-based proof of a certain convexity property of the twists of double-twist operators. 

Our general findings are related naturally to the picture of~\AldayMF, where 4d CFT correlators look like correlators in a 2d gapped local theory. In the original CFT this locality is realized in $\log s$. We reformulated their picture in terms of the OPE and that allowed us to extend the analysis of \AldayMF\ and compute the leading spin corrections exactly.

We proceeded by studying various examples and applications of these general properties. We have found that convexity is a recurring theme at large spin. We checked  that the general picture that we obtained is indeed realized in known theories (mostly weakly coupled theories or theories with gravity duals). 

We concluded that there should not be solutions of the crossing equations with $\tau^*_{s} > \tau_1 + \tau_2$ and we checked our picture against the available numerical results \ElShowkHT . We emphasized some predictions for the 3d critical $O(N)$ models that are relevant for phase transitions. One can set bounds on an infinite set of operators in the critical $O(N)$ models~\twistpr . In the case of the 3d Ising model one can do better~\pred . 

Another example we considered is the correction to the twists of chiral operators in $\CN=1$ SCFTs.  We found that to the order we computed the twist spectrum of chiral primaries remained flat at large spin (in principle it could have been concave). If we consider chiral operators that are not chiral primaries, then, as we have explained in section 4, the usual superconformal unitarity bound leads to convexity at large spin. 

For a generic CFT we found that there is a relation between the convexity/concavity of the large spin twists of double-twist operators and the mass/charge ratio. More precisely, assuming that the minimal twist operators are the stress tensor and a $U(1)$ current we found that
\eqn\chargeget{
{\cal O}\pa^s {\cal O} \ {\rm convexity}\leftrightarrow \Delta \geq  {(d-1) \over 2}  |q|~,
}
where we used the freedom to normalize the current to choose ${c_{T} \over \tau} =  {d (d+1) \over 2}$ (we can always do this for Abelian currents. For concavity the inequality works in the opposite direction.) 

Hence, by studying the dimensions and charges of arbitrary primary operators in various CFTs one can conclude about convexity/concavity of the large spin anomalous dimensions.

Our results  are also relevant for the high spin symmetry breaking in CFTs and through AdS/CFT to theories of quantum gravity in the bulk. As higher spin symmetry breaks, higher spin gauge bosons become massive through the Higgs mechanism. The minimal twists discussed above  correspond to almost conserved higher spin currents. 
For the case of parity preserving Vasiliev theory, from the property of additivity of the twists we infer that the masses of these gauge bosons at high spin follow a Regge-like formula, with the prefactor fixed by the dimension of the spin field. 

There are several concrete open questions which would be interesting to address. 

One obvious generalization would be to compute the next, subleading, orders in the  ${1 \over s}$ expansion. Another concrete problem that would be nice to address is the computation of the coefficients controlling the power law corrections for the operators $\CO_1 \pa^s \square^n \CO_2$. 
Finally, it would be interesting to derive convexity using the ideas of~\HofmanAR. It was noticed in~\HofmanAR\ that the Regge limit of the DIS amplitude we  considered in section 2 is related through a conformal transformation to the problem of measuring energy distributions in the final state created by some operator. Moreover, the fact that the energy distributions are integrable may constrain the Regge asymptotics of graviton deep inelastic scattering.  It would be interesting to understand whether this is indeed the case.

\vskip 1cm

\noindent {\bf Acknowledgments:}
We would like to thank  Thomas Dumitrescu, Tom Hartman, Juan Maldacena, Nathan Seiberg and David Simmons-Duffin for useful discussions. ZK was supported by NSF grant PHY-0969448, a research grant from Peter and Patricia Gruber Awards, a grant from the Robert Rees Fund for Applied Research, and by the Israel Science Foundation under grant number~884/11.  ZK  would also like to thank the United States-Israel Binational Science Foundation (BSF) for support under grant number~2010/629.
Any opinions, findings, and conclusions or recommendations expressed in this
material are those of the authors and do not necessarily reflect the views of the funding agencies.

\appendix{A}{Normalizations and Conventions}

Here we collect our normalization and conventions for two- and three-point functions as well as the normalizations of conformal blocks. We mostly follow the conventions of~\DolanUT\
\eqn\conventions{\eqalign{
r_{i j} &= (x_i - x_ j)^2, \cr
I^{\mu \nu}(x) &= \delta^{\mu \nu} - 2 {x^{\mu} x^{\nu} \over x^2}, \cr
Z^{\mu} &= {x_{13}^{\mu} \over r_{13}} - {x_{12}^{\mu} \over  r_{12}}.
}}
Then for the two- and three-point functions contracted with polarization tensors $C^{\mu}$ we have
\eqn\twopandthreep{\eqalign{
\la O_{l}\cdot C_1(x_1) O_{l}\cdot C_2 (x_2) \ra &=c_{O_{l} O_{l}} {\left(C_1^{\mu} I _{\mu \nu} C_2^{\nu} \right)^{l} \over r^{\Delta}_{12}}, \cr
\la {\cal O}_{\Delta_1}(x_1) {\cal O}_{\Delta_2} (x_2) O_{l}\cdot C_3(x_3) \ra &= c_{{\cal O} {\cal O} O_{l}} {(Z\cdot C)^{l} \over r_{12}^{{1 \over 2}\left( \Delta_1 + \Delta_2 - \Delta \right)} r_{13}^{\left( \Delta_1 + \Delta - \Delta_2 \right)} r_{23}^{\left( \Delta_2 + \Delta - \Delta_1 \right)}}.
}}

For the conformal block we have the following asymptotic behavior~\CostaDW\
\eqn\confblocks{
G_{d,\Delta,s}(z, \bar z) \sim_{z \to 0} (-{1 \over 2})^{s}  z ^{{\Delta - s \over 2}}  \bar z ^{{\Delta + s \over 2 }} \ _{2} F_{1} (\Delta + s, \Delta + s, 2 s + 2 \Delta, \bar z)~.
}
This leading behavior is independent of the number of space-time dimensions.

In the text various notions of twist, minimal twist etc. appeared. For convenience, we summarize here some of the notation and terminology we used.

\item{-} Conformal twist. It is given by $\tau = \Delta - s$ and governs the DIS experiment we considered. Unless stated otherwise, by twist we always mean the conformal twist.
\item{-} Collinear twist. It is given by $\tau^{{\rm coll}}=\Delta - s_{+-}$ and the light-cone OPE is arranged according to collinear twists of operators. 
\item{-} By $\tau_{s}^{*}$ we denote the conformal twist of the spin $s$ {\it minimal} conformal twist operator that appears in the OPE of some operator with its conjugate $\CO(x) \CO^\dagger(0)$.  This is again relevant for DIS.
\item{-} By $\tau_{\min}$ we denote the minimal twist among the operators different from the unit operator that appear in the t-channel of various four point functions. The spin of the corresponding operator could be arbitrary. 
\item{-} By $\tau_s$ we denote the twists of double-twist operators that appear in the s-channel and correspond via the bootstrap equations to  low twist operators in the t-channel.

\appendix{B}{Relating Spin to $\sigma$}

\subsec{A Naive Approach}

Consider the disconnected part of the four-point function of real scalar operators with dimension $\Delta$. In our notation for the cross ratios, the four point function takes the form
\eqn\fourpointfunction{
{\cal F} (z, \bar z) = 1 + (z \bar z)^{\Delta} + \left[ {z \bar z \over (1 - z)(1 - \bar z)} \right]^{\Delta}~.
}
As in the main text, we take $z$ to be small and consider the piece $z^{\Delta}$ only. 
This corresponds to focusing on the contributions of the leading collinear twist ``double-trace'' operators. 
This leading piece in the four-point function takes the form $z^\Delta f(\bar z)$ where
\eqn\doubletraceSiM{
 f(\bar z) = \bar z^{\Delta} \left[ 1 + {1 \over (1 - \bar z)^{\Delta}} \right].
}
We can decompose this function in terms of collinear conformal blocks 
\eqn\confblocks{
 f(\bar z) = \sum_{s=0}^{\infty} c_{s} \bar z ^{\Delta + s} \ _{2} F_{1} (\Delta + s, \Delta + s, 2 s + 2 \Delta, \bar z)~,
}
where the $c_s$ are known to be (see e.g.~\HeemskerkPN)
\eqn\coeff{
c_s = (1 + (-1)^s) {\Gamma (\Delta + s)^2 \Gamma (2 \Delta + s -1) \over \Gamma (s+1) \Gamma (\Delta)^2 \Gamma(2 \Delta + 2 s -1)}~.
}
This  expansion is valid in any number of dimensions even though the full conformal blocks for generic $d$ are not known. The simplification is due to the fact we consider small $z$ and thus collinear conformal blocks, which are much simpler than the full conformal blocks.

Let us now substitute $1 - \bar z = e^{- 2 \sigma} $. We would like to study where does the contribution come from in the sum~\confblocks\ for some given $\sigma$. In other words, we would like to find the typical spin of operators that dominate the contribution for some given $\sigma$. 
It is helpful to replace the sum over spins by an integral and also utilize the integral representation of the hypergeometric function. We get for $f(\bar z)$
\eqn\tostartwith{
\int_0^{1} d t \int_{\Delta}^{\infty} d \hat s \ {2 (2 \hat s -1) \Gamma (\Delta + \hat s -1) \over \Gamma (\Delta)^2 \Gamma(1+ \hat s - \Delta)} {\left[ { (1 - e^{- 2 \sigma})  t (1 - t) \over 1 - (1 - e^{- 2 \sigma}) t}\right]^{\hat{ s} } \over t (1 - t)}~,}
We expect the large $\sigma \gg 1$ limit to be dominated by the high spin operators. 
Thus, we expand to the leading order in $\hat s$
\eqn\prefexpA{
{2 (2 \hat s -1) \Gamma (\Delta + \hat s -1) \over \Gamma (\Delta)^2 \Gamma(1+ \hat s - \Delta)} = {4 \hat s^{2 \Delta - 1} \over \Gamma (\Delta)^2} + O({1 \over \hat s})~,
}
so that we get for the integral
\eqn\tostartwith{
{4 \over  \Gamma (\Delta)^2} \int_0^{1} d t \int_{\Delta}^{\infty} d \hat s  {e^{(2 \Delta - 1) \log \hat s + \hat s \log \left[ { (1 - e^{- 2 \sigma}) t (1 - t) \over 1 - (1 - e^{- 2 \sigma}) t}\right]} \over t (1 - t)}~.
}
We perform the integral over $\hat s$ by the saddle point approximation. We get for the extremum value $\hat s_0$
\eqn\extremumins{
\hat s_0 = - {2 \Delta - 1 \over \log \left[ { (1 - e^{- 2 \sigma}) t (1 - t) \over 1 - (1 - e^{- 2 \sigma}) t}\right]} .
}
After evaluating the integral we get
\eqn\integral{
{4 e^{1 - 2 \Delta} \sqrt{\pi} (2 \Delta -1)^{2 \Delta - {1 \over 2}} \over \sqrt{2 \Delta -1} \Gamma (\Delta)^2} \int_0^1 d t {\left(- \log {(e^{2 \sigma} - 1)(1 -t) t \over e^{2 \sigma} (1-t) + t }\right)^{- 2 \Delta} \over t (1 - t)} .
}
Next we use the saddle point approximation to evaluate the $t$ integral.
The extremal value for $t$ is 
\eqn\textremum{
t_0 = 1- \left( {2 \Delta - 1 \over 2 \Delta +1} \right)^{1 \over 2} e^{- \sigma} + O(e^{- 2 \sigma})~.}

Plugging it back to~\extremumins\ we see that the dominant spin, $s_{{\rm dom}}$, is
\eqn\domin{
\log s_{{\rm dom}} = \sigma + \log {(2 \Delta - 1)^{{ 3 \over 2 }} (2 \Delta + 1)^{ {1 \over 2} } \over 4 \Delta} + O(e^{- \sigma})~.}

The leading term in the relation \domin\ coincides with the identification of \AldayMF. Notice that for $\Delta \gg 1$ the expression~\domin\ further reduces to
\eqn\simplelargeDelt{
\log s_{{\rm dom}} \approx \sigma + \log \Delta~.
}

It is important to remark that the saddle point above is only localized when $\Delta\gg1$, hence, one cannot perform reliable computations (for finite $\Delta$) by the usual saddle point approximation method (one needs to re-sum various corrections). The purpose of this discussion is to motivate the claim that for $\bar z\rightarrow 1$ the important terms in the sum~\confblocks\ are those with large spin, according to the relation $\log s_{{\rm dom}} \approx \sigma$. 

In the next subsection we will present a more rigorous approach to the problem. This approach allows us to control the sum~\confblocks\ efficiently. The following discussion is crucial for reproducing the main results of section 3.

\subsec{A Systematic Approach}

In the previous subsection we learned that the spins dominating the s-channel expansion which  reproduces the unit operator in the t-channel are given by  $s\sim {1\over (1-\bar z)^{\half}}$. We denote $\epsilon=1-\bar z$ and plug this into the hypergeometric function. Below we will discuss the leading terms are $\epsilon\rightarrow 0$. 
From the integral representation of the hypergeometric function we see that it behaves as follows in this scaling limit ($A$ is an arbitrary coefficient that stays finite as $\epsilon\rightarrow0$)
\eqn\intrepx{_2F_1\left({A\over \sqrt\epsilon},{A\over \sqrt\epsilon},2{A\over \sqrt\epsilon},1-\epsilon\right)={\Gamma({2A\over\sqrt \epsilon})\over \Gamma^2({A\over\sqrt \epsilon})  }\int_0^1 dt (1-t)^{{A\over \sqrt\epsilon}-1}t^{{A\over \sqrt\epsilon}-1}(1-t+\epsilon t)^{-{A\over \sqrt\epsilon}}~.   }
We approximate the pre-factor using the Stirling formula as $\sqrt{{A\over 4\pi}}{1\over \epsilon^{1/4}} e^{{2A\log(2)\over \sqrt \epsilon}}$ and so
\eqn\intrepxi{_2F_1\left({A\over \sqrt\epsilon},{A\over \sqrt\epsilon},2{A\over \sqrt\epsilon},1-\epsilon\right)=  \sqrt{{A\over 4\pi}}{1\over \epsilon^{1/4}} 4^{{A\over \sqrt \epsilon}}     \int_0^1 dt (1-t)^{{A\over \sqrt\epsilon}-1}t^{{A\over \sqrt\epsilon}-1}(1-t+\epsilon t)^{-{A\over \sqrt\epsilon}} ~.}
Now we need to take the $\epsilon\rightarrow 0$ limit of this integral.

From the analysis of the previous subsection we are inspired to make a change of variables $1-t=\lambda\sqrt\epsilon$ and find 
\eqn\intrepxi{\eqalign{ _2F_1\left({A\over \sqrt\epsilon},{A\over \sqrt\epsilon},2{A\over \sqrt\epsilon},1-\epsilon\right)=  \sqrt{{A\over 4\pi}}{1\over \epsilon^{1/4}}   4^{{A\over \sqrt \epsilon}} \int_0^{1/\sqrt\epsilon} {d\lambda\over \lambda} (1-\lambda\sqrt\epsilon)^{{A\over \sqrt\epsilon}-1}(1+{\sqrt\epsilon\over \lambda} )^{-{A\over \sqrt\epsilon}}   ~.  } }
We take the $\epsilon\rightarrow 0$ limit and get 
\eqn\intrepxii{\eqalign{& _2F_1\left({A\over \sqrt\epsilon},{A\over \sqrt\epsilon},2{A\over \sqrt\epsilon},1-\epsilon\right)=  \sqrt{{A\over 4\pi}}{1\over \epsilon^{1/4}}   4^{{A\over \sqrt \epsilon}} \int_0^{\infty} {d\lambda\over \lambda} e^{-\lambda A-{A\over \lambda}}   ~.  } }
This is a nicely convergent integral, given by the modified Bessel function of the second kind. Thus
\eqn\intrepxiii{\eqalign{& _2F_1\left({A\over \sqrt\epsilon},{A\over \sqrt\epsilon},2{A\over \sqrt\epsilon},1-\epsilon\right)\rightarrow \sqrt{{A\over \pi}}{4^{{A\over \sqrt \epsilon}} \over \epsilon^{1/4}}  K_0(2A)  ~.  } }

We can now use this limiting expression and reconsider~\confblocks\ using this approximation for the hypergeometric function. We find 
\eqn\sintii{f(\bar z)={4\over \Gamma^2(\Delta) \epsilon^{\Delta}}\int_{0}^\infty dA   A^{2\Delta-1} K_0(2A) ~.}
It is encouraging to see that the dependence on $\epsilon$ is correct in the small $\epsilon$ limit. Now let us consider the pre-factor. This would be another test for the fact that the procedure above indeed captures the important contributions in the small $\epsilon$ limit.  
We use the integral $\int_{0}^\infty dA   A^{2\Delta-1}  K_0(2A) ={1\over 4}\Gamma^2(\Delta)$. Plugging this back into~\sintii\ we find 
\eqn\sintfinal{f(\bar z)={1\over\epsilon^{\Delta}}\left(1+\CO(\epsilon)\right)~.}
This is precisely the correct asymptotics for small $\epsilon$, including the pre-factor.

\subsec{The Coefficient $c_{\tau_{\min}}$ in~(3.12)}

In the main text in section 3 we have seen that since the correction to the anomalous dimension is of the form $e^{- \tau_{\min}}$ and $\log s_{{\rm dom}} = \sigma$,
we expect the anomalous twist at large $s$ to take the form ${c_{\tau_{\min}} \over s^{\tau_{\min}} }$. Here we would like to compute $c_{\tau_{\min}}$ exactly. The idea is that the small corrections to the anomalous twists in the s-channel reproduce the contribution of the operator $\CO_{min}$ in the t-channel. This contribution in the t-channel is given by 
\eqn\allprefactorstoget{
-{\Gamma (\tau_{\min} +  2 s_{\min}) \over (-2)^{s_{\min}} \Gamma \left({\tau_{\min} + 2 s_{\min} \over 2}\right)^2}  { C_{{\cal O} {\cal O} {\cal O}_{\tau_{\min}}}^2 \over  C_{{\cal O} {\cal O} }^2  C_{{\cal O}_{\tau_{\min}}{\cal O}_{\tau_{\min}}}}  \log z \ \   e^{(2 \Delta- \tau_{\min}) \sigma}~.
}

It is easy to match this in the s-channel because of the $\log z$. We reproduce this logarithm by summing the corrections to the twists in the s-channel  $z^{\Delta + { \delta \tau_{s} \over 2} } \sim  { \delta \tau_{s} \over 2} z^{\Delta} \log z$. Setting $\delta \tau_{s} = {c_{\tau_{\min}} \over s^{ \tau_{\min} } }$, we get the basic equation that determines $c_{\tau_{\min}}$
\eqn\equation{\eqalign{
-&{2 \over c_{\tau_{\min}} }{\Gamma (\tau_{\min} +  2 s_{\min}) \over (-2)^{s_{\min}} \Gamma \left({\tau_{\min} + 2 s_{\min} \over 2}\right)^2} { C_{{\cal O} {\cal O} {\cal O}_{\tau_{\min}}}^2 \over  C_{{\cal O} {\cal O} }^2  C_{{\cal O}_{\tau_{\min}}{\cal O}_{\tau_{\min}}}}   e^{(2 \Delta- \tau_{\min}) \sigma} \cr
&=  \sum_{s=0}^{\infty} {c_{s} \over s^{\tau_{\min}} } \bar z ^{\Delta + s} \ _{2} F_{1} (\Delta + s, \Delta + s, 2 s + 2 \Delta, \bar z), \cr
}}
where $\bar z = 1- e^{-2 \sigma}$ and the equality should be understood in terms of {\it leading pieces in $\sigma$}. 

Solving for $c_{\tau_{\min}}$ we get
\eqn\solvforc{\eqalign{
c_{\tau_{\min}} &= -2 {\Gamma (\tau_{\min} +  2 s_{\tau_{\min}}) \over (-2)^{s_{\tau_{\min}}} \Gamma \left({\tau_{\min} + 2 s_{\tau_{\min}} \over 2}\right)^2}  { C_{{\cal O} {\cal O} {\cal O}_{\tau_{\min}}}^2 \over  C_{{\cal O} {\cal O} }^2  C_{{\cal O}_{\tau_{\min}}{\cal O}_{\tau_{\min}}}} f(\tau_{\min}, \Delta)~, \cr
f(\tau_{\min}, \Delta)&= \lim_{\sigma \to \infty} {  e^{(2 \Delta- \tau_{\min}) \sigma} \over \sum_{s=\Lambda}^{\infty} {c_{s} \over s^{\tau_{\min}} } \bar z ^{\Delta + s} \ _{2} F_{1} (\Delta + s, \Delta + s, 2 s + 2 \Delta, \bar z)}~.
}}
and since the sum is dominated by the large spins we expect the result to be independent of $\Lambda$.

We compute $f(\tau_{\min}, \Delta)$ by performing the sum in the denominator carefully. We first switch to an integral and repeat the procedure of the previous subsection, including the double scaling limit for the hypergeometric function~\intrepxiii.  Doing the integral we finally find \eqn\answer{
f(\tau_{\min}, \Delta) = {\Gamma(\Delta)^2 \over \Gamma(\Delta - {\tau_{\min} \over 2})^2}~.
}

\subsec{The Most General (Scalar) Case}

We are interested in generalizing the computation of the correction $c_{\tau_{\min}}$ to the case when not all the operators are identical and $\Delta_1 \neq \Delta_2$. This is relevant for a variety of applications in the main body of the text. 
More precisely, we are considering the following correlation function $\la {\cal O}_1(x_1) {\cal O}_2 (x_2) {\cal O}_2(x_3) {\cal O}_1 (x_4) \ra$. What we call the s-channel is the OPE $12 \to 34$. What we call the t-channel is the OPE $23 \to 14$. Let us write the OPE expansion in each of the channels.

For the s-channel OPE we have
\eqn\schannelOPE{
\la {\cal O}_1(x_1) {\cal O}_2 (x_2) {\cal O}_2^{\dagger}(x_3) {\cal O}_1^{\dagger} (x_4) \ra = \left({ x_{24} x_{13} \over x_{14}^2} \right)^{\Delta_1 - \Delta_2} \sum_{\Delta , s} {p_{\Delta , s} g_{\Delta , s} (z, \bar z) \over x_{12}^{\Delta_1 + \Delta_2} x_{34}^{\Delta_1 + \Delta_2}}~.
} And for the t-channel OPE we have 
\eqn\schannelOPE{
\la {\cal O}_2^{\dagger}(x_3) {\cal O}_2 (x_2) {\cal O}_1(x_1) {\cal O}_1^{\dagger} (x_4) \ra = \sum_{\Delta , s} {\tilde p_{\Delta , s} g_{\Delta , s} (1-z, 1-\bar z) \over x_{23}^{2 \Delta_2} x_{14}^{2 \Delta_1}}~.
}
The crossing equation takes the form
\eqn\crossingequation{
\sum_{\Delta , s} p_{\Delta , s} g_{\Delta , s} (z, \bar z) ={z^{{\Delta_1 + \Delta_2 \over 2 }} \bar z^{{\Delta_1 + \Delta_2 \over 2 }} \over (1-z)^{\Delta_2} (1 - \bar z)^{\Delta_2}} \sum_{\Delta , s} \tilde p_{\Delta , s} g_{\Delta , s} (1-z, 1-\bar z)~.
}

As before  we focus on the unit operator in the t-channel. It is dominant when $\bar z \to 1$ so that the RHS at leading order takes the form ${z^{{\Delta_1 + \Delta_2 \over 2 }} \over (1 - \bar z)^{\Delta_2}}$ where we again keep only the leading piece in the $z$ expansion. This is so that we only have to deal with the collinear conformal blocks instead of the full ones. 

The collinear conformal blocks in this case are slightly different from the case $\Delta_1=\Delta_2$, so that we get the following equation
\eqn\equation{
\lim_{\bar z \to 1} \sum_{s = \Lambda}^{\infty} c_s \bar z ^s \ _{2} F_1 (s + \Delta_2, s+\Delta_2, 2 s + \Delta_1 + \Delta_2) = {1 \over  (1 - \bar z)^{\Delta_2} }~.
}
We know that the coefficients of  the theory of generalized free fields solve this problem. They could be easily found using the results of~\FitzpatrickDM\ 
\eqn\threepgen{
c_s = {\Gamma ( \Delta_1 + s) \Gamma (\Delta_2 + s) \Gamma (\Delta_1 + \Delta_2 + s -1) \over \Gamma (s+1) \Gamma (\Delta_1) \Gamma (\Delta_2) \Gamma (\Delta_1 + \Delta_2 + 2 s -1)}~.}

The limit $\bar z\rightarrow 1$ in~\equation\ is again dominated by large spins with the scaling as before. 
We have a similar double scaling limit of the hypergeometric function
\eqn\approx{
\ _{2} F_1 ({s\over \sqrt{\eps}} + \Delta_2, {s\over \sqrt{\eps}}+\Delta_2, 2 {s\over \sqrt{\eps}} + \Delta_1 + \Delta_2) \sim 2^{\Delta_1 + \Delta_2} \eps^{{\Delta_1 - \Delta_2 \over 2}} 4^{{s \over \sqrt{\eps}}} {\sqrt{s} K_{\Delta_2 - \Delta_1} (2 s)  \over \sqrt{\pi} \eps^{1 \over 4}}~.
}
Switching from the sum~\equation\ to an integral we get
\eqn\integral{
{4 (1- \bar z)^{- \Delta_2} \over \Gamma( \Delta_1) \Gamma (\Delta_2)} \int_{0}^{\infty} d s \ s^{\Delta_1 + \Delta_2 - 1} K_{\Delta_2 - \Delta_1}(2 s ) = {1 \over (1- \bar z)^{\Delta_2}}~.
}
This is analogous to~\sintii\ and \sintfinal. It serves to check that we have understood correctly how to resum  the hypergeomtric functions on the LHS of~\equation\ in this subtle limit.

To compute the correction to anomalous dimension we need a slightly different integral
\eqn\integral{
{4 \over \Gamma( \Delta_1) \Gamma (\Delta_2)} \int_{0}^{\infty} d s \ s^{\Delta_1 + \Delta_2 - \tau_{\min} - 1} K_{\Delta_2 - \Delta_1}(2 s ) = {\Gamma (\Delta_1 - { \tau_{\min} \over 2 } ) \over \Gamma (\Delta_1)} {\Gamma (\Delta_2 - { \tau_{\min} \over 2 } ) \over \Gamma (\Delta_2)}~.
}
So that the crossing equation becomes
\eqn\crossingeq{
c_{\tau_{\min}}{\Gamma (\Delta_1 - { \tau_{\min} \over 2 } ) \over \Gamma (\Delta_1)} {\Gamma (\Delta_2 - { \tau_{\min} \over 2 } ) \over \Gamma (\Delta_2)}  = 2 f^2 {\Gamma (\tau_{\min} +  2 s_{\min}) \over (-2)^{s_{\min}} \Gamma \left({\tau_{\min} + 2 s_{\min} \over 2}\right)^2}~,
}
and we have for the correction to the twist of $\CO_1 \pa^s \CO_2$ in this case
\eqn\solcrossdiff{\eqalign{
c_{\tau_{\min}} &= 2 f^2 {\Gamma (\tau_{\min} +  2 s_{\min} ) \over (-2)^{s_{\min}} \Gamma \left({\tau_{\min} + 2 s_{\min} \over 2}\right)^2} {\Gamma (\Delta_1 ) \over \Gamma (\Delta_1  - { \tau_{\min} \over 2 } )} {\Gamma (\Delta_2 ) \over \Gamma (\Delta_2  - { \tau_{\min} \over 2 })}~, \cr
f^2 &=  { C_{{\cal O}_1 {\cal O}_1^{\dagger} {\cal O}_{\tau_{\min}}} C_{{\cal O}_2^{\dagger} {\cal O}_2 {\cal O}_{\tau_{\min}}} \over  C_{{\cal O}_1 {\cal O}_1^{\dagger} } C_{{\cal O}_2 {\cal O}_2^{\dagger} }  C_{{\cal O}_{\tau_{\min}}{\cal O}_{\tau_{\min}}}}~.
}} 
If we take $\CO_2=\CO_1^\dagger$, the coefficient $c_{\tau_{\min}}$ is positive for any intermediate operator $\CO_{\min}$.

\appendix{C}{OPE of Chiral Primary Operators in SCFT}

Here we discuss $d=4$ $\CN=1$ SCFTs (see also~\PolandWG\ and references therein). The superconformal algebra contains as a bosonic sub-algebra the conformal algebra times $U(1)_R$.  Performing radial quantization, we label representations by the quantum numbers  $\left(\Delta, j_1, j_2, R\right)$, where $\Delta$ is the dimension of the operator.  

Superconformal primaries are annihilated by $S_\alpha$ and $\bar S_\alphadot$. This implies that they are annihilated by special conformal transformations as well, hence, they are primaries in the usual sense.  The complete representation can be constructed by acting with $Q$ and $\bar Q$ on the superconformal primary.

The quantum numbers of $Q_\alpha$ are $(1/2,1/2,0,-1)$ and of $\bar Q_\alphadot$ are $(1/2,0,1/2,1)$.
Given a superconformal primary in the representation $(\Delta, j_1, j_2, R)$, we can act on it with $Q_\alpha$ and find states in the representation $(\Delta+\half, j_1-\half, j_2, R-1)\oplus (\Delta+\half, j_1+\half, j_2, R-1)$. Similarly, we can act with $\bar Q_\alphadot$ and find states in 
$(\Delta+\half, j_1, j_2-\half, R+1)\oplus (\Delta+\half, j_1, j_2+\half, R+1)$.

In both cases the irreducible representation with the smaller norm is then one with the smaller spins. Demanding therefore that the states above have non-negative norm we find two inequalities 
\eqn\twoineq{\eqalign{&\Delta+2\delta_{j_1,0}-\left(2+2j_1-{3r\over 2}\right)\geq 0~,\cr&
\Delta+2\delta_{j_2,0}-\left(2+2j_2+{3r\over 2}\right)\geq 0}~.}
If one of the inequalities above is saturated, the superconformal primary is annihilated by some supercharges.

In the special case that either $j_1=0$ or $j_2=0$ (or both), there is more information that can be extracted by considering states at level 2. (This is similar to the case $j_1=j_2=0$ in the ordinary conformal group.) 
Let us take for example $j_1=0$. Then states at level 2 transform in $(\Delta+1,0, j_2, R-2)$.  The norm of these states is such that the window 
\eqn\windon{-{3r\over 2}<\Delta<2-{3r\over 2} }
is {\it excluded}. The norm of this state at level 2 precisely vanishes when $\Delta=-3r/2$ or $\Delta=2-3r/2$.
Similarly, if $j_2=0$, the disallowed range is ${3r\over 2}<\Delta<2+{3r\over 2}$.

Chiral primaries are annihilated by all the $\bar Q_\alphadot$, which means they carry $j_2=0$, and thus satisfy $\Delta=3r/2$.

We will now use these facts to analyze the operator product expansion of two chiral primaries $\Phi(x) \Phi(0)$. It is well known that there are no singularities in this OPE. Let us examine all the possible primaries that can appear in this OPE. (The operators that appear on the right hand side are not necessarily superconformal primaries.) We denote $R(\Phi)\equiv R_\Phi=2\Delta_\Phi/3$.

\item{\bf A} Chiral primaries, i.e. states with $ j_2=0$. Those must have $R=2R_\Phi$ and thus $\Delta=2\Delta_\Phi$. So this is just the operator $\Phi^2$ in the OPE. It has $j_1=j_2=0$.

\medskip 

The other operators in the OPE must still be chiral (because the LHS is chiral) but they cannot be chiral primaries. So we have to write general supeconformal descendants that are chiral (annihilated by $\bar Q$) and are primaries of the usual conformal group.  So we can have 

\item{\bf B} $\bar Q^{( \alphadot_1} \CO^{\alphadot_2...\alphadot_l), (\alpha_1,...,\alpha_l)}$ with $\CO$ a superconfromal primary. This is a conformal primary, and to make sure it is chiral we need the superconformal primary to obey $\bar Q_{ \alphadot} \CO^{(\alphadot...\alphadot_l), (\alpha_1,...,\alpha_l)}=0$. This first order condition guarantees that $\bar Q^{( \alphadot_1} \CO^{\alphadot_2...\alphadot_l), (\alpha_1,...,\alpha_l)}$ is a chiral field. By matching the $R$-charge we also find $R_\CO=2R_\Phi-1$. The shortening condition and the $R$-charge imply immediately from~\twoineq\  that $\Delta_\CO=2+2j_2+3R_\Phi-3/2$. We dropped the term $\delta_{j_2,0}$ because we know that $\CO$ has to have half-integer spin in the $\alphadot$ indices. (In other words, l is even.) 
Finally, we substitute $j_2=l/2-1/2$ and find $\Delta_\CO=l+2\Delta_\Phi-1/2$, and hence the dimension of $\bar Q^{( \alphadot_1} \CO^{\alphadot_2...\alphadot_l), (\alpha_1,...,\alpha_l)}$ is $l+2\Delta_\Phi$. This means that the twist of these operators is precisely $2\Delta$.\foot{Another closely related class of operators one could think of would be obtained by antisymmetrizing the spinor index of $\bar Q$ with some superconformal primary $\CO$ and imposing a first order equation that the symmetrization gives zero. (This equation is necessary for chirality.) This is clearly inconsistent, since, as we explained, among the descendants, the state of lowest norm is always the state of smallest spin.}

\item{\bf C}  We can have $\bar Q^2 \CO^{(\alphadot_1...\alphadot_l), (\alpha_1,...,\alpha_l)}$ for $\CO$ superconformal primary. We immediately find that $R_\CO=2R_\Phi-2$ and plugging this into the unitarity bounds we obtain $\Delta(\bar Q^2 \CO)\geq 1+2+l+3R_\Phi-3=l+2\Delta$ and $\Delta(\bar Q^2 \CO)\geq 1+2+l-3R_\Phi+3=l+6-2\Delta$. Either way, the twists of these operators are clearly not smaller than $2\Delta$.\foot{One can in fact argue the twist must be strictly larger than $2\Delta$. Suppose the twist were $2\Delta$. Then the second inequality in~\twoineq\ would have been saturated. In this case $\bar Q_\alphadot \CO^{(\alphadot...\alphadot_l), (\alpha_1,...,\alpha_l)}=0$. But in this case $\bar Q^2 \CO^{(\alphadot_1...\alphadot_l), (\alpha_1,...,\alpha_l)}=0$ and thus there is no such operator in the OPE.}

\medskip

\appendix{D}{DIS for Traceless Symmetric Representations}

Consider a symmetric traceless operator of spin $s$, ${\cal O}_{\mu_1 ... \mu_{ s} }(y)$. We can contract it with a light like 
complex polarization vector $\zeta^{\mu}$, $\zeta^2 = 0$. We denote the result as ${\cal O}(\zeta,y)$. The object we are interested in is 
\eqn\scattamplitude{\eqalign{
A (\nu, q^2,\zeta) &={i \over \pi} \int d^d y e^{i q y} \la P | {\rm T} \left( {\cal O}(\zeta^*,y) {\cal O} (\zeta,0) \right) | P \ra~,\qquad
\nu = 2q\cdot P~.}}
The imaginary part of $\CA$ is related to the total cross section for DIS. 

Let us consider the most general possible form for the operator product expansion
\eqn\opescal{
{\cal O}(\zeta^*,y) {\cal O} (\zeta,0)= \sum_{s=0,2,4...}\ \sum_{\alpha\in \CI_s} f_{s}^{(\alpha)}(y,\zeta, \zeta^*)^{\mu_1 ... \mu_s} {\cal O}_{\mu_1 ... \mu_s}^{(\alpha)}(0)~,
}
where the symbols $s,\CI_s$ represent the same objects as in section~2. If the operator ${\cal O}_{\mu_1 ... \mu_{ s} }(x)$ is conserved that would constrain the allowed structures in~\opescal.
In the OPE~\opescal\ we retain only primary symmetric traceless operators on the right hand side. The other operators are discarded because they do not contribute to DIS as long as the target particle $|P\rangle$ is a scalar. (We will henceforth assume the target is a scalar particle for simplicity.)

We substitute the OPE expansion~\opescal\ in~\scattamplitude\  and evaluate the expectation values of the operators  according to
\eqn\VEVs{\langle P|\CO^{\alpha}_{\mu_1...\mu_s}(0)|P\rangle=\CA_{n}^{(\alpha)}\left(P_{\mu_1}P_{\mu_2}\cdots P_{\mu_s}-traces\right)~.}
The kinematics is a little more complicated than in the scalar case of section~2 because of the polarization vector $\zeta$. However, there is a simple choice for $\zeta$ which makes the problem virtually isomorphic to the scalar case. We can always pick 
\eqn\specialzeta{\zeta\cdot P=0~.}
In this case, the contribution from the $P_{\mu_1}P_{\mu_2}\cdots P_{\mu_s}$ term in~\VEVs\ automatically selects the same kinematic structure as in the scalar case, and all the arguments from section~2 go through.\foot{Again, we need to be careful that the ``trace'' terms in~\VEVs\ do not overwhelm contributions coming from lower spins (with a different power of $\nu$). In section~2 we have already analyzed this for the case $f_{s}^{(\alpha)}(y,\zeta, \zeta^*)^{\mu_1 ... \mu_s}\sim y^{\mu_1}\cdots y^{\mu_s}$. Here we could obtain new structures, such as
$f_{s}^{(\alpha)}(y,\zeta, \zeta^*)^{\mu_1 ... \mu_s}\sim \zeta^{\mu_1}(\zeta^*)^{\mu_2}y^{\mu_3}\cdots y^{\mu_s}$. Due to~\specialzeta, this could contribute only when dotted into the ``trace'' terms, e.g. $g_{\mu_1\mu_2}P_{\mu_3}\cdots P_{\mu_s}$. A simple calculation shows that this would scale like (at large $-q^2$) $x^{-s+2}(q^2)^{-\half\tau_s^*+\Delta_\CO-d/2}$. Comparing this to the already existing contribution from spin $s-2$, $x^{-s+2}(q^2)^{-\half\tau_{s-2}^*+\Delta_\CO-d/2}$, we see that again it is sufficient that the twists $\tau^*_s$ are nondecreasing, as found in~\nondec. } One finds the same sum rules and obtains convexity under the same assumptions as in section 2.

\appendix{E}{The ${1 \over s^2}$ Correction in the Critical $O(N)$ Model}

Here we would like to show how one can use the general formula for the coefficient $c_{\tau_{\min}}$ in order to compute the ${1 \over s^2}$ tem in~\currentslNb. 
The correction $1/s^2$ arises from the exchange of the $\sigma^2$ operator in the t-channel.

To contrast~\currentslNb\ with our formula~\correctionformx\ we need to compute the three-point function $\la \sigma_i (x_1) \sigma_j (x_2) \sigma^2 (x_3) \ra $. 
For the two-point functions we have~\GiombiWH
\eqn\twopf{\eqalign{
\la \sigma_i (x) \sigma_j (0) \ra &= {\delta_{i j} \over N} {\Gamma({d \over 2} - 1) \over 4 \pi^{d \over 2}} {1 \over x^{d-2}} = {\gamma_{s} \over N}  {1 \over x^{d-2}}~, \cr
\la \sigma^2(x) \sigma^2(0)  \ra &= {\gamma_{\phi^2} \over N} {1 \over x^4}~, ~~~
\gamma_{\phi^2} = {2^{d+2} \sin \left( {\pi d \over 2} \right) \ \Gamma ({d -1  \over 2}) \over \pi^{3 \over 2} \Gamma({d \over 2} - 2)}~.
}}

\ifig\figthirt{The diagram for the three-point function $\la \sigma_i (x_1) \sigma_i (x_2) \sigma^2(x_3) \ra$.} {\epsfxsize1.8in\epsfbox{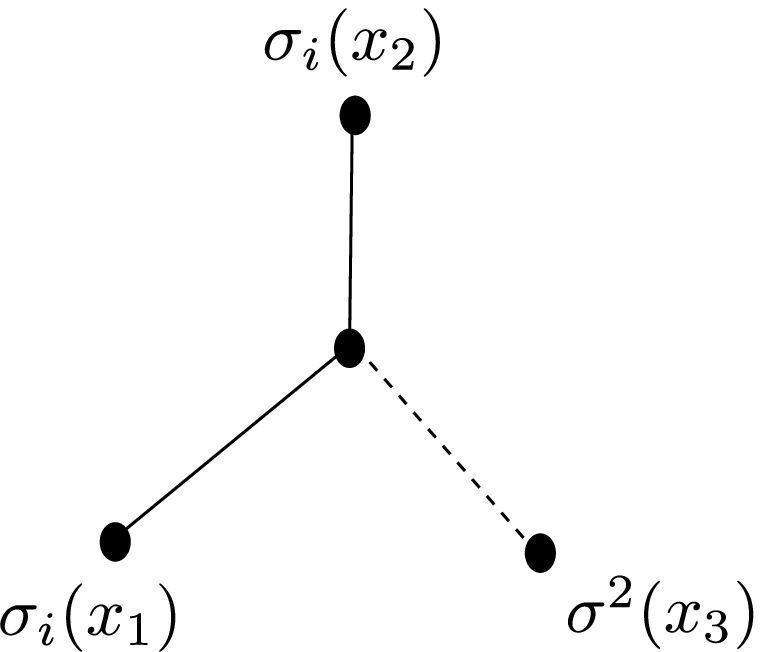}}

The three-point function $\la \sigma_i (x_1) \sigma_i (x_2) \sigma^2(x_3) \ra$ at leading order is given by the diagram~\figthirt, where the interaction vertex is $-{N \over 2} \sigma_i^2$. So that we get
\eqn\correction{\eqalign{
\la \sigma_i (x_1) \sigma_i (x_2) \sigma^2(x_3) \ra &= - {\gamma_{\sigma^2} \gamma_{s}^2 \over N^2} \int d^d x_0 {1 \over x_{10}^{d-2} x_{20}^{d-2} x_{30}^4 } \cr
&=  - {\gamma_{\sigma^2} \gamma_{s}^2 \over N^2} {\pi^{d \over 2} \Gamma ({d \over 2} - 2) \over \Gamma ({d -2 \over 2})^2} {1 \over x_{12}^{d-4} x_{13}^{2} x_{23}^{2}}~.
}}
(We used the standard formula~\KazakovKM, and there is no summation over $i$ ). Thus, we get
\eqn\threep{
{C_{\sigma \sigma \sigma^2}^2 \over C_{\sigma^2 \sigma^2} C_{\sigma \sigma}^2 } = {1 \over N} \gamma_{\sigma^2}  \gamma_{s}^2  {\pi^{d} \Gamma ({d \over 2} - 2)^2 \over \Gamma ({d -2 \over 2})^4}~.
}

Now we have everything we need to apply~\correctionformx. (Notice that in this case $\tau_{\min} > 2 \Delta$, but we can think about our formula in terms of an analytic continuation.) Thus we set $s_{\min} = 0$, $\tau_{\min}= 2$ and $\Delta = {d - 2 \over 2}$ to get
\eqn\answer{
c_{\sigma^2} = {2^{d-1} \Gamma ({d -1 \over 2}) \sin ( {\pi d \over 2} ) \over N \pi^{3/2} \Gamma({d \over 2} - 2)}~,
}
which can be easily checked to reproduce the result~\currentslNb\ precisely.

\listrefs

\end